\documentclass[useAMS,usenatbib]{mnras}
\usepackage{times}
\usepackage{graphicx, latexsym, amsmath, amssymb, amscd, psfrag, cancel}
\usepackage{epsfig}
\usepackage{color}
\title[The Complete Local Volume Groups Sample III]{The Complete Local Volume Groups Sample - III. Characteristics of group central radio galaxies in the Local Universe}

\author[Konstantinos Kolokythas et al.]{Konstantinos Kolokythas,$^{1}$\thanks{e-mail: kkolok@iucaa.in}  Ewan O'Sullivan,$^{2}$ Huib Intema,$^{3,4}$ 
\newauthor{Somak Raychaudhury,$^{1,5,6}$ Arif Babul,$^{7,8}$ Simona Giacintucci,$^{9}$ Myriam Gitti$^{10,11}$}  \\ \\
$^{1}$Inter-University Centre for Astronomy and Astrophysics, Pune University Campus, Ganeshkhind, Pune, Maharashtra 411007, India\\
$^{2}$Harvard-Smithsonian Center for Astrophysics, 60 Garden Street, Cambridge, MA 02138, USA\\
$^{3}$International Centre for Radio Astronomy Research, Curtin University, Bentley, WA 6102, Australia\\
$^{4}$Leiden Observatory, Leiden University, Niels Bohrweg 2, 2333 CA Leiden, The Netherlands\\
$^{5}$School of Physics and Astronomy, University of Birmingham, Birmingham B15~2TT, UK\\
$^{6}$Department of Physics, Presidency University, 86/1 College Street, Kolkata 700073, India\\
$^{7}$Department of Physics and Astronomy, University of Victoria, Victoria, BC V8P 1A1, Canada\\
$^{8}$Center for Theoretical Astrophysics and Cosmology, Institute for Computational Science, University of Zurich, Winterthurerstrasse 190, 8057 Zurich, Switzerland\\
$^{9}$Naval Research Laboratory, 4555 Overlook Avenue SW, Code 7213, Washington, DC 20375, USA\\
$^{10}$Dipartimento di Fisica e Astronomia, Universit\'a di Bologna, via Gobetti 93/2, 40129 Bologna, Italy \\
$^{11}$INAF, Istituto di Radioastronomia di Bologna, via Gobetti 101, 40129 Bologna, Italy}

\newcommand{\kms}{~km~s$^{-1}$}

\newcommand\kmsmpc{km~s$^{-1}$~Mpc$^{-1}$}

\begin{document}

\date{Accepted 2019 July 22. Received 2019 July 17; in original form 2019 May 31}

\pagerange{\pageref{firstpage}--\pageref{lastpage}} \pubyear{2019}
\maketitle

\label{firstpage}
 \begin{abstract}
Using new 610~MHz and 235~MHz observations from the Giant Metrewave Radio Telescope (GMRT) in combination with archival GMRT and Very Large Array (VLA) survey data we present the radio properties of the dominant early--type galaxies in the low-richness sub-sample of the Complete Local-volume Groups Sample (CLoGS; 27 galaxy groups) and provide results for the radio properties of the full CLoGS sample for the first time. We find a high radio detection rate in the dominant galaxies of the low-richness sub-sample of 82\% (22/27); for the full CLoGS sample the detection rate is 87\% (46/53). The group-dominant galaxies exhibit a wide range of radio power, 10$^{20}$ $-$ 10$^{25}$~W~Hz$^{-1}$ in the 235 and 610~MHz bands, with the majority (53\%) presenting point-like radio emission, 19\% hosting currently active radio jets, 6\% having remnant jets, 9\% being diffuse and 13\% having no detected radio emission. The mean spectral index of the detected radio sources in the 235$-$610~MHz frequency range is found to be $\alpha_{235}^{610}\sim$0.68, and $\alpha_{235}^{1400}\sim$0.59 in the 235$-$1400~MHz one. In agreement with earlier studies, we find that the fraction of ultra-steep spectrum sources ($\alpha>$1.3) is $\sim$4\%, mostly dependent on the detection limit at 235~MHz. The majority of point-like systems are found to reside in dynamically young groups, whereas jet systems show no preference between spiral-rich and spiral-poor group environments. The mechanical power of the jet sources in the low--richness sample groups is estimated to be $\sim$10$^{42}$ $-$ 10$^{44}$ erg s$^{-1}$ with their black hole masses ranging between 2$\times$10$^{8}$ $-$ 5$\times$10$^{9}$ M$_{\odot}$. We confirm previous findings that, while radio jet sources tend to be associated with more massive black holes, black hole mass is not the decisive factor in determining jet activity or power.

 \end{abstract}
 

 
 \section{Introduction} 
 
 The bulk of galaxies and baryonic matter in the local Universe is found in galaxy groups \citep{GellerHuchra83,Fukugita98,Eke05}. Typically, they extend less than a Mpc with a mass range of 10$^{12.5}$ $-$ 10$^{14}$ M$_{\odot}$ \citep{HuchraGeller82}, about an order of magnitude less massive than galaxy clusters. Galaxy groups exhibit shallow gravitational potential wells, with their members being at close distances and at low relative velocities. These parameters are essential in driving the transformation processes (e.g., mergers and tidal interactions) of galaxy evolution \citep[e.g.,][]{Alonso12}. Galaxies are thought to be `pre-processed' in the group environment before they become parts of clusters \citep[e.g.][]{vandenBosch14,Haines17} hence the association between rich cluster galaxies and their evolution is intimately linked to the group environment \citep[e.g.][]{Bekki99,MossWhittle00}. Since the properties of galaxies and their evolution depends on their local environment, galaxy groups are an ideal environment where the study of galaxy formation and evolution is of uttermost significance \citep[e.g.,][]{Forbes06,Sun12}.

 Many groups maintain extensive halos of hot gas with short central cooling times \citep[e.g.,][]{OSullivan17}, which can fuel both star formation and active galactic nuclei (AGN). However, the cool gas can either be removed from the central region via AGN outflows \citep{Alexander10,Morganti13} or heated up, eventually suppressing star formation and galaxy growth. The imprints of such a heating mechanism (`feedback') capable of balancing the radiative losses in central dominant early-type galaxies, are evident via the interaction of radio-loud AGN activity and the surrounding hot X-ray gas, leading to excavation of cavities \citep[e.g.,][]{McNamara05,McNu07}. Their low masses mean that galaxy groups are the environment in which AGN feedback mechanism may have the greatest impact on the evolution of the hot diffuse gas.
 
 
 It has been suggested that in most groups and clusters, feedback operates in a near-continuous `bubbling' mode where thermal regulation is relatively gentle \citep{Birzan08,Birzan12,Panagoulia14}. However, there are cases that AGN feedback can also manifest via extreme AGN outbursts, potentially shutting down the central engine for long periods \citep[e.g., in Hydra A, MS 0735+7421, NGC~4261, NGC~193 \& IC~4296;][]{Rafferty06,Gitti12,Gitti07,OSullivan11,Kolokythas15,Kolokythas18,Grossova19}. Such powerful outbursts may have a significant effect on the development of groups, and it is therefore vital to study the energy output of radio galaxies in an unbiased set of low-mass systems in order to understand the impact of AGN in the group regime.
 
 


It is well known that galaxies at the centres of evolved groups are generally early-types, lacking strong ongoing
star formation \cite[e.g.,][]{Vaddi16}, with their properties depending strongly on the group halo mass \citep{DeLuciaBlaizot07,SkibbaSheth09,Gozaliasl18}, and the galaxy density (i.e., compact vs. loose groups).
Brightest group early-type galaxies (BGEs) as well as brightest cluster galaxies (BCGs), exhibit different surface brightness profiles and scaling relations from field or satellite massive galaxies \cite[e.g.,][]{Graham96,Bernardi07} and play an important role in the evolution of galaxy groups.  They are also found to exhibit different morphologies, star formation rates, radio emission, and AGN properties compared to satellite galaxies of the same stellar mass. Their privileged position near or at the centres of the extended X-ray emitting hot intra-group medium (IGM) and dark matter halo makes them suitable laboratories for constraining cosmological models, for studying the growth history of massive galaxies  \cite[e.g.,][]{Linden07,Liu09,Stott10,Liang16}, their super-massive black holes (e.g., \citealt{Rafferty06,Sabater19}), and the connection between galaxy properties such as stellar kinematics and the larger environment \citep{Loubser18}.


The radio detection rate of massive galaxies is found to be $\sim30$\% \citep[e.g.,][]{Best05,Shabala08}, but this is dependent on the sensitivity of the survey used and the redshift range considered. Studies of the BGEs/BCGS of groups and clusters in the local Universe finding detection rates of $\sim$80$-$90\%  \citep[e.g.,][]{Magliocchetti07,Dunn10,Kolokythas18} and this can even rise to 100\% for the most massive systems \citep{Sabater19}. It has been suggested that, in addition to fuelling of AGN by cooling from the IGM, the connection between radio AGN and higher galaxy densities \citep{Lilly09,Bardelli10,Malavasi15} may be driven by large scale merging (e.g., infall of groups into clusters) or by `inter-group' galaxy-galaxy interactions and mergers  \citep{Miles04,TaylorBabul05}. In such mergers and interactions gas can be channelled to the central AGN resulting in radio emission and the launching of jets. However, although AGNs dominate the radio emission from these massive galaxies, star formation may contribute in the less radio luminous objects, whose radio morphologies are often unresolved \citep[e.g.,][]{Smolcic17}.



In this paper we present results from the study of the radio properties of the dominant galaxies of the 27-group low--richness subset of the Complete Local-Volume Groups Sample (CLoGS), including new Giant Metrewave Radio Telescope (GMRT) 235 and 610~MHz observations of 25 systems. The CLoGS sample and the X-ray properties of the high--richness groups are described in more detail in \citet[][hereafter Paper~I]{OSullivan17} while the radio properties of the BGEs in the high--richness sub-sample are described in \citet[hereafter Paper~II]{Kolokythas18}. This paper continues the work described in paper~II, presenting the properties of the central radio sources in the low--richness sub-sample and discussing the whole CLoGS sample for the first time. We examine the connection between the group environment and the dominant radio galaxies, the contribution of star formation on the radio emission of pointlike radio sources, and provide a qualitative comparison of the radio emission that BGEs exhibit in high and low--richness group sub-samples. The paper is organized as follows: In Section 2 we briefly present the sample of galaxy groups, in Section 3 we describe the GMRT observations and the radio data analysis, in Section 4 we present the radio detection statistics of the BGEs, and in Section 5 their radio properties including information on the contribution of star formation on the radio emission. Section 6 contains the discussion of our results for the CLoGS sample as a whole, focusing on the detection statistics, the properties of the radio sources, their environment and their energetics. The summary and the conclusions are given in Section 7. Radio images and information on the central galaxies of this sample work are presented in Appendices~\ref{AppA} to \ref{AppC}. Throughout the paper we adopt the $\Lambda$CDM cosmology with $H_o=70$ \kmsmpc, $\Omega_m$ = 0.27, and $\Omega_\Lambda$ = 0.73. The radio spectral index $\alpha$ is defined as $S_\nu \propto \nu^{-\alpha}$, where $S_\nu$ is the flux density at the frequency $\nu$. In general we quote 1$\sigma$ uncertainties, and the uncertainties on our radio flux density measurements are described in Section~\ref{sec:GMRTanalysis}.



\section{The Complete Local-Volume Groups Sample}

The Complete Local-Volume Groups Sample (CLoGS) is an optically selected sample of 53 groups in the local universe (D$\le$80~Mpc), that is collected from the relatively not deep, all-sky Lyon Galaxy Group catalog (LGG; \citealt{Garcia93}). Paper~I provides a detailed description of the sample, its selection criteria, and the X-ray properties of the high--richness sub-sample, and we therefore only provide a brief summary of the selection here. The sample is statistically complete in the sense that, to the completeness limit of the LGG sample, it contains every group which meets our selection criteria. It is intended to be a representative survey of groups in the local universe including studies of their radio, X-ray and optical properties.


Constraints on the selection of CLoGS groups include: i) a membership of at least 4 galaxies, ii) the presence of $\geq$1 luminous early-type galaxy (L$_B$ $>$ 3$\times$10$^{10}$ L$_\odot$) and iii) a declination of $>$-30$^\circ$ in order to certify that the groups are observable from the GMRT and Very Large Array (VLA). The group membership was extended and refined using the HyperLEDA catalog \citep{Paturel03}, based on which the group mean velocity dispersion and richness $R$ parameter were estimated ($R$; number of member galaxies with log L$_B$ $\geq$ 10.2). Systems with $R>10$ were not included in the sample as they were already known galaxy clusters, and systems with $R$ = 1 were also excluded, as they were too poor to provide results on their physical parameters that would be reliable. From this process, we obtained a 53--group statistically complete sample, that was divided into two sub-samples: i) the 26 high-richness groups with $R$ = 4$-$8 (see Paper~II) and ii) the 27 low-richness groups with $R$ = 2$-$3.

Distances to the CLoGS group-dominant galaxies are estimated from their recession velocities, corrected for Virgocentric flow. Inaccuracies in these distances will affect calculated values such as luminosities, but since we mainly consider relations between quantities presented on logarithmic scales, these differences will be small and will not significantly alter our findings (e.g., a distance error of 20 per cent would give a luminosity error of only 0.2~dex).

\begin{table*}
 \caption{Details of the new GMRT observations analyzed for this study, along with information on archival data analyzed by the authors or by \citet[][marked \textit{a}]{Simona11} and \citet[][marked \textit{b}]{Giacintucci12}. For each source the first line displays the details for the 610~MHz and the second line for the 235~MHz. The columns give the LGG (Lyon Groups of Galaxies) number for each group, the BGE name, observation date, frequency, time on source, beam parameters and the rms noise in the resulting images.}
 \label{GMRTtable}
 \begin{center}

\begin{tabular}{|c|c|c|c|c|c|c|}
 
 \hline \hline
  Group Name & BGE       & Observation & Frequency &    On source      &   Beam, P.A.                  & rms \\
 
    LGG      &           &    Date    &  (MHz)    &  Time (minutes)    & (Full array, $''\times'',{}^{\circ}$) & (mJy beam$^{-1}$) \\

 \hline
  6  &NGC 128    &  2011 Nov &    610    &     157     &  $5.32\times4.06$, -2.96   & 0.16 \\ 
      &          &  2011 Nov  &   235    &     157      &  $16.15\times12.07$, 74.34   & 1.90 \\ 
  12   &NGC 252  &  2011 Nov  &   610  &     171      &  $5.46\times4.78$, -44.72   & 0.06 \\ 
      &          &  2011 Nov  &   235  &     171      &  $16.16\times14.17$, 77.73   & 0.60 \\ 
  78  &NGC 1106  &  2011 Nov  &    610    &   171    &  $5.44\times4.08$,  -2.47   & 0.10 \\ 
      &          &  2011 Nov  &    235    &  171   &  $14.83\times10.31$, -9.06   & 0.60 \\ 
  97  &NGC 1395  &  2011 Nov  &   610    &    168   &  $9.59\times3.80$, 37.03   & 0.10 \\ 
      &          &  2011 Nov  &   235     &    168  &  $28.74\times9.20$, 39.36  & 0.70 \\ 
 113  &NGC 1550  &  2011 Dec  &    610    &   187 &  $5.66\times3.96$, 88.52   & 0.04 \\ 
      &          &  2011 Dec  &    235    &   187   &  $14.28\times10.65$, 88.65  & 0.45 \\ 
 126  & NGC 1779 &  2011 Dec  &    610    &    124      &  $4.68\times3.77$,  34.90   & 0.05 \\ 
      &          &  2011 Dec  &    235    &    124      &  $12.67\times9.68$, 25.13   & 0.45 \\ 
 138  & NGC 2292 &  2011 Dec  &   610    &    137    &  $7.58\times3.59$, 36.19   & 0.05 \\ 
      &          &  2011 Dec  &    235    &     137   &  $20.11\times9.22$, 31.86  & 0.42 \\ 
 167  & NGC 2768 &  2012 Jan  &   610    &   209   &  $6.26\times3.65$, 49.03   & 0.03 \\ 
      &          &  2012 Jan  &    235    &   209   &  $16.31\times9.49$, 36.10  & 0.30 \\ 
 177  & NGC 2911 &  2012 Jan  &    610    &   196   &  $5.22\times3.79$, 60.15  & 0.10$^{1}$\\ 
      &          &  2012 Jan  &    235    &   196    &  $11.58\times10.21$, 39.92  & 0.30 \\ 
 232  & NGC 3613 &  2012 Jan  &    610    &   212    &  $5.79\times3.63$, 33.96  & 0.03 \\ 
      &          &  2012 Jan  &    235    &   212    &  $16.09\times8.77$, 22.22  & 0.26 \\ 
 255  & NGC 3923 &  2012 Mar  &    610    &   156   &  $7.33\times4.02$,  -16.06 & 0.05 \\ 
      &          &  2012 Mar  &    235    &   156   &  $20.77\times10.24$, -10.22  & 0.45 \\
 329  & NGC 4956 &  2012 Apr  &    610    &  199   &  $5.24\times3.96$, 24.60   & 0.05 \\ 
      &          &  2012 Apr  &    235    &  199   &  $12.54\times12.04$, 11.54  & 0.48 \\ 
 341  & NGC 5061 &  2012 May  &    610    &   200  &  $6.54\times3.68$, 28.89   & 0.05 \\ 
      &          &  2012 May  &    235    &   200  &  $22.42\times9.75$, 38.61  & 0.50 \\ 
 350  & NGC 5127 &  2012 Apr  &    610    &   200  &  $6.56\times4.42$, -30.97  & 0.25 \\ 
      &          &  2012 Apr  &    235    &   200  &  $13.51\times11.68$, -27.35  & 0.65 \\ 
 360  & NGC 5322 &  2012 May  &    610    &   199  &  $5.07\times4.00$,  -0.06   & 0.05 \\ 
      &          &  2012 May  &    235    &   199  &  $15.30\times11.31$, -49.08  & 0.70 \\ 
 370  & NGC 5444 &  2012 May  &    610    &  160   &  $4.42\times3.99$, 62.22   & 0.20 \\ 
      &          &  2012 May  &    235    &  160   &  $12.90\times10.79$, -61.85   & 0.80 \\ 
 376  & NGC 5490 &  2012 May  &    610    &  160   &  $4.14\times3.55$,  34.12   & 0.25 \\ 
      &          &  2012 May  &    235    &  160   &  $11.88\times8.69$, 35.44   & 1.85 \\ 
 383  & NGC 5629 &  2012 May  &    610    &  227  &  $4.95\times3.96$,  80.80   & 0.05 \\ 
      &          &  2012 May  &    235    &  227  &  $12.58\times10.71$, 83.01   & 0.40 \\ 
 398  & NGC 5903 &  2012 May  &    610    &  181  &  $5.78\times3.81$, 39.02  & 0.08 \\   
      &          &  2012 May  &    235    &  181  &  $20.19\times9.54$, 45.97   & 0.60 \\ 
 457  & NGC 7252 &  2012 Aug  &    610    &  199  &  $5.18\times4.36$,  20.12   & 0.10 \\ 
      &          &  2012 Aug  &    235    &  199  &  $14.25\times10.03$, 19.52   & 0.60 \\ 
 463  & NGC 7377 &  2012 Aug  &    610    &  217  &  $5.41\times3.92$, 26.75   & 0.04 \\ 
      &          &  2012 Aug  &    235    &  217  &  $15.68\times9.97$, 17.05   & 0.40 \\ 
  \hline

  Archival data &   &           &          &      &          &      \\

 \hline
 14 &NGC 315$^{a}$  & 2008 Feb  & 610  & 380  &   $5.20\times5.00$, 61.00  &   0.10  \\ 
    &            &   2008 Aug    &    235    &  280   &   $15.0\times15.0$, 0.00   &  0.70    \\  
 23  & NGC 524       & 2006 Mar  &    610  &  248  &  $5.52\times4.15$, 89.20   & 0.18    \\
    &               & 2006 Mar  &    235  &  248  &  $11.79\times9.89$, 66.70  & 2.10   \\
100 &  NGC 1407$^{b}$ & 2009 Nov &  610 &  270   &     $8.20\times4.40$, 42.00  & 0.05   \\
    &               & 2009 Nov &  235 &  270   &    $16.10\times10.90$, 36.00    &    0.25 \\
205 & NGC 3325      & 2004 May &   610   & 63  &  $10.52\times3.78$, 60.46  &  0.31 \\
    &               & 2004 May &   235   & 63  &  $32.57\times12.12$, 57.37  &  38.0  \\
236 & NGC 3665      &  2009 Feb  &    610    &   248   &  $5.08\times4.10$, 35.38   & 0.17 \\ 
      &             &  2009 Feb  &    235    &   248   &  $12.04\times9.53$, 32.99  & 0.78 \\ 
314 & NGC 4697      & 2006 Jan  &    610    &   347   &  $5.44\times4.46$, -74.58   & 0.07 \\ 
    &               & 2006 Jan  &    235    &   347   &  $14.00\times11.32$, -64.90  & 1.14 \\ 

 \hline
 \end{tabular}
 \end{center}
  $^1$ The rms around the source area here is 0.2~mJy

 \end{table*}

\section{OBSERVATIONS AND DATA ANALYSIS}

\subsection{GMRT observations} 

Excluding six systems (NGC~315, NGC~524, NGC~1407, NGC~3325, NGC~3665 and NGC~4697; see Table~\ref{GMRTtable}) for which archival data were available, the low--richness sample of galaxy groups were observed using the GMRT in dual 235/610~MHz frequency mode during observing cycle 21, from 2011 Nov $-$ 2012 Aug. Each target was observed at both 235 and 610~MHz with the upper side band correlator (USB) for an average of $\sim$4 hours on source. The total observing bandwidth at both frequencies is 32~MHz with the effective bandwidth at 235~MHz being $\sim$16 MHz.
 
At 610~MHz the data were obtained in 512 channels with a spectral resolution of 65.1 kHz for each channel, whereas at 235~MHz the data were obtained in 256 channels and a spectral resolution of 130.2 kHz for each channel. A detailed summary of the observations can be found in Table~\ref{GMRTtable}.

\subsection{Data analysis} 
\label{sec:GMRTanalysis}

The data were processed using the \textsc{spam} pipeline\footnote{For more information  on how to download and run SPAM see \url{http://www.intema.nl/doku.php?id=huibintemaspam}} \citep{Intema14}. \textsc{spam} is a \textsc{python} based extension to the NRAO Astronomical Image Processing System (\textsc{aips}) package which includes direction-dependent calibration, radio frequency interference (RFI) mitigation schemes, along with imaging and ionospheric modeling, adjusting for the dispersive delay in the ionosphere. We provide here only a brief description of the \textsc{spam} pipeline to account for the data analysis procedure followed. For more details regarding the \textsc{spam} pipeline and the algorithms of the \textsc{spam} package see \citet{Intema09,IntemaTGSS}.

The \textsc{spam} pipeline is run in two parts. In the first, pre-processing stage, the pipeline converts the raw LTA (Long Term Accumulation) format data collected from the observations into pre-calibrated visibility data sets for the total of the pointings observed (UVFITS format). The second stage converts these pre-calibrated visibility data of each pointing into a final Stokes~I image (FITS format), via several repeated steps of (self)calibration, flagging, and wide-field imaging.

The final full resolution images, corrected for the GMRT primary beam pattern, provide a field of view of $\sim 1.2^{\circ}\times1.2^{\circ}$ at 610~MHz and $\sim 3^{\circ}\times3^{\circ}$ at 235~MHz having a mean sensitivity for the data we observed (1$\sigma$ noise level) of $\sim$0.09~mJy/beam at 610~MHz and $\sim$0.64~mJy/beam at 235~MHz (see Table~\ref{GMRTtable}). The measured sensitivities achieved from the analysis at both frequencies are in line with expectations from previous GMRT experience. The theoretical (thermal) noise values for our observations are 29~$\mu$Jy/beam for 610~MHz and 80~$\mu$Jy/beam for 235~MHz\footnote{Calculated using the rms noise sensitivity equation in \S 2.1.6 from the GMRT Observer's Manual: http$://$gmrt.ncra.tifr.res.in/gmrt$\_$hpage/Users /doc/manual/Manual$\_$2013/manual$\_$20Sep2013.pdf}. We note that the theoretical sensitivity is dependent on the square root of the time on source, and that there is therefore a variation in the sensitivity between different targets at the same frequency, as well as a difference in quality between the older archival hardware correlator data and the newer observations using the software correlator. The main source of the residual noise in our final images comes either from calibration uncertainties in the form of phase errors from rapidly varying ionospheric delays (especially at the lowest frequencies) or from dynamic range limitation due to limited data quality, calibration and image reconstruction, that is mostly revealed by the presence of bright sources in the field. We also note that the full resolution of the GMRT is $\sim6''$ at 610~MHz and $\sim13''$ at 235~MHz with the \textit{u-v} range at 610~MHz being $\sim0.1 - 50$ k$\lambda$ and at 235~MHz $\sim0.05 - 21$ k$\lambda$.



The flux density scale in the images was set from the available flux calibrators in each observing session (3C~48, 3C~286 and 3C~147) using the models from \citet{ScaifeHeald12}. We adopt a flux density uncertainty of 5\% at 610~MHz and 8\% at 235~MHz, representing the residual amplitude calibration errors \citep{Chandra04}. 



Out of the 27 BGEs in the CLoGS low--richness sub-sample, 25 were analyzed in this study at both 610 and 235~MHz using the \textsc{spam} pipeline. The GMRT observations and images for NGC~315 at both 235 and 610 MHz are drawn from the earlier study of \citet{Simona11} and for NGC~1407 from the detailed follow-up work of \citet{Giacintucci12}, with the analysis of the observations being described in detail in those studies. In addition, we note that the GMRT data for NGC~5903 system were also analysed by the standard procedure followed as described in \citet{Kolokythas15, Kolokythas18} and the results are found to be consistent with results from the \textsc{spam} pipeline output.

1400~MHz data were drawn primarily from the NRAO VLA Sky Survey (NVSS, \citealt{Condon98}) and the study of \citet{Brown11}, that included measurements from the NVSS, Green Bank Telescope (GBT), and Parkes Radio Telescope, and from \citet{Condon02} for NGC~252, NGC~1106 and NGC~5127.

\section{Radio detections of brightest group galaxies in the low--richness sample} 

The radio sources detected in the BGEs of each group were identified from the 235 and 610~MHz GMRT images and the available NVSS catalog data. After that, we determined their morphology, calculated their flux densities (see Table~\ref{Sourcetable}) and their largest linear size. The GMRT images of the CLoGS low--richness groups at both frequencies along with more detailed information on these sources can be found in Appendix~\ref{AppA} and \ref{AppB}.




We find a radio detection rate of 82\% (22 of 27 BGEs) for the low--richness CLoGS sample, considering both GMRT frequencies and the 1.4~GHz surveys, with five galaxies being undetected at any of our three radio frequencies (NGC~2292, NGC~3325, NGC~4956, NGC~5444 and NGC~5629). 

Of these 5 undetected systems, NGC~5444 has previously been identified with a 4C source, but our higher resolution data shows that it is not related to the BGE (for more details see Section~\ref{AppA20} and Figure~\ref{fig:5444} for the radio image). NGC~3325 has a very high upper limit due to the high noise level from the analysis of the short archival observations available (see Table~\ref{Sourcetable}).

\begin{figure}
\centering
\includegraphics[width=0.48\textwidth]{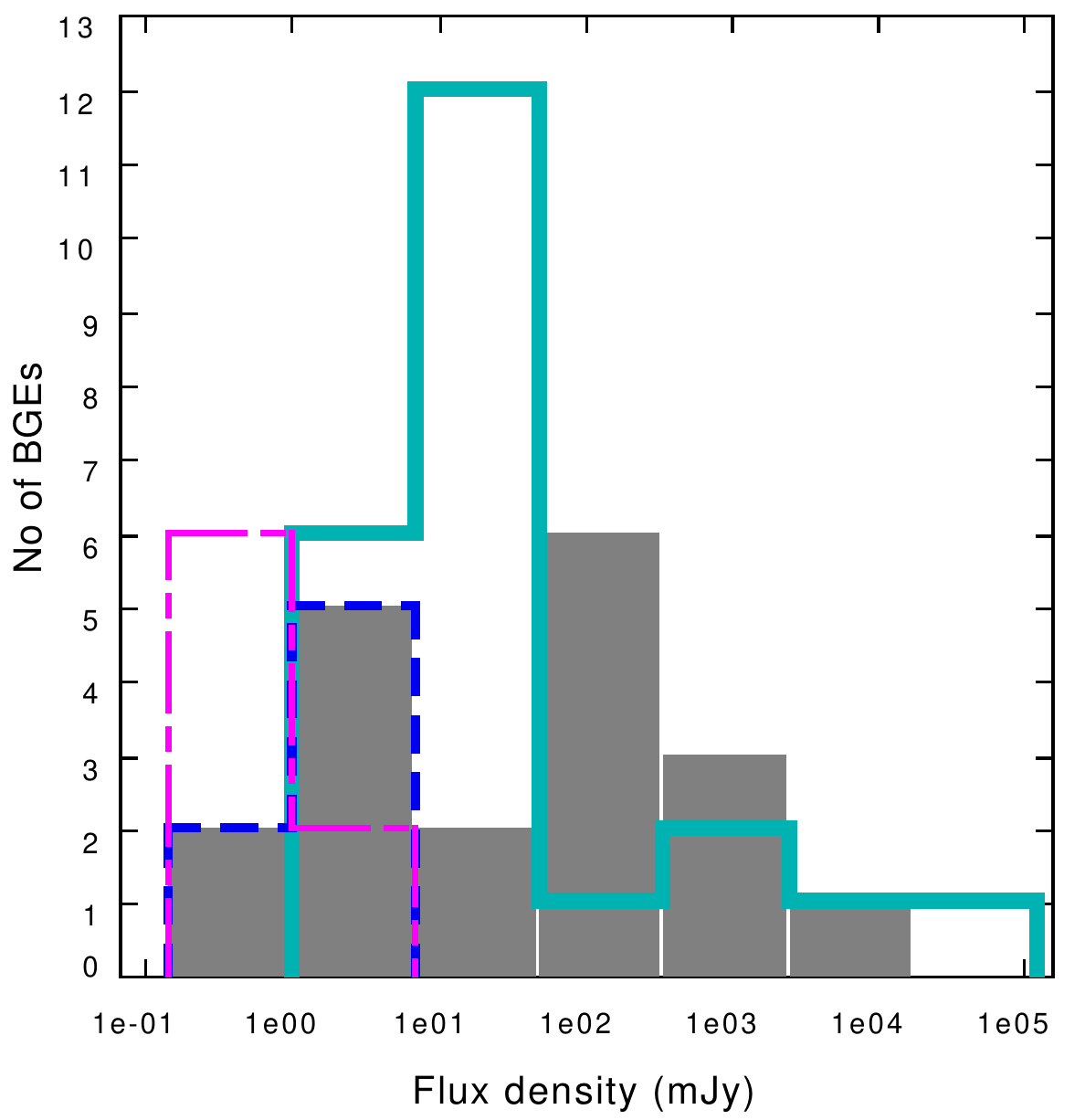}
\caption{Flux density distribution of the CLoGS BGEs at 610~MHz, comparing the low--richness sub-sample (grey columns) with the high--richness systems (cyan solid line) from Paper~II. The upper limits of the non detected BGEs in the low--richness sub-sample are shown in magenta dot-dashed line whereas the equivalent upper limits in the high--richness sub-sample are shown in blue dashed line.}
\label{610Flux}
\end{figure}

\begin{figure}
\centering
\includegraphics[width=0.48\textwidth]{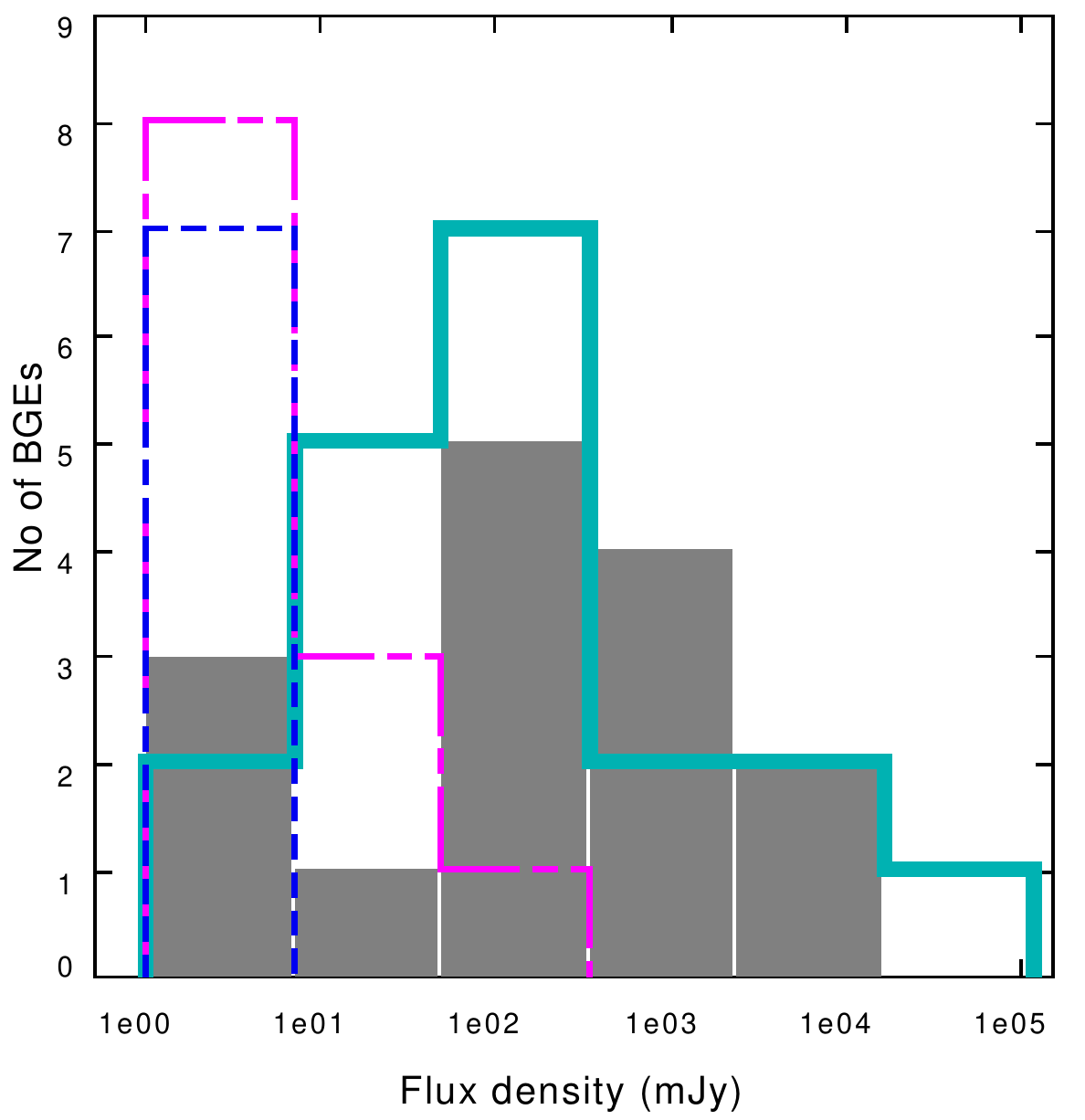}
\caption{Same as in figure~\ref{610Flux}, but for the flux density distribution at 235~MHz. }
\label{235Flux}
\end{figure}

If we take into consideration the detection statistics only from the GMRT data, we find that 70\% of the BGEs are detected at 235 and/or 610~MHz (19 of 27 BGEs) with 3 BGEs (NGC~128, NGC~3613 and NGC~4697) being detected only at 1.4~GHz data. Figure~\ref{610Flux}, presents the flux density distribution of all 53 CLoGS BGEs along with their upper limits (both high and low--richness sub-samples) at 610~MHz with figure~\ref{235Flux} presenting the distribution at 235~MHz. We observe that the majority of the BGEs have flux densities in the range $10-100$~mJy, but that the 610~MHz flux density distribution of the low--richness sub-sample is shifted towards lower flux densities compared to the high--richness systems. 

An upper limit of 5$\times$ the r.m.s. noise level in each image is used for the undetected radio sources, with the limits falling in the range $\sim$0.3$-$10~mJy for both sub-samples and GMRT frequency ranges. However, while in figure~\ref{610Flux} at 610~MHz we observe a similar number of radio non-detected BGEs (upper limits) between both sub-samples, from figure~\ref{235Flux} at 235~MHz the number of radio non detected BGEs in the low--richness sub-sample is found to be almost twice than the number of non detections in the high--richness sub--sample.

As in paper~II, the limiting sensitivity of our sample is estimated based on the typical noise level of our images and the maximum distance for the observed groups, which is 78~Mpc. The mean r.m.s from the analysis of the low--richness sub-sample (excluding the archival data) is $\sim$90 $\mu$Jy~beam$^{-1}$ at 610~MHz and $\sim$640~$\mu$Jy beam$^{-1}$ at 235~MHz. This means that the limiting sensitivity in the low--richness sub-sample using the automated pipeline SPAM is $>3.3\times10^{20}$ W Hz$^{-1}$ at 610~MHz or $>2.3\times10^{21}$ W Hz$^{-1}$ at 235~MHz which is similar to the sensitivity power achieved for the high--richness sub-sample ($>2.9\times10^{20}$ W Hz$^{-1}$ at 610~MHz or $>2.2\times10^{21}$ W Hz$^{-1}$ at 235~MHz; Paper~II). We note here that the limiting sensitivities are representing on average the total low--richness sub-sample and there may be individual radio sources detected at $>$5$\sigma$ level of significance at lower powers in nearby groups. The equivalent limit for NVSS 1400~MHz power sensitivity at CLoGS groups distance and level of significance for comparison is $>1.7\times10^{21}$ W Hz$^{-1}$. 




\begin{table*} 
\caption{Radio flux densities, radio power and spectral indices of the sources in the low--richness sub-sample. The columns list the BGE name, redshift, flux density of each source at 235 and 610~MHz, the 235$-$610~MHz spectral index, the flux density at 1.4~GHz (drawn from the literature), the 235$-$1400~MHz spectral index, and the radio power at 235, 610 and 1400~MHz. All upper limits shown here from our analysis are 5 $\times$ r.m.s. Five galaxies show no radio emission detected at 235, 610 or 1400~MHz.
The references for the 1.4~GHz flux densities and the GMRT measurements from previous works are listed at the bottom of the table. \label{Sourcetable}}
\begin{center}
\begin{tabular}{lccccccccc}
\hline 
 Source& Redshift&S$_{235 MHz}$ & S$_{610 MHz}$ & $\alpha_{235 MHz}^{610 MHz}$ & S$_{1.4GHz}$ & $\alpha_{235 MHz}^{1400 MHz}$ & P$_{235 MHz}$ & P$_{610 MHz}$ & P$_{1.4 GHz}$\\ 
      & ($z$) & $\pm8\%$ (mJy) & $\pm5\%$ (mJy) &   ($\pm$0.04)      &   (mJy)     &      & \scriptsize{(10$^{23}$ W Hz$^{-1}$)} & \scriptsize{(10$^{23}$ W Hz$^{-1}$)}  & \scriptsize{(10$^{23}$ W Hz$^{-1}$)}  \\
\hline
 NGC 128   &  0.0141 & $\leq$9.5 &  $\leq$0.8 &  - & 1.5$\pm0.5^a$ & - & $\leq$0.041 & $\leq$0.003 & 0.006 \\
 NGC 252   &  0.0165 & $\leq$3.0 & 0.9 &  -   & 2.5$^b$   &  -   &  $\leq$0.019  & 0.006 & 0.016\\
 NGC 1106  &  0.0145 & 435.0   & 221.0& 0.71 & 132$\pm$4$^b$ & 0.67$\pm$0.04 & 2.122  & 1.078 & 0.645 \\
 NGC 1395  &  0.0057 & $\leq$3.5 &   3.6 &  - & 1.1$\pm0.5^a$ & - &  $\leq$0.002 & 0.002 & 0.0006\\
 NGC 1550  &  0.0124 &   223.0   &    62.0  &  1.34 &  17$\pm2^a$  &   1.44$\pm$0.06   & 0.752  & 0.209 & 0.057\\
 NGC 1779  &  0.0111 &    4.6   &   2.6 &   0.60   &  5.4$\pm0.6^c$ &  -0.09$\pm$0.05    &  0.011   &  0.006 & 0.013\\
 NGC 2292  &  0.0068 & $\leq$2.1 &  $\leq$0.3 &   -    &   -    &      -     &  $\leq$0.002  & $\leq$0.0003 & -\\
 NGC 2768  &  0.0045 & 13.1  & 11.5  &  0.14 & 14$\pm1^a$   & -0.04$\pm$0.07 &  0.008   &   0.007 & 0.009\\
 NGC 2911  &  0.0106 & 69.0   & 58.0 &  0.18 & 56$\pm2^a$  & 0.12$\pm$0.04 &  0.166 &  0.139 & 0.135 \\
 NGC 3613  &  0.0068 & $\leq$1.3 &  $\leq$0.2 & - & 0.3$\pm0.3^a$ & - &  $\leq$0.002 &  $\leq$0.0002 & 0.0004\\
 NGC 3923  &  0.0058 &  5.0    &  2.1        &  0.91 & $1\pm0.5^a$   & 0.90$\pm$0.22 &  0.002  & 0.001 & 0.0005\\
 NGC 4956  &  0.0158 & $\leq$2.4  & $\leq$0.3  & - & - & - & $\leq$0.014  &  $\leq$0.002 & -\\
 NGC 5061  &  0.0069 & $\leq$2.5 & 0.8 &   -   & -   &   - &  $\leq$0.002  &  0.001  & - \\
 NGC 5127  &  0.0162 & 5690  & 1630  &  1.32 & 1980$^b$ & 0.59$\pm$0.04 & 35.464 & 10.159 &  12.278\\
 NGC 5322  &  0.0059 &  134.6 & 106.0  &  0.25 & 79.3$\pm2.8^c$   & 0.30$\pm$0.04 & 0.135  &  0.106 &  0.080\\
 NGC 5444  &  0.0131 & $\leq$12$^d$  &  $\leq$1$^d$ & - & - \large{$^{*}$}  & - &  $\leq$0.051 &  $\leq$0.004 & -\\
 NGC 5490  &  0.0162 &  1140 & 815  &  0.36  &  1300$\pm100^a$   &  -0.07$\pm$0.05 &  6.804   &  4.864  &  7.839\\
 NGC 5629  &  0.0150 &  $\leq$2.0  & $\leq$0.3  &  - & - & - &  $\leq$0.011 &  $\leq$0.002 & -\\
 NGC 5903  &  0.0086 &  1830 & 970  &  0.68 & 321.5$\pm16.0^d$  & 0.99$\pm$0.04 & 2.829  &  1.500 &  0.498\\
 NGC 7252  &  0.0160 & 76.6  & 41.5  &  0.66 &  25.3$\pm1.2^c$  & 0.63$\pm$0.04 &  0.397  & 0.215 &  0.132\\
 NGC 7377  &  0.0111 &  2.7  &   2.1  &  0.27 &  3.4$\pm0.5^c$   &  -0.13$\pm$0.07   &  0.007   &  0.005 &  0.009 \\
\hline
 \multicolumn{9}{l}{Archival data/Previous work}\\
\hline
 NGC 315$^e$ & 0.0165   &   15411 &   $\geq$2500 &  $\leq$ 1.91   & 1800$\pm100^a$ & 1.20$\pm$0.04 & 99.691  & 16.172 &  11.474\\
 NGC 524  &  0.0080  &  $\leq$10.5 & 2.0  &  -  & 3.1$\pm0.4^a$  & - &  $\leq$0.015  &  0.003 &  0.004\\
 NGC 1407$^f$&  0.0059  &   945    &   194     &   1.66     & 38$\pm2^f$      &   1.80$\pm$0.04  &  0.603   & 0.124 &  0.024\\ 
 NGC 3325    &  0.0189  &  $\leq$191  &  $\leq$1.55   &   -    &    -  &     -     &    $\leq$1.449    & $\leq$0.012 & - \\
 NGC 3665    &  0.0069  &  225.2  &  149.0 &  0.43  & 113.2$\pm3.8^c$   &  0.39$\pm$0.04 &  0.275  & 0.182 & 0.139\\
 NGC 4697    &  0.0041  &   $\leq$5.7  & $\leq$0.35  &  -  &  $0.6\pm0.5^a$   & - &  $\leq$0.002 &  $\leq$0.0001 &  0.0002\\
\hline
\end{tabular}
\end{center}
$^a$ \citet{Brown11}, \mbox{ }$^b$ \citet{Condon02}, \mbox{ }$^c$ \citet{Condon98}, \mbox{ }$^d$ Upper limit was calculated from the maximum peak flux density on the position of the BGE as the phase calibration errors are too large due to a very bright 4C radio source nearby. The detection of the system in this case is limited by the dynamical range due to the very strong background source right next to the position of the BGE,\mbox{ }$^e$ \citet{Simona11},\mbox{ }$^f$ \citet{Giacintucci12} \mbox{ }\mbox{ }\mbox{ } \large{$^{*}$}\footnotesize{\citet{Condon98} (NVSS) and \citet{Brown11} report for NGC~5444 a flux density at 1.4~GHz of $\sim$660$\pm20$ which is a 4C nearby source}
\end{table*}

\section{Radio properties of the brightest group galaxies}
We examine here the radio properties of the detected radio sources in the central galaxies of the low--richness sample. As in the high--richness sample, the sources present a range in size, power and morphology. The GMRT images of the groups along with more details for these sources are presented in Appendices~\ref{AppA} and \ref{AppB}.

\subsection{Radio morphology} 

The radio structures found in the central galaxies have extents ranging from a few kpc (galactic scale) to hundreds of kpc (group scale). Table~\ref{radiomorph} lists the radio morphology for each source. We apply the radio morphology scheme described in Paper~II, classifying our radio sources from our own GMRT 235 and 610~MHz analysis, the 1.4~GHz NVSS survey and from earlier studies on some of our BGEs. The purpose of this classification is to distinguish between the different classes of radio emission and provide information on their environment. Systems may belong to more than one category; when that is the case, the system is classified according to its most extended and prominent state/class.

The morphological classes are: 
\begin{enumerate}
\item point-like (unresolved) sources,
\item diffuse sources with no clear jet-lobe structure,
\item small-scale jets ($<$20~kpc),
\item large-scale jets ($>$20~kpc), and
\item non-detections.
\end{enumerate}
A subset of the large-scale jet sources have been shown by previous studies to be \textit{remnant} systems, where the AGN is quiescent, but emission from the aging jets or lobes is still visible. 

Table~\ref{radiomorph} summarizes the radio properties of the BGEs in the low--richness sample, showing their largest radio linear size, their radio morphology, their power at 235~MHz (see \S~5.3 below for more details) and an estimate of the energy injected into the IGM by the inflation of radio lobes by jet sources  (P$_{\mathrm{cav}}$) calculated from the radio power at 235~MHz using the scaling relation from \citet[][see also $\S6.4$]{OSullivan11}: 

\begin{equation}
\label{Pcav}
{\mathrm{log\, P}}_{\mathrm{cav}}=0.71 \,(\pm0.11)\, \mathrm{logP}_{235}+1.26\, (\pm0.12),
\end{equation}

where P$_{\mathrm{cav}}$ and P$_{235}$ are in units of 10$^{42}$ erg~s$^{-1}$ and 10$^{24}$ W~Hz$^{-1}$.



We find that in the low--richness sample 14 central galaxies exhibit point-like radio emission, 2 host small-scale jets, 5 host large-scale jets (of which 1, NGC~1407, is a remnant jet; see below for details) with 5 BGEs being undetected at the frequencies we investigated, and only 1 system presenting a diffuse radio structure. 

NGC~3665 and NGC~5322 are the two systems that exhibit a small-scale jet morphology 
and both present symmetrical structures and thin, straight jets (see Figures~\ref{fig:3665} and \ref{fig:5322}) with similar sizes (see Table~\ref{radiomorph}). The four currently-active large-scale jet systems are all Fanaroff-Riley type I (FR I; \citealt{FanaroffRiley74}) radio galaxies with their jet/lobe components extending from several tens of kpc (e.g., 33~kpc; NGC~1550) to several hundreds of kpc (1200~kpc; NGC~315) away from their host galaxy. 

Looking at the large-scale jet systems in more detail, NGC~1550 presents an unusually asymmetric FR~I radio morphology with the eastern lobe roughly half as far from the optical centre of the galaxy as its western counterpart (see Figure~\ref{fig:1550}). Our GMRT data provides a considerably clearer view of the radio morphology than previous studies \citep[e.g., ][]{Dunn10}, revealing the previously unresolved lobes and a sharp z-bend in the western jet which produces an offset between the east and west jet axes. We will discuss this system in greater detail in a forthcoming paper (Kolokythas et al., in prep.).

In NGC~5490 we also observe an asymmetric morphology with the eastern jet detected to $\sim$50~kpc while the opposite side of the source breaks up into detached clumps of radio emission (see Figure~\ref{fig:5490}). NGC~5127 on the other hand, presents a normal symmetric double jet morphology (see Figure~\ref{fig:5127}) with NGC~315 hosting the giant FR~I radio galaxy B2~0055+30 which has been extensively investigated at multiple frequencies and angular resolutions \citep[e.g.,][]{Bridle76,Bridle79,Venturi93,Mack97,Worrall07,Simona11}. NGC~315 presents two asymmetric jets, with one (northwest) being brighter and appearing bent backwards and the opposite (southeast) jet appearing much fainter and intermittent with the radio source having a total linear size of $\sim$1200~kpc \citep[see][]{Simona11}. 

NGC~1407 is the only radio galaxy in the low--richness sample classed as remnant radio source, as the radio spectral age analysis in the study of \citet{Giacintucci12} revealed that a faint $\sim$300 Myr old, ultra-steep spectrum ($\alpha = 1.8$) radio plasma of $\sim$80~kpc surrounds the central jet source, which is a typical product of former AGN activity and  characteristic of a dying radio galaxy.


The only diffuse radio source in the low--richness sample, appears in the galaxy group NGC~5903 (see Figure~\ref{fig:5903}). It has also been examined by several earlier studies at many different frequencies \citep[e.g.,][]{GopalKrishna78,Gopal12,OSullivan5903}. It hosts a $\sim$75~kpc wide diffuse, steep-spectrum ($\alpha_{150}^{612}\sim1.03$; \citealt{OSullivan5903}) radio source whose origin may be through a combination of AGN activity and violent galaxy interactions (see~\ref{AppA23}). We note that NGC~5903 could also be categorised as a remnant system, but due to its complex origin and current state we consider it more conservative to class this system as diffuse.


\subsection{Radio Spectral index} 


Where possible, we estimate the spectral indices of each radio source in the frequency ranges 235$-$610~MHz and 235$-$1400~MHz. The radio spectral index of a source will, over time, steepen owing to synchrotron and inverse Compton losses, provided that there is no new source of electrons and no additional energy injection. Spectral index is therefore, in the absence of complicating factors, an indicator of source age.


Examining the radio sources in the low--richness sample, we find that only just over half the BGEs (15/27 galaxies; 56\%) are detected at both GMRT frequencies at 235 and 610~MHz. At this frequency range we find that four sources present steep radio spectra of $\alpha_{235}^{610}>$1 (NGC~1550, NGC~5127, NGC~315 and NGC~1407) with NGC~315 and NGC~1407 presenting ultra-steep spectra of $\alpha_{235}^{610}\geq$1.91 and $\alpha_{235}^{610}=$1.66 respectively (see Table~\ref{Sourcetable}). However, in the study of \citet{Simona11} the flux density of NGC~315 at 610~MHz is quoted as underestimated, therefore the spectral index calculated must be considered an upper limit. For the same galaxy, \citet{Mack97}, using the WSRT at the same frequency (609~MHz) reported a flux density of 5.3~Jy, thus reducing the spectral index value of the radio source in NGC~315 to $\alpha_{235}^{609}=$1.11. For NGC~1407, \citet{Giacintucci12} reported that the radio spectral index value at the 235$-$610~MHz range was calculated after subtracting  contributions in the flux densities at both frequencies from an inner double (young jets associated with the NGC~1407 AGN) and two point radio sources whose positions fall within the large-scale diffuse emission of the old radio lobes.

The rest of the 11 radio sources present $\alpha_{235}^{610}$ that ranges from very flat values of $\sim$0.1 to typical radio synchrotron spectra of $\sim$0.9. We find that the total mean spectral index value at the GMRT frequency range, leaving the two overestimated steep spectrum outliers out (NGC~315 and NGC~1407),  is $\alpha_{235}^{610}=0.60\pm0.16$ (for 13/15 radio sources). 

Examining the radio spectral index distribution between 235 and 1400~MHz we find that four radio sources (NGC~1779, NGC~2768, NGC~5490 and NGC~7377), three of which exhibit a relatively flat spectral index in the 235$-$610~MHz range (all except NGC~1779 which has $\alpha_{235}^{610}=0.60$), have flux densities greater at 1400~MHz than at 235~MHz, giving an inverted spectral index in the 235$-$1400~MHz range (see Table~\ref{Sourcetable}). Three of these four systems present weak point-like radio sources at both frequencies (NGC~1779, NGC~2768 and NGC~7377), hence the deviation from a powerlaw of the observed flux density can be attributed to self-absorption of the lower frequency emission at 235~MHz as a result of the `cosmic conspiracy' (see \citealt{Cotton80}).

For the FR~I radio jet system in NGC~5490, the observed inverted spectral index in the 235$-$1400~MHz range ($\alpha_{235}^{1400}=-0.07$) arises primarily from the difference in detected morphology between the two frequencies. The bright source close to NGC~5490 limits the sensitivity we were able to achieve at 235~MHz and probably prevented  us from detecting all the corresponding radio emission from the source that has been picked up at 1400 MHz. The mean value of $\alpha_{235}^{1400}$ that we calculate for the 15/27 BGEs in the low-richness sample is $0.57\pm0.16$. 

In Table~\ref{radiospixlow} we list the mean values of the radio spectral indices for the point-like, small-scale and large-scale jet radio morphologies in the low--richness sample. We neglected to include values for remnant jets and diffuse sources since these classes consist of only 1 system each. We find that large-scale jet systems present the steepest mean indices ($\alpha_{235}^{610}=1.23\pm0.08$ and $\alpha_{235}^{1400}=0.79\pm0.08$) in the low--richness sample, with small-scale jet systems exhibiting the flattest spectra, having similar values at both frequency ranges we examined. The point-like radio sources are found to exhibit typical radio spectrum indices with the mean values being comparable within the uncertainties at both frequency ranges. In large-scale jet systems, the steeper mean spectral index seen in the 235$-$610~MHz frequency range can be attributed to low number statistics and detection sensitivity between the two frequency ranges, as NGC~315 presents an upper limit for $\alpha_{235}^{610}$ due to the flux density underestimation at 610~MHz \citep{Simona11} and NGC~5490 presents an inverted spectrum at 235$-$1400~MHz reducing the mean at this frequency range (see below for more on NGC~5490).


For the systems analysed in this paper where it was possible to separate their extended emission from the main core at both 235 and 610~MHz, we calculated the spectral indices for their core and the extended components separately. We created images with matched resolution, cellsize, \textit{uv} range and image pixel size at these two frequencies. Table~\ref{corespix} lists the spectral index values for the 4 systems that meet these criteria. In NGC~3665 we find that the spectral index does not differ between the two components. For two systems (NGC~1550 and NGC~5127) the spectral index in their cores is flatter, indicating either aging along their jets, self-absorption, or the presence of free-free emission. For NGC~5490 we find that the core (which is clearly detected) presents a fairly typical spectral index value of 0.55$\pm$0.04 but that the extended emission has a very flat value of 0.02$\pm$0.04. The extended low-surface brightness emission presents different morphology between 235~MHz and 610~MHz, so the flat spectral index we measure is likely biased, and probably does not reflect the true index of the extended component. A more accurate measurement would require higher signal-to-noise ratio data and low resolution images to trace the extended emission at both frequencies. 

\begin{table*}
 \caption{Morphological properties of CLoGS low--richness groups and their central radio sources. For each group we note the LGG number, the BGE name, the angular scale, the largest linear size (LLS) of the radio source, measured from the 235 MHz radio images unless stated otherwise, the radio morphology class, and lastly the energy output of any radio jets, estimated from the 235~MHz power using equation~\ref{Pcav}. The errors on the energy output were calculated based on the errors from scaling relation~\ref{Pcav}.
}
 \label{radiomorph}
\begin{center}
\begin{tabular}{llcccc}
\hline 
Group & BGE & Scale &  LLS & Radio morphology &  Energy output (radio) \\ 
      &     & (kpc/$''$) & (kpc) &         &  (10$^{42}$ erg s$^{-1}$)  \\
\hline 
LGG 6   & NGC 128   & 0.291 &  $\leq$3$^a$ &   point  &  -  \\
LGG 12  & NGC 252   & 0.349 &    $\leq$3   &  point    &  -   \\
LGG 14  & NGC 315   & 0.354 &  1200$^b$ &  large-scale jet & 93.13$^{+64.97}_{-38.27}$   \\
LGG 23 & NGC 524    & 0.165 &  $\leq$2$^c$  &  point & -      \\
LGG 78 & NGC 1106   & 0.310 &    $\leq$14 &  point   & -      \\
LGG 97 & NGC 1395   & 0.102 &    $\leq$3  &  point   & -     \\
LGG 100 & NGC 1407  & 0.112 &   80$^b$  &  remnant  & -      \\
LGG 113 & NGC 1550  & 0.257 &    33     &  large-scale jet & 2.89$^{+0.02}_{-0.03}$      \\
LGG 126 & NGC 1779  & 0.218 &    $\leq$4&  point  & -       \\
LGG 138 & NGC 2292  & 0.145 &    -   &  -  & -    \\
LGG 167 & NGC 2768  & 0.112 &    $\leq$3&  point  & -     \\
LGG 177 & NGC 2911  & 0.218 &    $\leq$7&  point  & -       \\
LGG 205 & NGC 3325  & 0.388 &    -      &    -    & -     \\
LGG 232 & NGC 3613  & 0.156 &  $\leq$3$^a$ &  point & -       \\
LGG 236 & NGC 3665  & 0.156 &    16  & small-scale jet & 1.42$^{+0.16}_{-0.18}$   \\ 
LGG 255 & NGC 3923  & 0.097 &    $\leq$3&  point  & -    \\
LGG 314 & NGC 4697  & 0.087 &    $\leq$3$^a$& point &  -    \\
LGG 329 & NGC 4956  & 0.344 &    -      &    -    &  -      \\ 
LGG 341 & NGC 5061  & 0.136 &    $\leq$2&  point   &  -     \\
LGG 350 & NGC 5127  & 0.349 &    244    &  large-scale jet & 44.74$^{+23.06}_{-15.22}$     \\  
LGG 360 & NGC 5322  & 0.141 &    15$^c$ &  small-scale jet & 0.86$^{+0.15}_{-0.19}$      \\
LGG 370 & NGC 5444  & 0.291 &    -      &  -    & -     \\ 
LGG 376 & NGC 5490  & 0.344 &    120 (30 at 610 MHz) & large-scale jet & 13.84$^{+3.65}_{-2.89}$   \\
LGG 383 & NGC 5629  & 0.325 &     - &    -  & -   \\ 
LGG 398 & NGC 5903  & 0.175 &    65$^d$ &  diffuse & -     \\
LGG 457 & NGC 7252  & 0.320 &  $\leq$14    &  point   & -    \\ 
LGG 463 & NGC 7377  & 0.223 &   $\leq$5    &  point   & -     \\ 
\hline 
\end{tabular}
\end{center}
$^a$ Measured from the 1.4~GHz image, $^b$ \citet{Simona11}, $^c$ Measured from the 610~MHz image, $^d$ \citet{OSullivan5903} \\

\end{table*}

\begin{table}
 \centering
 \caption{Mean spectral indices $\alpha_{235}^{610}$ and $\alpha_{235}^{1400}$ for the different radio morphology classes in the low--richness sample. The last column shows the number of sources used in calculating the means.}
\begin{tabular}{lccc}
 \hline
 Radio Morphology  &  Mean $\alpha_{235}^{610}$ & Mean $\alpha_{235}^{1400}$ & No of sources\\
 \hline
 Point-like &  0.50$\pm$0.11 & 0.29$\pm$0.11  & 7 \\
 Small-scale jet& 0.34$\pm$0.06  &0.35$\pm$0.06 &  2  \\
 Large-scale jet & 1.23$\pm$0.08 & 0.79$\pm$0.08  & 4 \\ \hline
 \end{tabular}
 \label{radiospixlow}
 \end{table}


\begin{table}
 \centering
 \caption{Spectral indices in the 235$-$610~MHz range for the cores and extended emission of those radio sources in our sample whose morphology is resolved at both frequencies. Columns show the group LGG number, the BGE name, and the spectral index $\alpha_{235}^{610}$ of the core and the surrounding emission in each source.}
\begin{tabular}{llcc}
 \hline
 Group   & BGE &  Core $\alpha_{235}^{610}$    &  Surrounding $\alpha_{235}^{610}$  \\
            
 \hline
 LGG 113 & NGC 1550 &   0.94$\pm$0.04  & 1.36$\pm$0.04   \\
 LGG 236 & NGC 3665 &   0.39$\pm$0.04  & 0.44$\pm$0.04     \\
 LGG 350 & NGC 5127 &   0.77$\pm$0.04  & 1.33$\pm$0.04   \\
 LGG 376 & NGC 5490 &   0.55$\pm$0.04  & 0.02$\pm$0.04   \\

 \hline
 \end{tabular}
 \label{corespix}
 \end{table}

\subsection{Radio power in CLoGS BGEs}
As in paper~II, we calculate the radio power P$_{\nu}$, for the low--richness sample BGEs as:

\begin{equation}
\label{Power610}
P_{\nu}=4 \pi D^2 (1+\rm z)^{(\alpha-1)}S_{\nu},
\end{equation}

where D is the distance to the source, $\alpha$ is the spectral index in the 235$-$610~MHz regime, $z$ the redshift and S$_{\nu}$ is the flux density of the source at frequency $\nu$. For the systems that were detected at only one frequency the typical spectral index value of $\alpha = 0.8$ was used for the calculation \citep{Condon92}.

\begin{figure}
\centering
\includegraphics[width=0.48\textwidth]{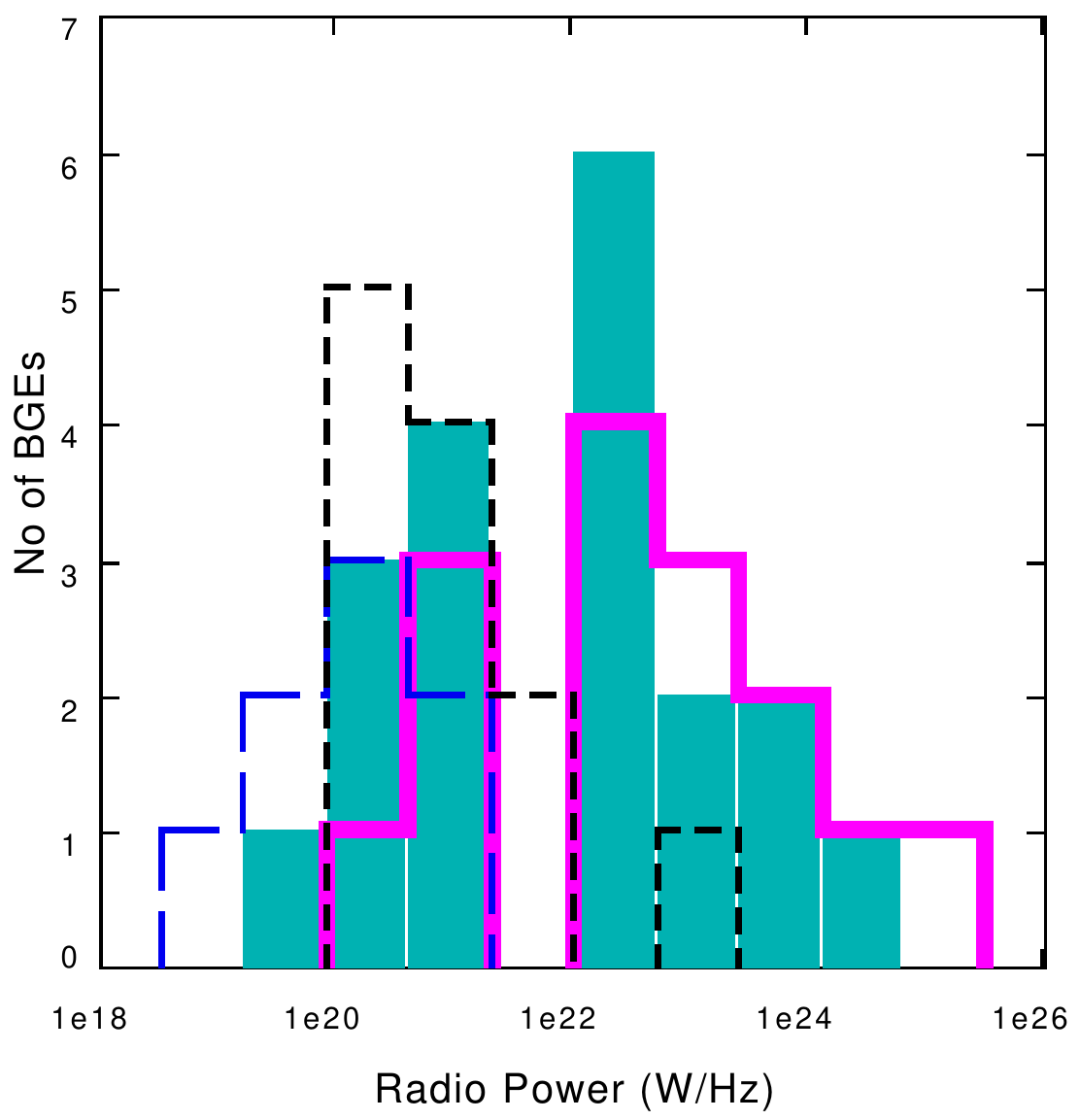}
\caption{Radio power distribution of CLoGS BGEs in low--richness sample at 610~MHz (cyan columns) and 235~MHz (solid magenta line), with the upper limits at 610~MHz (blue dashed line) and at 235~MHz (black dotted line) also shown here.}
\label{Lradio}
\end{figure}

We find that in the low--richness sample the radio power of the detected radio sources in the BGEs is in the range 10$^{19}$ $-$ 10$^{25}$ W Hz$^{-1}$. Figure~\ref{Lradio}, presents the radio power distribution of the BGEs from the low--richness sub-sample at both 235 and 610~MHz. The majority of the galaxies in figure~\ref{Lradio} have low radio powers in the range between 10$^{20}$ $-$ 10$^{21}$ W Hz$^{-1}$ and 10$^{22}$ $-$ 10$^{23}$ W Hz$^{-1}$ with only one BGE exhibiting power in the range $10^{21}-10^{22}$ W Hz$^{-1}$ (NGC 1779). The CLoGS low--richness sample contains two high-power sources with P$_{235MHz}>10^{24}$ W Hz$^{-1}$, both of which host large-scale bright radio jets (NGC~315 and NGC~5127). The radio power upper limits of the non detected radio sources at 610~MHz extend to lower values (10$^{19}$ $-$ 10$^{20}$ W Hz$^{-1}$) than at 235~MHz, with the majority of the radio power upper limits at both frequencies seen between $\sim$5$\times$10$^{19}$ $-$ 5$\times$10$^{21}$ W Hz$^{-1}$. Only NGC~3325 presents an upper limit of an order of magnitude higher ($\sim$10$^{23}$ W~Hz$^{-1}$) compared to rest of the BGEs, due to the very high noise level in the image produced from the available archival data. We note that previously reported typical values of radio power at 235 and 610~MHz in BCGs range between $\sim10^{23}- 5\times10^{26}$ W Hz$^{-1}$ \citep[see e.g. Table 2,][]{Yuan16}. The equivalent radio power statistics for the full sample will be discussed in \S~\ref{disc}.

\subsubsection{Radio-loudness in low--richness CLoGS sample BGEs}  

Following \citet{Best05}, we consider as radio-loud systems with P$_{1.4GHz}$ $>$ 10$^{23}$~W~Hz$^{-1}$. In the low--richness sample we find that three BGEs have radio power higher than this value: NGC~315, NGC~5127 and NGC~5490 (see Table~\ref{radioloud23}). While NGC~5903 and NGC~1106 have radio powers greater than 10$^{23}$~W~Hz$^{-1}$ at 610~MHz their corresponding flux density at 1.4~GHz is lower than the radio-loud limit, and we do not include them in the radio-loud category. In addition, the sensitivity limit of the NVSS survey and \citet{Brown11} study guarantees that all the undetected systems at 1.4~GHz have radio powers below this limit. In \S~\ref{disc} we discuss the equivalent statistics for the radio-loud systems in the full CLoGS sample.

\begin{table}
 \centering
 \caption{Radio-loud AGN (P$_{1.4GHz}>$10$^{23}$ W Hz$^{-1}$) in the BGEs of the CLoGS low--richness sample. Columns show the LGG number, the BGE name and the radio power at 1.4~GHz.}
\begin{tabular}{lcc}
 \hline
 Group   & BGE &     P$_{1.4GHz}$        \\
         &     & (10$^{23}$ W Hz$^{-1}$) \\   
 \hline
  LGG 14    & NGC 315  &   11.5$\pm$0.6    \\             
  LGG 350   & NGC 5127 &   12.3$\pm$0.6    \\             
  LGG 376   & NGC 5490 &    7.8$\pm$0.6    \\
 \hline
 \end{tabular}
 \label{radioloud23}
 \end{table}


\subsubsection{235~MHz radio power vs largest linear size}

It is well known that the linear size of FR~I radio galaxies is proportional to their radio power \citep{Ledlow02}. As in Paper~II, we examine the relation between the 235~MHz power of our sources against their largest linear size (LLS). The relation was examined for the radio sources that were morphologically resolved, with their largest linear size being calculated across the longest extent of the detected radio emission, at the frequency at which the emission is most extended. 

The resolved radio sources of the CLoGS low--richness central galaxies span a broad spatial scale from $\sim$15~kpc (small scale jets; NGC~5322) to $>1200$~kpc (large scale jets; NGC~315) with the corresponding 235~MHz radio powers being in the range of $\sim10^{21}$ W~Hz$^{-1}$ to $\sim10^{25}$ W~Hz$^{-1}$.

\begin{figure}
\centering
\includegraphics[width=0.49\textwidth]{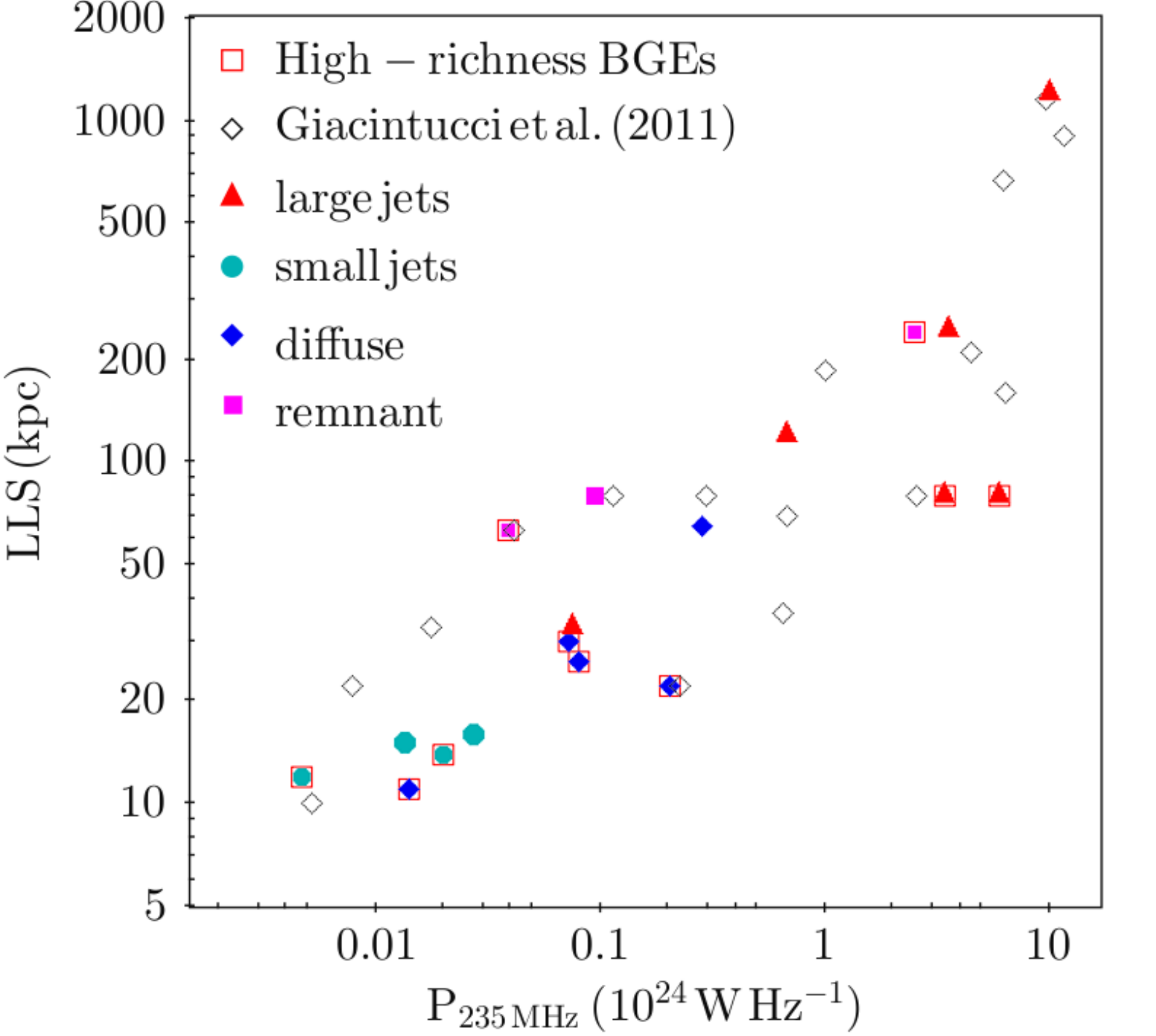}
\caption{Radio power at 235 MHz plotted against the largest linear size for the full sample of CLoGS radio sources. Different symbols indicate the radio morphology of the radio sources. The points circumscribed by square box are representing BGEs from the high--richness sample, while those without the box are from the low--richness one. Open diamonds indicate systems in the \citet{Simona11} sample, for comparison. }
\label{LLS235Power}
\end{figure}

Figure~\ref{LLS235Power} shows the relation for the full CLoGS sample, along with the group systems from the study of \citet{Simona11}. We find that our CLoGS radio sources are in agreement with the linear correlation between size and power found by \citet{Ledlow02} and \citet{Simona11}. In addition, our CLoGS BGEs range of luminosities and size distribution is in line with the measured values of radio sources from the recent LOFAR AGN study of \citet{Hardcastle16} at 150~MHz over a greater redshift range (0$-$0.8) (see their fig. 18).  We find that there is no significant difference in the spread of our group radio sources between the low and high--richness sample and confirm across group and cluster environments the linear correlation for the radio sources is valid and stands for almost 4 orders of magnitude at 235~MHz.



\subsection{Star formation contribution in CLoGS radio sources}

\begin{table*}
 \caption{Star formation rates calculated from FUV, and expected radio power at 610~MHz owing to star formation for the CLoGS low--richness sample point-like radio sources. For each system we list the SFR$_{FUV}$, the expected radio power at 610~MHz P$_{610~expected}$ from the calculated SFR$_{FUV}$, the radio power at 610~MHz and 1.4~GHz (P$_{610~MHz}$, P$_{1.4~GHz}$), the molecular mass values (M$_{H2}$) drawn from \citet{OSullivanCO18} and the radio morphology of the source. 
}
 \label{SFR}
\begin{center}
\begin{tabular}{cccccccc}
\hline 
Group &  BGE &  SFR$_{FUV}$  &  P$_{610~expected}$ & P$_{610~MHz}$ &   P$_{1.4~GHz}$  & M$_{H2}$ & Morphology \\
LGG   &      & (10$^{-2}$ M$_{\odot}$ yr$^{-1}$) &  (10$^{21}$ W Hz$^{-1}$)&   (10$^{21}$ W Hz$^{-1}$) &  (10$^{21}$ W Hz$^{-1}$) & (10$^8$ M$_{\odot}$)  &  \\
\hline 
 6  & NGC 128   & 4.9   & 0.13  & 1.25$^a$ & 0.64 & 1.76$\pm$1.5 & point \\
 12 & NGC 252   & 30.7  & 0.69  &  0.60  & 1.55 & 6.29$\pm$0.79 & point  \\
 23 & NGC 524   & 2.7   & 0.05  &  0.30  & 0.43 & 1.90$\pm$0.23 & point  \\
 78 & NGC 1106  & 26.7  & 0.94  &  107.80 & 64.50 & 6.81$\pm$0.33 & point    \\
 97 & NGC 1395  & 4.2   & 0.07  &  0.20 & 0.06 & $<$0.27 & point  \\
126 & NGC 1779  & -     & -     & -  & 1.31 & 4.57$\pm$0.60 & point    \\
167 & NGC 2768  & 2.6   & 0.06  &  0.70  & 0.89 & 0.18$\pm$0.01 & point    \\
177 & NGC 2911  & 3.2   & 0.11  &   13.90  & 13.53 & 2.66$\pm$0.31 & point \\
232 & NGC 3613  & 1.4   & 0.02  &   0.07$^a$ & 0.04 & $<$0.46 & point  \\
255 & NGC 3923  & 4.2   & 0.06  &   0.10  & 0.05 & $<$0.29 & point    \\
314 & NGC 4697  & 3.1   & 0.04  &   0.05$^a$  & 0.02 & $<$0.07 & point  \\
341 & NGC 5061  & 2.7   & 0.04  &   0.08    &  0.04$^b$ & $<$0.43&  point  \\
457 & NGC 7252  & 36.8  & 1.29  &   21.50  & 13.16 & 58.00$\pm$8.70 &  point\\
463 & NGC 7377  & 6.0   & 0.13  &   0.50  & 0.86 & 4.74$\pm$0.44 & point    \\

\hline 
\end{tabular}
\end{center}
$^a$ The P$_{610~MHz}$ was calculated by extrapolating the 610~MHz flux density from the available 1.4~GHz emission, using a spectral index of 0.8\\
$^b$ The P$_{1.4~GHz}$ was calculated by extrapolating the 1400~MHz flux density from the available 610~MHz emission, using a spectral index of 0.8\\
\end{table*}

\begin{figure}
\centering
\includegraphics[width=0.48\textwidth]{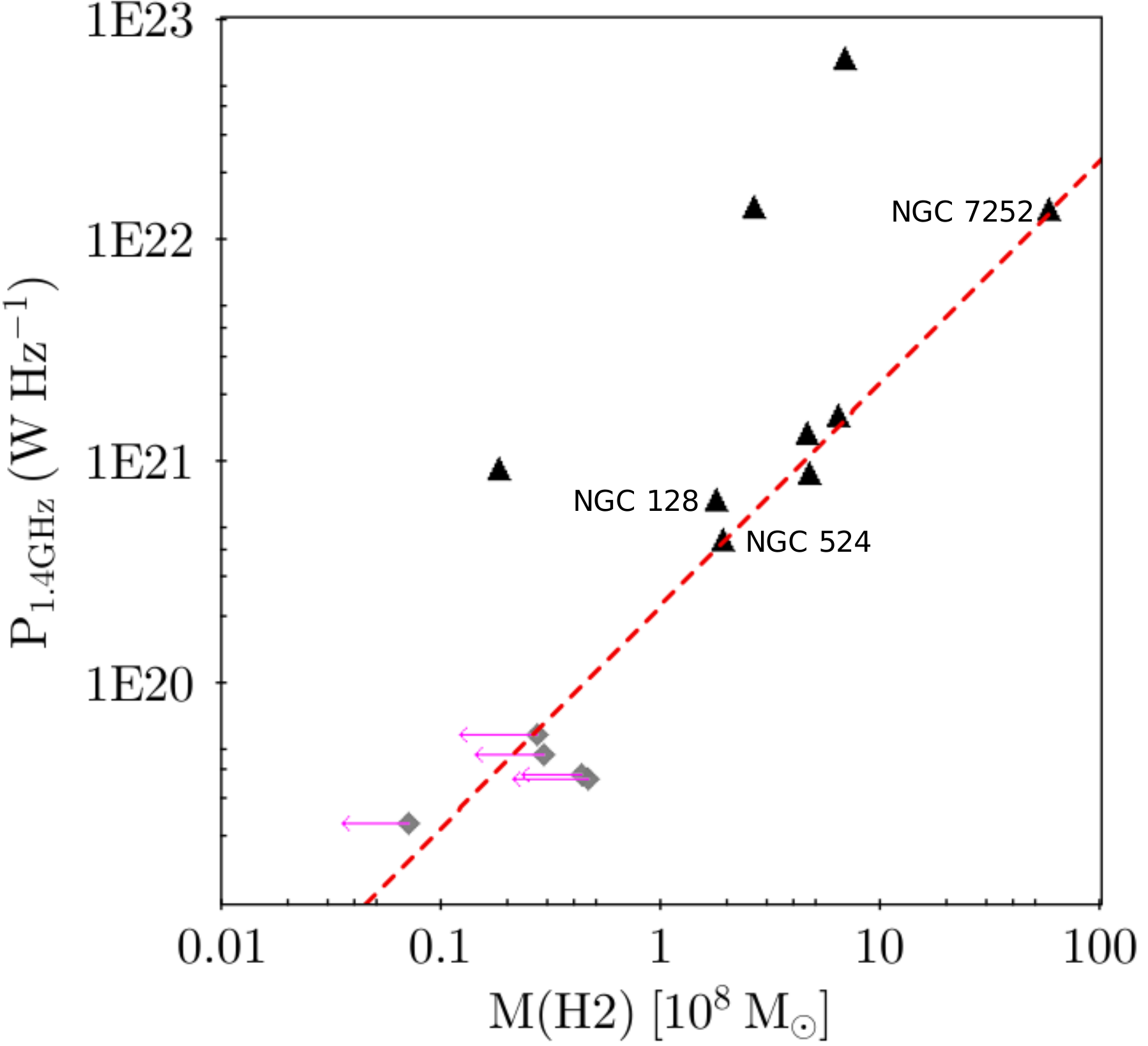}
\caption{Radio power at 1.4~GHz in relation to molecular gas mass for the 14 point-like BGEs in the low--richness sample. Black triangles indicate systems with detected molecular gas and rhombus with arrows indicates systems with 3$\sigma$ upper limits on molecular gas mass. The dashed red line represents the expected radio emission from star formation, assuming the molecular gas mass to star formation rate scaling relation of \citet{GaoSolomon04}.  }
\label{MH2}
\end{figure}

Although the source engine of radio jet systems is most certainly an AGN, it is not clear whether that is the case for the majority of the radio sources in the low--richness sample, which present unresolved (point-like) radio morphology. Unless a luminous radio source is present (P$_{1.4GHz}>$10$^{23}$ W Hz$^{-1}$), it is possible in these systems to confuse a compact central star forming region with AGN emission, as the beam sizes correspond to areas a few kiloparsecs across. Therefore, we examine the possibility of stellar contribution to the radio emission of the low--richness sample BGEs and in particular for those systems which exhibit low radio luminosities. We note here that in NGC~5903, the only system in the low--richness sample that presents a diffuse radio structure, the radio emission likely arises from a combination of AGN jet activity and galaxy interactions driven, and star formation cannot be a significant contributor to the radio luminosity of this system (see \citealt{OSullivan5903}). 

We estimate the expected contribution to the radio flux density from star formation using two methods. Firstly, we use Far-Ultraviolet (FUV) fluxes from the Galaxy Evolution Survey GR6 catalog\footnote{$http://galex.stsci.edu/GR6/?page=mastform$} \citep{Martin05} to estimate the star formation rate (SFR) from the calibration of \citet{Salim07}. We then calculate the expected 610~MHz radio emission from star formation at that rate using the relation of \citet{Garn09} (more details on the star formation in CLoGS BGEs will be presented in a future paper; Kolokythas et al., in prep.). 

Table~\ref{SFR} lists the calculated SFR from FUV and the estimated contribution to 610~MHz radio power, for the galaxies where star formation could potentially affect the radio flux density measurements in the low--richness sample (point-like systems). The only exception is NGC~1779, where no FUV flux was available. We consider that star formation dominates the radio emission if the expected radio power at 610~MHz of a BGE is $>$50\% of its actual P$_{610 MHz}$ value. We find that in 6/13 systems, the luminosity from the detected radio emission exceeds by over an order of magnitude the expected radio luminosity from the FUV SFR, making these galaxies AGN dominated. In three galaxies, (NGC~1395, NGC~3613 and NGC~7377) we find that star formation may contribute 20$-$40\% in the radio emission, in two galaxies may contribute significantly (50$-$60\%; NGC~3923 and NGC~5061) and in two systems (NGC~252 and NGC~4697) it appears to be the dominant source of radio emission with the expected radio luminosity from the FUV SFR exceeding or being similar to the observed radio luminosity. A plausible explanation to why the expected radio luminosity from the FUV SFR exceeds the observed radio luminosity, is that different phases of star formation can be traced by radio and FUV wavelengths - FUV from the largest young stars, radio from supernovae - therefore, a change in SFR over time could lead to a disagreement.


However, using as star formation estimate the FUV wavelength we probe star formation from only young massive stars and cannot account for internal extinction due to dust. In addition, the calibration from \citet{Salim07} is only valid when the SFR remains constant over the lifetime of the UV-emitting stars ($<$10$^8$ yr). Thus the expected radio from SFR$_{FUV}$ can hold for systems that do not present extreme recent starbursts or dust obscured star forming regions. NGC~7252 is likely an example of dust-obscured star formation in a post-starburst merger (see \ref{AppA24}). We find SFR$_{FUV}$$\sim$0.04~M$_\odot$~yr$^{-1}$, considerably less than the SFR estimated from the integrated far infrared (FIR) flux \citep[8.1~M$_\odot$~yr$^{-1}$,][]{OSullivanCO}. However, as noted by \citet{Georgeetal18}, radio and H$\alpha$ SFR estimates for NGC~7252 \citep{Schweizeretal13} agree within a factor of $\sim$2 with the FIR rate, and the implied degree of extinction is not unreasonable for such a gas-rich system. Modelling of the star formation history of the galaxy suggests peak SFRs of tens of M$_\odot$~yr$^{-1}$ \citep{ChienBarnes10} within the last 200$-$600~Myr, and the discrepancy may also indicate a more recent decline in SFR, with the FUV tracing the youngest stars. While the high radio/FUV ratio might have been an indicator of a radio AGN, the low limit on the AGN X-ray luminosity \citep{Nolan04} and its radio morphology strongly suggest the radio emission from this system is dominated by star formation rather than an active nucleus \citep[though there is evidence that the opposite was true in the past,][]{Schweizeretal13}.


In order to test whether the star formation contribution in the radio emission of point-like BGEs is consistent with typical SFRs in star forming galaxies, we also examine the expected radio power from star formation at 1.4~GHz, by converting molecular gas mass to SFR using the scaling relation of \citet{GaoSolomon04}. Following \citet{OSullivanCO18} in figure~\ref{MH2} we show the relation between the radio power at 1.4~GHz and the molecular gas mass (M$_{H2}$) for the 14 point-like radio BGEs (see also Table~\ref{SFR}). The dashed red line indicates the expected radio emission from star formation for the measured M$_{H2}$. We find that along with NGC~7252, five more BGEs (NGC~128, NGC~252, NGC~524, NGC~1779 and NGC7377) fall close to the line for the expected radio emission from star formation along with the systems that present upper limits in molecular gas mass. The systems with upper limits (NGC~1395, NGC~3613, NGC~3923, NGC~4697 and NGC~5061) are seen about the line, indicating that they may be star formation dominated or have a significant contribution from star formation to their radio emission. In NGC~4697 the expected radio emission from SFR$_{FUV}$ is $>$50\% hence we consider that this galaxy's radio emission is star formation dominated. The radio emission from three BGEs (NGC~1106, NGC~2768 and NGC~2911) is clearly far above the line, suggesting that their radio emission is AGN dominated.




While estimating the expected SFR from the molecular gas mass is not a direct measurement like that provided by the FUV flux, we find that both methods suggest that three systems are AGN dominated:  NGC~1106, NGC~2768 and NGC~2911. We find inconsistent results between the two methods for NGC~524, NGC~7252 and NGC~128, which were classed as AGN dominated from their SFR$_{FUV}$, but both methods produce similar results for the systems where star formation may make a significant contribute to the radio emission. We know that NGC~7252 is a starburst system and we hence we consider its radio emission star formation dominated as well as in NGC~524 as is shown by figure~\ref{MH2}. For NGC~128, where P$_{610 MHz}$ was estimated by the 1.4~GHz emission using a spectral index of 0.8, we consider that star formation may contribute significantly to the radio emission as is indicated by figure~\ref{MH2}. 

We therefore in summary, consider that in the low--richness sample, star formation most likely dominates the radio emission in 5/14 of the BGEs with point-like radio emission. It is possible that star formation may have a significant impact on our measurements of radio properties in six more (NGC~128 and the five BGEs with upper limits on molecular gas mass). We find that with certainty, the radio emission is AGN dominated in 3/14 point-like BGEs.

\begin{table}
 \centering
 \caption{Radio morphology occurrence for the BGEs in total CLOGS sample. The numbers for the high--richness sample are drawn from Paper~II.}
\begin{tabular}{lccc}
 \hline
 Radio Morphology    &  Low--richness  &  High--richness & CLoGS total  \\
            
 \hline
 Point-like          & 52\% (14/27) &  54\% (14/26) &  53\% (28/53) \\
 Non-detection       &  19\% (5/27) & 8\% (2/26)   &  13\% (7/53)  \\ 
 Large-scale jet     &  15\% (4/27) & 8\% (2/26)  &  11\% (6/53) \\  Diffuse emission    &  4\% (1/27)  & 14\% (4/26)   &  9\% (5/53) \\ 
 Small-scale jet     &  7\% (2/27)  & 8\% (2/26)  &  8\% (4/53)\\      
 Remnant jet         &  4\% (1/27)  & 8\% (2/26)   &  6\% (3/53)  \\    
 Overall             &  82\% (22/27)&  92\% (24/26)  &  87\% (46/53)\\      
 \hline
 \end{tabular}
 \label{Morphvshighlow}
 \end{table}

\section{DISCUSSION}
\label{disc}
\subsection{Detection statistics in CLoGS}

Brightest group and cluster galaxies (BGGs and BCGs) in the local Universe are more likely to be detected in radio than the wider galaxy population. However, matching group and cluster samples to perform a comparison one has to be prudent, as very often different selection criteria have been applied (redshift, luminosity etc). The most recent statistical study by \citet{Hogan15} using different X-ray selected cluster samples including systems out to redshift z$\sim$0.4, reports a radio detection percentage for BCGs of $60.3\pm7.7\%$ (from the ROSAT ESO Flux Limited X-ray/Sydney University Molonglo Sky Survey (REFLEX-SUMSS) sample, \citealt{Bohringer04}), $62.6\pm5.5\%$ (REFLEX-NVSS sample), and $61.1\pm5.5\%$ from the extended Brightest Cluster Sample (eBCS; \citealt{Ebeling00}). 
BCGs in the X-ray selected galaxy clusters of \citet{Ma13} and \citet{Kale15} (from the Extended Giant Metrewave Radio Telescope -GMRT- Radio Halo Survey; EGRHS; \citealt{Venturi07, Venturi08}) present a radio detection rate of $\sim$52\% and $\sim$48\% respectively, from the NVSS ($>3$~mJy) and NVSS/FIRST catalogs.


In what may be the most relevant studies for CLoGS, \citet{Dunn10} used a sample of nearby early-type galaxies, most of which are in the centres of groups or clusters, and found a detection rate of $\sim$81\% (34/42) using NVSS and SUMSS data. \citet{Bharadwaj14} report a $\sim$77\% (20/26) detection rate for the X-ray selected groups sample of \citet{Eckmiller11} using the NVSS, SUMSS and VLA Low frequency Sky Survey (VLSS, 74~MHz) radio catalogs. 

In the CLoGS low--richness sample, considering only the 1.4~GHz data, we find a detection rate of $\sim$78\% (21/27 BGEs) with a similar detection rate of $\sim$81\% (21/26 BGEs) being found for the high--richness sample in Paper~II. In total, the detection percentage at 1.4~GHz for the complete CLoGS sample is found to be $\sim$79\% (42/53 BGEs) which is in good agreement with the general trend of the high detection rate of brightest galaxies in the local Universe found by the studies mentioned above. We also note that CLoGS brightest group galaxies present a higher detection rate compared to those from clusters, as the average detection rate value for the BCGs from \citet{Hogan15} is $\sim$61.3$\pm$11\%. However, we note here that the latter study included cluster systems at a much greater volume (z$\sim$0.4) than our CLoGS groups ($\sim$20x farther) along with systems from the southern hemisphere, where the detection limit from SUMSS is $\sim$6~mJy ($\sim$2.5x higher than NVSS). Hence, we treat this only as indicative of the relative detection statistics between groups and clusters.

The total radio detection rate for the CLoGS sample at any of the three radio frequencies (235, 610 and 1400~MHz) is $\sim$87\% (46/53 BGEs) which drops to $\sim$79\% (42/53 BGEs) if we consider only the GMRT data. In a comparison between the CLoGS two group sub-samples, the radio detection rate was found to be slightly higher for the high--richness sample (24/26 BGEs, $\sim$92\%; Paper~II) than for the low--richness groups ($\sim$82\% 22/27 BGEs), but the difference is not  statistically significant.

Examining the radio morphology appearance in CLoGS sample as a total, we find that more than half of the radio sources ($\sim$53\%; 28/53) present point-like radio emission, followed by the BGEs with non-detections ($\sim$13\%; 7/53). Large-scale jets are present in $\sim$11\% (6/53) of the BGEs, $\sim$9\% (5/53) host a diffuse radio source, $\sim$8\% (4/53) a small-scale jet, and $\sim$6\% (3/53) a remnant of an old radio outburst. In Table~\ref{Morphvshighlow} we show the occurrence of radio morphology in total and in both sub-samples. While both sub-samples present a similar rate for small-scale and remnant jet morphologies, large-scale jet emission appears at higher rate in high-richness sample, with the most prominent difference between the two sub-samples being the higher rate of undetected radio sources presented in the low--richness sample ($\sim$19\%) compared to the high--richness one ($\sim$8\%). 

We find that, in total, 6/53 of our CLoGS BGEs can be considered as radio-loud following the definition by \citet{Best05} (P$_{1.4GHz}>10^{23}$~W~Hz$^{-1}$). This gives a percentage of $\sim$11\% radio loud galaxies for CLoGS groups which is similar to the $\sim$13\% fraction of radio loud galaxies found in low-mass clusters and groups (10$^{13}<M_{200}<10^{14.2}$) by \citet{LinMohr07}.

We find that $\sim$9$\pm4$\% (5/53) of the CLoGS BGEs are currently hosts of double-lobed radio galaxies (note that this excludes the lobeless double-jet system NGC~5490). This is slightly higher than the $\sim$8$\pm5$\% (2/26; Paper~I) found from the high--richness sub-sample but comparable within uncertainties. \citet{MeiLin10} find that the equivalent percentage of BCGs having double-lobed radio morphology, using a catalog of $\sim$13000 clusters from the maxBCG sample, is $\sim$4\%. 
It should also be noted that the remnant radio sources in our sample (NGC~1167, NGC~1407 and NGC~5044) all appear to be the remains of large-scale jet-lobe systems.  Although this could suggest that double-lobed radio sources are more common in galaxy groups than clusters, this result needs to be treated with caution, since it is unclear how sensitive the cluster samples are to such steep spectrum remnant sources.


\begin{figure}
\centering
\includegraphics[width=0.49\textwidth]{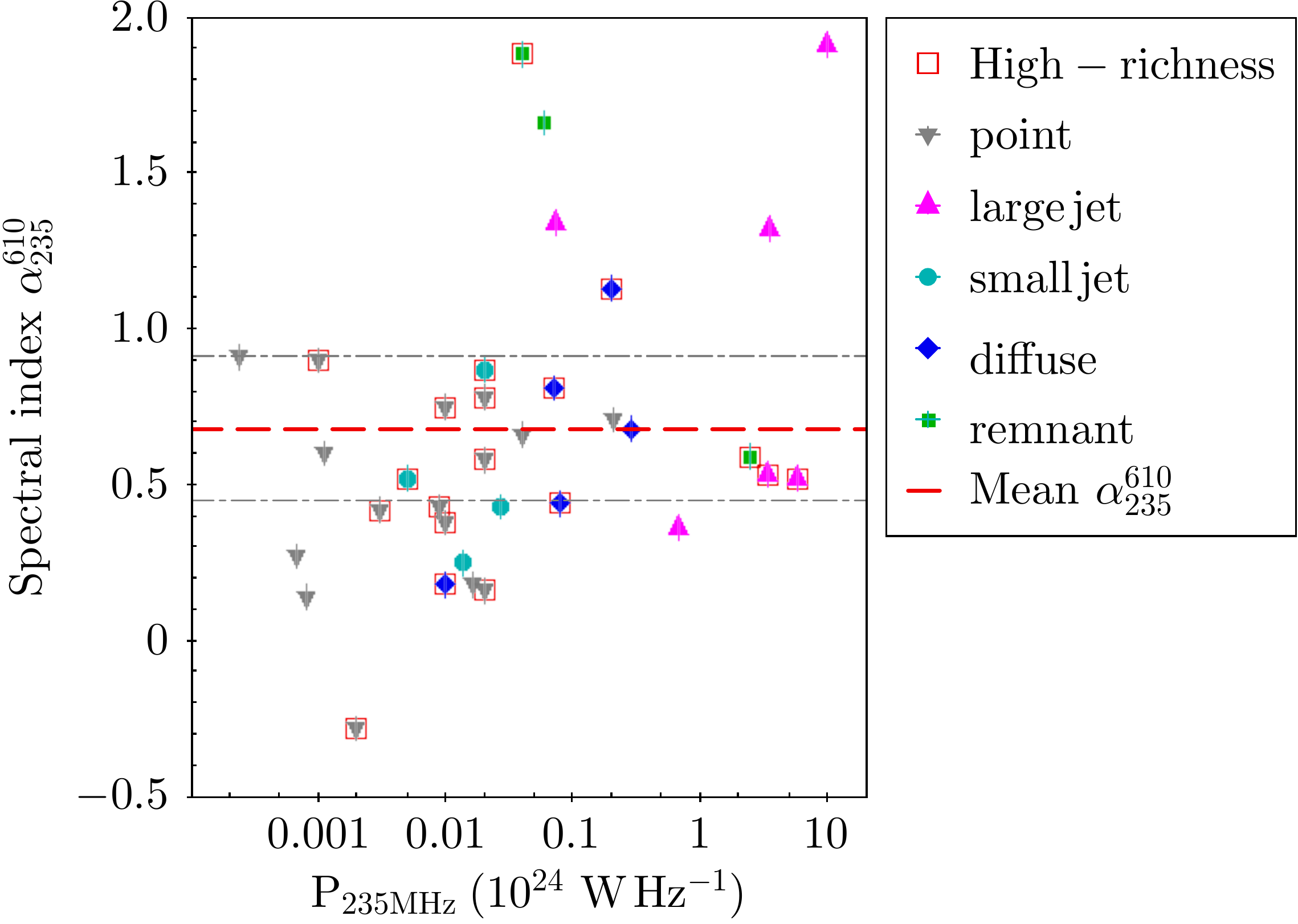}
\caption{Spectral index $\alpha$ of CLoGS different radio morphologies in the frequency range 235$-$610~MHz, in relation to radio power at 235~MHz (P$_{235 MHz}$). The radio morphology of each source is indicated by the symbols, with the groups from the high--richness sample being marked by open squares. 
}
\label{spixp235}
\end{figure}

\subsection{Spectral index of CLoGS BGEs}

The spectral index of a radio source provides insight on the age and the nature of the mechanisms that give rise to radio emission \citep{LisenfeldVolk00}. In AGN, the most important mechanisms at high frequencies are inverse Compton losses and synchrotron emission which is produced by relativistic electrons spiralling down magnetic field lines. The cores of AGN usually present a flat radio spectrum as a result of self-absorption at low frequencies rather than a flat electron energy distribution, with a steep spectrum being observed from the extended emission (lobes) of an AGN. 






 For extended radio galaxies that lie in the centre of cool-core clusters and groups, the spectral index distribution is generally found to be $1-2$ in the 235$-$610~MHz frequency range \citep{Simona11}. The spectral indices of these sources are thought to be steep owing to the confinement of the relativistic radio plasma by the surrounding intra-cluster medium  \citep[e.g.,][]{Fanti95,Murgia99}.  \citet{Bornancini10}, using a sample of maxBCG clusters and radio data from the Faint Images of the Radio Sky at Twenty-cm survey (FIRST; 1400~MHz), the NVSS (1400~MHz), the Westerbork Northern Sky Survey (WENSS; 325~MHz) and the Green Bank 6~cm Survey (GB6; 4850~MHz), calculated the mean value of the spectral index for BCGs between 325~MHz and 1.4~GHz to be $\alpha_{325}^{1400}=0.65$. However, the authors did not select by radio morphology, and the inclusion in the spectral index calculation of radio point sources as well as jet-lobe systems probably explains the flatter mean index, showing that different morphological types of radio sources have an impact on the mean value of the spectral index.

In the CLoGS sample we are able to measure the 235-610~MHz spectral index in 34/53 BGEs. The mean spectral index for these radio sources is $\alpha_{235}^{610}=0.68\pm0.23$, with a standard deviation of 0.55, whereas in the 235$-$1400~MHz frequency range, the mean radio spectral index for 33/53 BGEs is $\alpha_{235}^{1400}=0.59\pm0.26$ with a standard deviation of 0.43. The uncertainties on the 1400~MHz flux for two galaxies, NGC~3923 and NGC~5982, were excluded from the error calculation of the mean $\alpha_{235}^{1400}$ as their high uncertainties provided for their low flux densities by \citet{Brown11} are equivalent to their value ($\leq$1~mJy see~Table~\ref{Sourcetable} and Paper~II) which was increasing the average error on the mean by over a factor of 2. 

The mean 235$-$610~MHz spectral index for CLoGS sample is significantly flatter than the value found by \citet{Simona11}, probably owing to the inclusion of sources with a wide range of morphologies in our study. If we take into consideration the spectral indices only from the CLoGS large$-$scale and remnant jet systems in order to match with the extended group radio sources examined by the sample of \citet{Simona11}, we find that $\alpha_{235}^{610}=1.13\pm0.12$, in much better agreement. The mean $\alpha_{235}^{1400}$ of $0.59\pm0.26$ found for all our CLoGS systems is very close to the mean spectral index found by \citet{Bornancini10} ($\alpha_{325}^{1400}=0.65$) using a similar frequency range. 

In the recent study of \citet{deGasperin18}, using data extracted from the NVSS (1400 MHz) and a re-imaged version of the TGSS (TIFR GMRT Sky Survey, 147~MHz; \citealt{IntemaTGSS}), the weighted average spectral index distribution $\alpha_{147}^{1400}$ for over a million extra-galactic radio sources was found to be 0.79 with a standard deviation of 0.24 (for $S_\nu \propto \nu^{-\alpha}$). This is in agreement with previous results on extra-galactic radio sources which found  mean spectral indices $\alpha_{147}^{1400}$ in the range $0.75-0.8$ \citep{Chandra10,Intema11}. Although these studies include radio sources from various environments and redshifts, the CLoGS mean spectral index is consistent (within uncertainties) with their mean spectral indices.



Table~\ref{radiospix} shows the mean spectral index values in the 235$-$610~MHz and 235$-$1400~MHz bands for the different radio morphology classes in the full CLoGS sample. We find that point-like systems present a mean spectral index of $\alpha_{235}^{610}=0.47\pm0.16$, diffuse radio sources a value of $\alpha_{235}^{610}=0.65\pm0.09$ with jet systems (both small and large scale jets) presenting a steeper index of $\alpha_{235}^{610}=0.81\pm0.13$. The mean 235-1400~MHz spectral index in jet systems is $\alpha_{235}^{1400}=0.62\pm0.14$ with the mean value for point-like systems being $\alpha_{235}^{1400}=0.43\pm0.26$. 


As expected, jet systems show a somewhat steeper mean spectral index than point-like sources. In figure~\ref{spixp235} we present the spectral index $\alpha_{235}^{610}$ distribution of CLoGS BGEs, in relation to their radio power at 235~MHz (P$_{235 MHz}$). We observe that point-like systems fall mainly about the calculated mean spectral index distribution. We also find that 50\% (3/6) of the large$-$scale jet systems exhibit steep $\alpha_{235}^{610}$ values with the other half having values close to the mean. While small$-$scale jet systems present small deviations in their spectral index values, we see that large$-$scale systems are divided between flat and very steep spectral $\alpha_{235}^{610}$ indices.

This divide in the large-scale jet sources might be explained by the different evolutionary stages of the jet systems (younger sources with flatter spectral indices, vs. older or remnant sources with aged electron populations) or by the properties of the environment that the jets propagate into (e.g., bending lobes or confinement of electrons due to higher density environment). From figure~\ref{spixp235} we also find a set of six radio sources that present a very steep spectrum, ($>1$), in the 235$-$610~MHz frequency range, with two of them being naturally remnant radio sources, which as expected present the largest deviation from the mean in our sample, another three sources being large$-$scale jet systems and the remaining one being of diffuse radio morphology. 


 \citet{Intema11} categorized as ultra-steep spectrum (USS) those sources with spectral indices steeper than 1.3, and found that the fraction of USS found in the 153$-$1400~MHz frequency range is 3.8\% (16/417). This is in good agreement with earlier studies (e.g., 3.7\%; \citealt{Sirothia09}). In the CLoGS sample, we find that only 2 radio sources present a spectral index steeper than 1.3 in the 235$-$1400~MHz frequency range (NGC~1407 and NGC~1550). This gives a USS fraction in good agreement with larger samples (3.8\%), but the result is strongly dependent on the detection limit at 235~MHz.


\begin{table}
 \centering
 \caption{Mean spectral indices $\alpha_{235}^{610}$  and $\alpha_{235}^{1400}$ for the different radio morphology classes in the 53-group CLoGS sample.}
\begin{tabular}{lcc}
 \hline
 Radio Morphology      &  Mean $\alpha_{235}^{610}$    &  Mean $\alpha_{235}^{1400}$  \\
            
 \hline
 Point-like            &  0.47$\pm$0.16    &    0.43$\pm$0.26       \\
 Small-scale jet       &   0.52$\pm$0.08   &    0.47$\pm$0.08       \\      
 Remnant jet           &   1.38$\pm$0.07   &    1.11$\pm$0.07       \\            
 Large-scale jet       &   1.00$\pm$0.10   &    0.72$\pm$0.11      \\                 
 Diffuse emission      &   0.65$\pm$0.09   &    0.72$\pm$0.10      \\ 
 \hline
 \end{tabular}
 \label{radiospix}
 \end{table}

\subsection{Environmental properties}

 We investigate the relation between the radio morphology of a BGE and its near environment by examining the fraction of spiral galaxies in each group. The number of spiral galaxies that a group possesses is known to be connected to its dynamical age, as older systems will have had more opportunities for galaxy interaction and/or merging to drive morphological transformation.

We classify as spiral-rich (i.e, dynamically young) galaxy groups with a spiral fraction F$_{sp}>0.75$ \citep{Bitsakis10}. In Table~\ref{Fsp} we present the fraction of spiral galaxies in each group for the low--richness sample. The morphologies of the galaxies were drawn from the  HyperLEDA\footnote{$http://leda.univ-lyon1.fr/$} catalogue, with morphological T-type $<$0 galaxies being classified as early-type, and T-type $\geq$ 0 (or unknown) galaxies classed as late-type. As in Paper~II, we define spiral fraction F$_{sp}$ as the number of late-type galaxies over the total number of group members.


We find that 19/27 (70$\pm$9\%) of the groups in the low--richness sample are spiral-rich and 8/27 (30$\pm$9\%) are spiral-poor, whereas for the high--richness sample in Paper~II we found that 11/26 (42$\pm$10\%) of the groups are spiral-rich and 15/26 (58$\pm$10\%) spiral-poor. Declining spiral fraction with richness may be an indication that galaxies undergo gradual morphological transformation if rich groups emerge over an extended period of time, or it may be a manifestation of galaxy downsizing, where galaxy-sized density peaks riding on larger-scale density fluctuations destined to form rich groups collapse early and rapidly to form preferentially spheroidal galaxies.

Examining the CLoGS sample in total, 30/53 (57\%) of the groups are spiral-rich and 23/53 (43\%) are spiral-poor. \citet{Bitsakis14}, using a sample of 28 Hickson compact groups \citep[HCGs,][]{Hickson92}, found that 46\% (13/28) were spiral-rich and 54\% (15/28) spiral-poor. Although \citet{Bitsakis14} used a sample of compact groups which are likely to have higher galaxy interaction rates and thus evolve more rapidly than our systems, the spiral-rich fraction found in \citet{Bitsakis14} is similar to our high--richness sample. This again demonstrates that richer groups are a conducive environment for galaxy evolution.

\begin{figure}
\centering
\includegraphics[width=0.49\textwidth]{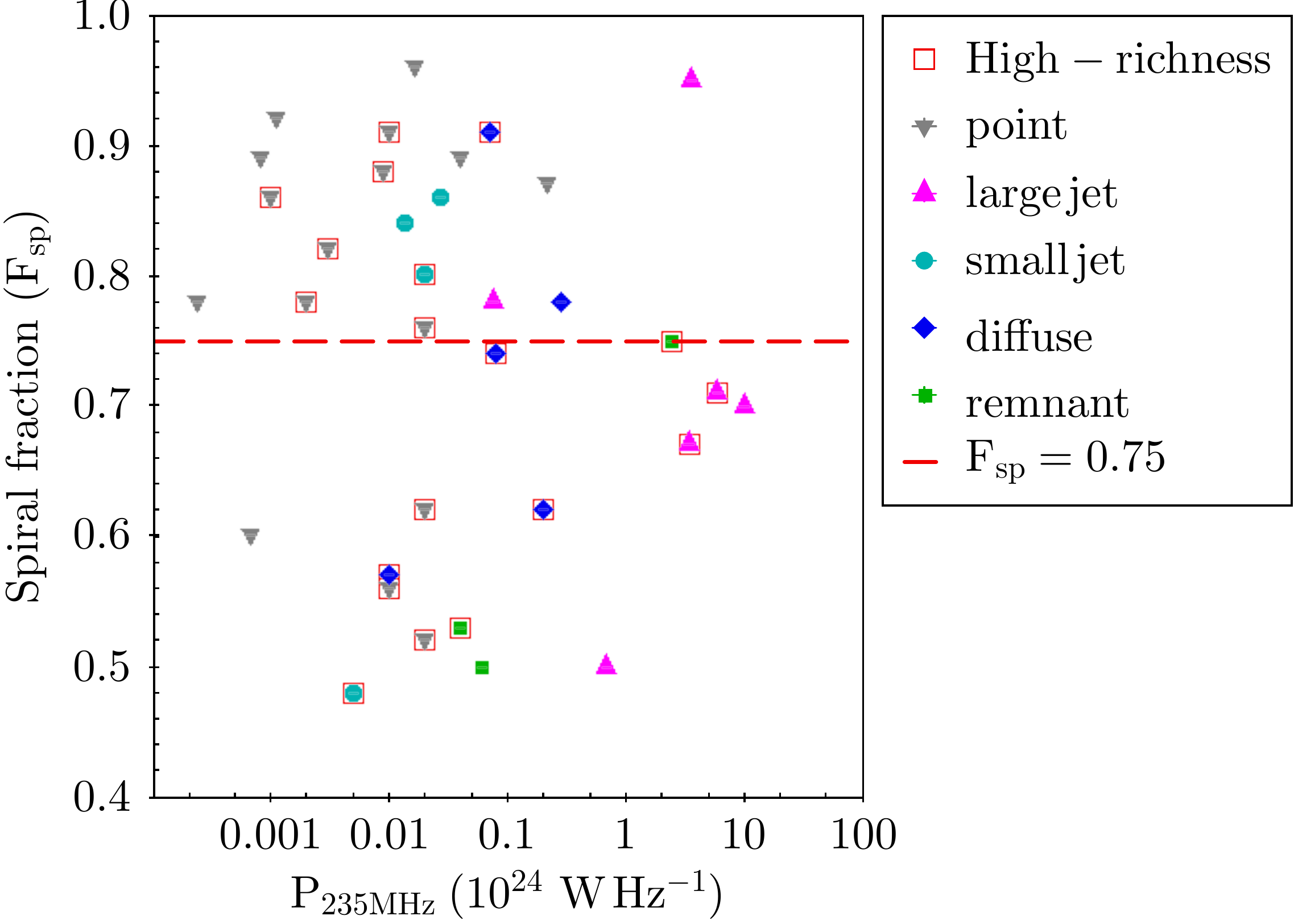}
\caption{Spiral fraction F$_{sp}$ plotted against 235~MHz radio power, with symbols indicating radio morphology of the BGE. Groups from the high--richness sample are marked with open squares.}
\label{Fspunknown}
\end{figure}


In Fig~\ref{Fspunknown} we show the relation between F$_{sp}$ and the radio power at 235~MHz (P$_{235 MHz}$) for all CLoGS systems whose BGEs are detected at 235~MHz. For the low--richness sample we find that all but one of the point-like systems appear in galaxy groups with higher spiral fractions (6/7; 86\%), with a mean spiral fraction of 0.84$\pm$0.14. Jet systems in the low-richness sub-sample are also found to reside more often in a spiral-rich environment (4/6; 67\%) having a mean spiral fraction of 0.77$\pm$0.17. 

Looking at CLoGS groups as a whole, we find that the majority (75\%) of radio point-like systems detected at 235~MHz (12/16) reside in groups with high spiral fractions (dynamically young) with a mean spiral fraction of 0.78$\pm$0.10. Jet systems on the other hand, are found to be equally distributed between spiral-rich (5/10; 50\%) and spiral-poor environments (5/10; 50\%). However, we find the mean F$_{SP}$ for jet sources (0.73$\pm$0.14) is similar to that for point radio sources. Given the large uncertainties, this result cannot be regarded as statistically significant and larger sample of groups are required in order to confirm it. However, the even appearance of jet systems in both spiral rich and poor groups suggests that the dynamical evolution of the galaxy population of a group has little influence on the radio jet activity of the BGE. This is of interest given the connection between jets and intra-group medium properties found in paper~I, where we found that more massive groups were more likely to have declining central temperature profiles and that the presence of jets was correlated with these cool cores and other evidence of rapid gas cooling. We might have expected that more dynamically evolved (i.e., spiral-poor) groups, which are likely to be more massive, would more commonly host central jet sources. Of the remnant jet sources, 2/3 (NGC~5044 and NGC~1407) reside in dynamically old groups (F$_{SP}\sim$0.5) while the third, NGC~1167, is the BGE of an X-ray faint group with a high spiral fraction (0.75). Diffuse radio sources, similar to jet systems, occur in groups with a wide range of spiral fractions (0.57$-$0.91) with 3/5 residing in high spiral fraction groups. 
Our X-ray analysis of the low-richness sample is not yet complete, and will be reported in a later paper, but our preliminary analysis suggests that 10/13 jet systems in the full CLoGS sample are hosted by X-ray bright groups with group-scale X-ray halos. As reported in paper~I, NGC~1167 is the BGE of an X-ray faint group with little or no hot gas, and in the low-richness sample we find two jet sources in systems with only galaxy-scale gas halos. This suggests that even in low mass, less evolved groups, a hot gas rich environment is preferable for jets \citep[e.g.,][]{McNu07}. We examine the properties of jet-hosting BGEs in more detail in the next section (\S~\ref{disc4}).



\begin{table}
\begin{minipage}{\linewidth}
 \caption{Spiral fraction for the low--richness CLoGS groups. Spiral fraction (F$_{sp}$) is defined as the ratio of the number of late-type and unknown morphology galaxies to the total number of group members.}
\centering
 \label{Fsp}
\begin{tabular}{|c|c|c|}
\hline 
 LGG & BGE   &    Spiral fraction   \\ 
       &    F$_{sp}$  \\
\hline 
LGG 6   & NGC 128   & 0.71 \\
LGG 12  & NGC 252   & 0.67 \\ 
LGG 14  & NGC 315   & 0.70 \\
LGG 23 & NGC 524    & 0.64 \\
LGG 78 & NGC 1106   & 0.87 \\
LGG 97 & NGC 1395   & 0.79 \\
LGG 100 & NGC 1407  & 0.50 \\
LGG 113 & NGC 1550  & 0.78 \\
LGG 126 & NGC 1779  & 0.92 \\
LGG 138 & NGC 2292  & 0.64 \\
LGG 167 & NGC 2768  & 0.89 \\
LGG 177 & NGC 2911  & 0.96 \\
LGG 205 & NGC 3325  & 0.96 \\
LGG 232 & NGC 3613  & 0.90 \\
LGG 236 & NGC 3665  & 0.86 \\
LGG 255 & NGC 3923  & 0.78 \\
LGG 314 & NGC 4697  & 0.87 \\
LGG 329 & NGC 4956  & 0.90 \\
LGG 341 & NGC 5061  & 0.86 \\
LGG 350 & NGC 5127  & 0.95 \\  
LGG 360 & NGC 5322  & 0.84 \\
LGG 370 & NGC 5444  & 0.79 \\
LGG 376 & NGC 5490  & 0.50 \\
LGG 383 & NGC 5629  & 0.95 \\
LGG 398 & NGC 5903  & 0.78 \\
LGG 457 & NGC 7252  & 0.89 \\
LGG 463 & NGC 7377  & 0.60 \\
\hline 
\end{tabular}
 \end{minipage}
 \end{table}

\subsection{Power output and properties of jet activity in CLoGS} 
\label{disc4}
As in Paper~II, the energy output of jet sources in the low--richness sample is estimated using the P$_\mathrm{cav}$ $-$ P$_{235}$ scaling relation by \citet{OSullivan11}. This relation estimates the cavity power of a jet system based on the BGE's radio power at 235~MHz allowing the use of radio data as a proxy to estimate the AGN mechanical power output of radio jet systems. It has the benefit of providing with an estimate of the mechanical power output in the absence of direct calculations of cavity power (P$_\mathrm{cav}$) from the X-rays. Table~\ref{radiomorph} shows the results from the calculations of the radio energy output. We find that the energy output values for the jet systems in the low--richness sample range from $\sim10^{42}-10^{44}$ erg~s$^{-1}$, which is common for galaxy groups, but roughly an order of magnitude greater than the range found for the high--richness sample. The giant radio galaxy NGC~315 produces the difference in the upper bound of the range, and given the limited scale of our sample, the low and high--richness sub-samples are likely comparable.


For the BGEs in the CLoGS sample that present either an ongoing jet radio emission (large and small$-$scale jets; 10/53) or a past one (remnant jets; 3/53) we examine whether a correlation exists between the mass of the central black hole (BH) and their radio activity at 235~MHz. We calculate the BH masses for all CLoGS systems using the correlation of \citet{FerrareseMerritt00}, estimating the mass of the BH from the central velocity dispersion of the host galaxy. This correlation is expressed as:

\begin{equation}
    log_{10}\Big(\frac{M_{BH}}{M_{\odot}}\Big)=\alpha+b log_{10}\Big(\frac{\sigma_o}{200~km~s^{-1}}\Big)
\end{equation}

where M$_{BH}$ is the mass of the black hole in solar masses (M$_\odot$) and $\sigma_o$ is the central velocity dispersion. The central velocities dispersions for 23/53 of our BGEs were drawn from the recent kinematic study of \citet{Loubser18} with the values for the remaining 30 obtained from the HyperLeda\footnote{$http://leda.univ-lyon1.fr/$} catalogue. We estimated the BH masses for all CLoGS BGEs (see~Table~\ref{MBHsigma}) using the values for the parameters  a = 8.39 $\pm$ 0.06 and b = 5.20 $\pm$ 0.36 given by the fit of the velocity dispersion$-$black hole mass relation in \citet{McConnellMa13}.

\begin{figure}
\centering
\includegraphics[width=0.49\textwidth]{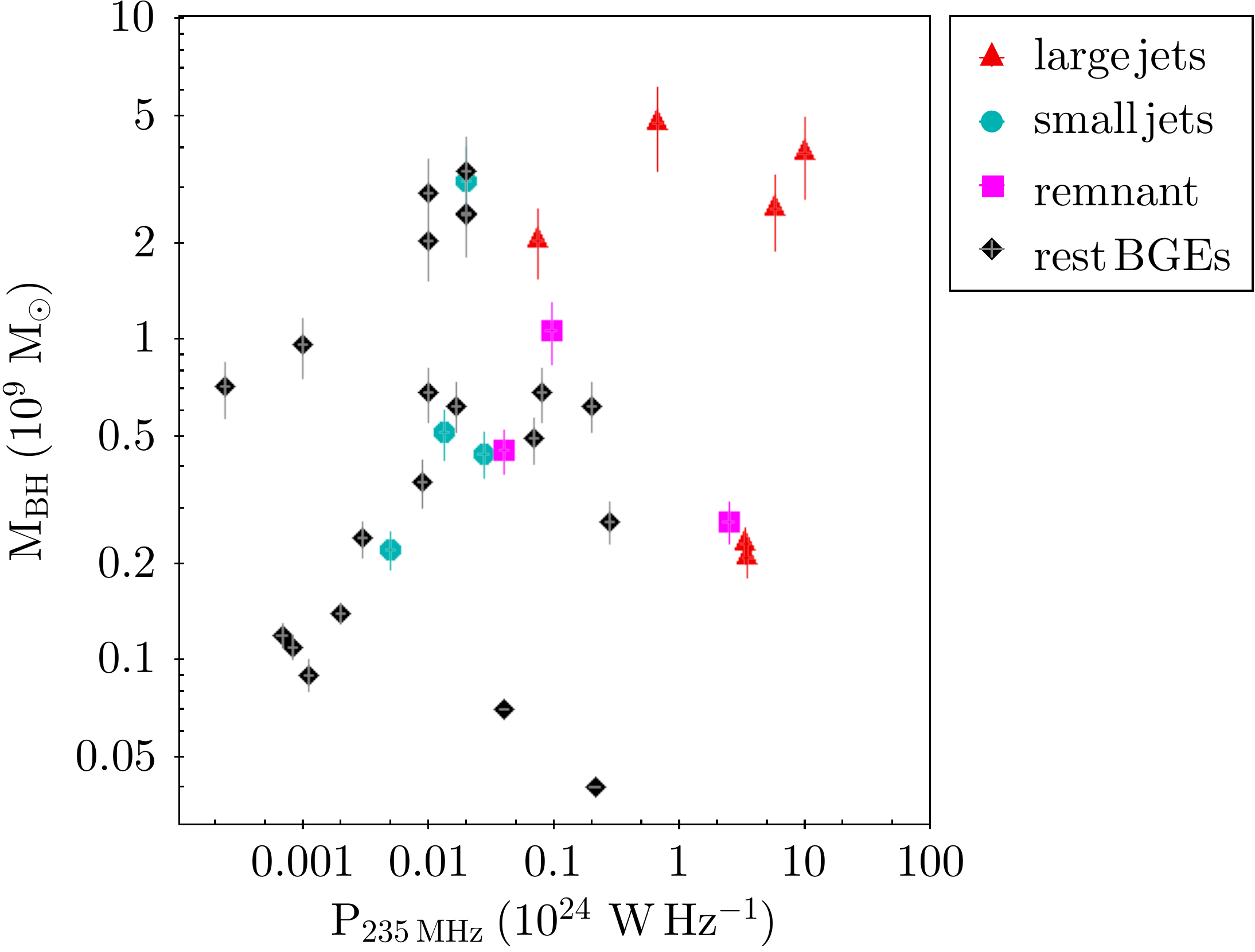}
\caption{Black hole mass (M$_{BH}$) of different radio morphologies, in relation to radio power at 235~MHz (P$_{235 MHz}$). The radio morphology of each group can be seen in different symbols, with the host BGEs other than jet systems being marked as `rest of BGEs' in triangles.}
\label{BHP235}
\end{figure}

We find that the BGE systems with 235~MHz detections (34/53) have BH masses in the range of $\sim$10$^8$ $-$ 5$\times$10$^9$~M$_\odot$, with the majority of the systems having masses $\leq$10$^9$~M$_\odot$. Figure~\ref{BHP235} shows the relation between the radio power at 235~MHz (P$_{235MHz}$) and the BH mass for different radio morphologies. Jet systems present in total a BH mass range of 0.2$\times$10$^9$ $-$ 5$\times$10$^9$~M$_\odot$, with an average of $\sim$2.3$\pm$2.7$\times$10$^9$~M$_\odot$ for the large$-$scale systems and $\sim$1.07$\pm$1.16$\times$10$^9$~M$_\odot$ for the small$-$scale. 

We observe that 4/6 large$-$scale jet systems present a BH mass $>$2$\times$10$^9$ M$_{\odot}$ with the hosts of 2/6 large$-$scale systems exhibiting a BH mass an order of magnitude less (NGC~193 and NGC~5127; $\sim$2$\times$10$^8$ M$_{\odot}$). The most extended large$-$scale radio jet system in CLoGS ($\sim$1200~kpc; NGC~315) presents a BH mass estimate of $\sim$3.8$\times$10$^9$~M$_\odot$. The recent study of \citet{Horellou18} reports a BH mass of $\sim$1.6$\times$10$^9$~M$_\odot$ for the BGE that hosts the giant radio source `Double Irony' ($\sim$1100~kpc), which is equivalent to the BH mass that we find for NGC~315. 
We observe that 5/8 systems with BH mass $\geq$10$^9$~M$_\odot$ are large$-$scale jet systems. For the systems that have no radio emission detected at any frequency (7/53) we find that their BH mass ranges $\sim$0.4$\times$10$^8$ $-$ 9$\times$10$^8$~M$_\odot$ with an average of 2.8$\pm$2.5$\times$10$^8$~M$_\odot$ which is on order of magnitude lower than the host BGEs of large$-$scale / jet systems.  

On the other hand, the majority of small$-$scale jet systems (3/4) presents BH masses $\sim$2$-$5$\times$10$^{8}$ M$_{\odot}$ with only one system, NGC~1060, which is known as a dynamically interacting group \citep{OSullivan17} having a BH mass of $\sim$3.1$\times$10$^{9}$ M$_{\odot}$. We also note that only 10 BGEs present BH masses $>$10$^{9}$ M$_{\odot}$ with all of them being known to be embedded in an extended X-ray environment (Paper~II and \citet{Sun03} for NGC~1550). 

From figure~\ref{BHP235} we observe a mild trend that suggests higher BH masses are linked with highest radio powers, as found by previous studies \citep[e.g.,][]{Franceschini98,Liu06}. However, we do not see a strong correlation between radio power at 235~MHz and the BH mass of the host BGEs in jet systems. There is a broad dispersion in radio power for a given BH mass for our CLoGS BGEs suggesting that a certain threshold value of BH mass may be necessary (but not on its own sufficient) to produce a powerful radio source, as suggested by a study of nearby field and group elliptical galaxies by \citet{Vaddi16}. This may indicate that jet activity depends more closely on environmental processes and the fuel that is available, rather than the mass of the host BGE's BH. 
In the diverse environment of galaxy groups we find that BH mass is not the main drive for the occurrence of jet emission, in agreement with the recent study of \citet{Sabater19} in a sample of nearby AGN in the LOFAR Two-meter Sky Survey (LoTSS). Environmental mechanisms such as interactions, radiative cooling of the IGM, and precipitation of cold molecular gas are most likely prominent drivers for jet emission in galaxy groups. In addition to the effects of the large scale environment, the efficiency of gas accretion and the BH spin likely also play a role in the large radio power dispersion seen for our group BGES at a given BH mass \citep[e.g.,][]{Baum95}.


 \section{Conclusions}
 \label{conclusions}
 
In this paper we have presented the GMRT radio observations of the brightest group early-type galaxies from the low--richness CLoGS sample at 235 and 610~MHz. Combining with the study of the high--richness sample from Paper~II we present results and statistics from the complete CLoGS sample. 
The CLoGS low--richness sample is found to have a radio detection rate of 82\% (22 of 27 BGEs) at 235, 610 or 1400~MHz, with three radio sources being characterized as radio loud (P$_{1.4~GHz}$ $>$10$^{23}$ W Hz$^{-1}$ - NGC~315, NGC~5127 and NGC~5490). The overall detection rate of the CLoGS sample as a whole is 87\% (46 of 53 BGEs), a result which confirms the high radio detection rate of dominant galaxies in groups or clusters in the nearby Universe. We find that CLoGS group-dominant radio galaxies exhibit radio powers in the range of 10$^{20}$ $-$ 10$^{25}$ W Hz$^{-1}$. 

Examining the radio morphology, we find that the majority of the groups in the CLoGS sample, $\sim$53\% (28/53) have a point-like radio morphology while $\sim$13\% (7/53) of groups are not detected in radio at any frequency. $\sim$25\% (13/53) of the groups are found to be hosting currently or recently active small/large scale jets and the remaining $\sim$9\% (5/53) host diffuse radio sources. The size of the extended CLoGS radio sources spans over 3 orders of magnitude from $\sim10-1200$~kpc.

For the very low-powered unresolved point-like radio sources in the CLoGS low--richness sample using the expected radio emission from SFR$_{FUV}$ and the expected radio emission from the conversion of molecular gas to star formation, we find that 3/14 are AGN dominated (NGC~1106, NGC~2768 and NGC~2911), and 5/14 sources (NGC~128, NGC~1395, NGC~3613, NGC~3923 and NGC~5061) probably have a significant contribution (20-50\%) to their radio emission from star formation. In 6/14 point-like radio sources (NGC~252, NGC~524, NGC~1779, NGC~4697, NGC~7252 and NGC7377) we find that the stellar population likely dominates the radio emission.

In the CLoGS low--richness sample only 15/27 galaxies are detected at both 610 and 235~MHz with 2/15 sources exhibiting ultra-steep radio spectra with $\alpha_{235}^{610}>1.3$ (NGC~315 and NGC~1407). The mean spectral index value for 13/15 sources is  $\alpha_{235}^{610}=0.60\pm0.16$ with the equivalent mean value in the 235$-$1400~MHz frequency range measured to be $\alpha_{235}^{1400}=0.57\pm0.16$. In the full CLoGS sample we are able to measure a spectral index for 34/53 BGEs, and the mean spectral index is $\alpha_{235}^{610}=0.68\pm0.23$. In the 235$-$1400~MHz frequency range the mean spectral index for 33/53 BGEs is measured to be $\alpha_{235}^{1400}=0.59\pm0.26$. Jet systems are found to exhibit steeper mean $\alpha_{235}^{610}$ than point-like systems. Roughly 4\% of our sources have ultra-steep spectral indices ($\alpha_{235}^{610}>$1.3), in good agreement with the earlier studies of  \citet{Intema11} and \citet{Sirothia09}.

Considering the group environments in which the BGEs are located, we find that 57\% (30/53) of the CLoGS groups are spiral-rich with 43\% (23/53) being spiral-poor. Comparing the mean spiral fraction (F$_{sp}$) to BGE radio morphology, we find that the majority (75\%) of radio point-like systems are found in dynamically young group systems (mean spiral fraction of 0.78$\pm$0.10), with jet systems appearing equally in both dynamically young and old group systems, meaning that groups may form/host jets at any stage of their evolution. Our three remnant jet systems are also split between spiral-rich and spiral-poor groups, and BGEs which host diffuse, non-jet radio sources appear in groups with a wide range of spiral fractions.

We estimate the mechanical power output of the jet sources in the CLoGS low--richness sample from their 235~MHz luminosity, and find it to be in the range $\sim10^{42}-10^{44}$ erg~s$^{-1}$. This is typical for galaxy groups, but about an order of magnitude greater than the range of powers measured in our high-richness sub-sample. A combined X-ray and radio analysis will be needed to explore the reliability and origin of this difference. 

Lastly, for all radio detected CLoGS groups, we find only a mild positive trend between the BH mass of the BGE and its radio power at 235~MHz. We do not find a correlation between the central BH mass and jet activity. Although a certain minimum black hole mass is necessary to form jets, we also suggest that it is not the most essential prerequisite for the formation of a jet, in agreement with other recent studies \citep[e.g.,][]{Vaddi16,Sabater19}.


 \section{Acknowledgments}
The GMRT staff are gratefully acknowledged for their assistance during data acquisition for this project. GMRT is run by the National Centre for Radio Astrophysics of the Tata Institute of Fundamental Research. E. O'Sullivan acknowledges support for this work from the National Aeronautics and Space Administration through Chandra Awards GO6-17121X and GO6-17122X, issued by the Chandra X-ray Observatory Center, which is operated by the Smithsonian Astrophysical Observatory for and on behalf of NASA under contract NAS8-03060, and through the Astrophysical Data Analysis programme, award NNX13AE71G. S. Giacintucci acknowledges support for basic research in radio astronomy at the Naval Research Laboratory by 6.1 Base funding.
Some of this research was supported by the EU/FP7 Marie Curie award of the IRSES grant CAFEGROUPS (247653). We acknowledge the use of the HyperLeda database (http://leda.univ-lyon1.fr). This research has made use of the NASA/IPAC Extragalactic Database (NED) which is operated by the Jet Propulsion Laboratory, California Institute of Technology, under contract with the National Aeronautics and Space Administration. 


 \clearpage


\appendix

\section{Information on individual groups}
\label{AppA}
In this appendix we provide notes on the 25/27 individual groups of the low--richness sample analyzed in this study, the BGEs and their radio sources, where relevant. The GMRT data for NGC~315 and NGC~1407 at both 235 and 610~MHz are drawn from the earlier studies of \citet{Simona11} and \citet{Giacintucci12}.






\subsection{LGG 6 / NGC 128} 
\label{AppA1}

NGC~128 has a peculiar (`peanut-shape') bulge morphology that kinematic studies attribute to the current strong dynamical interactions with its close neighbours NGC 127 and NGC 130 \citep{Jarvis90}. A tidal bridge of cold gas links NGC~128 and NGC~127, and the three galaxies have narrow spread in systemic velocity ($\leq$200~km~s$^{-1}$; \citealt{DOnofrio99}).

\begin{figure*}
\vspace{5.8cm}
\centering
\includegraphics[width=1.00\textwidth]{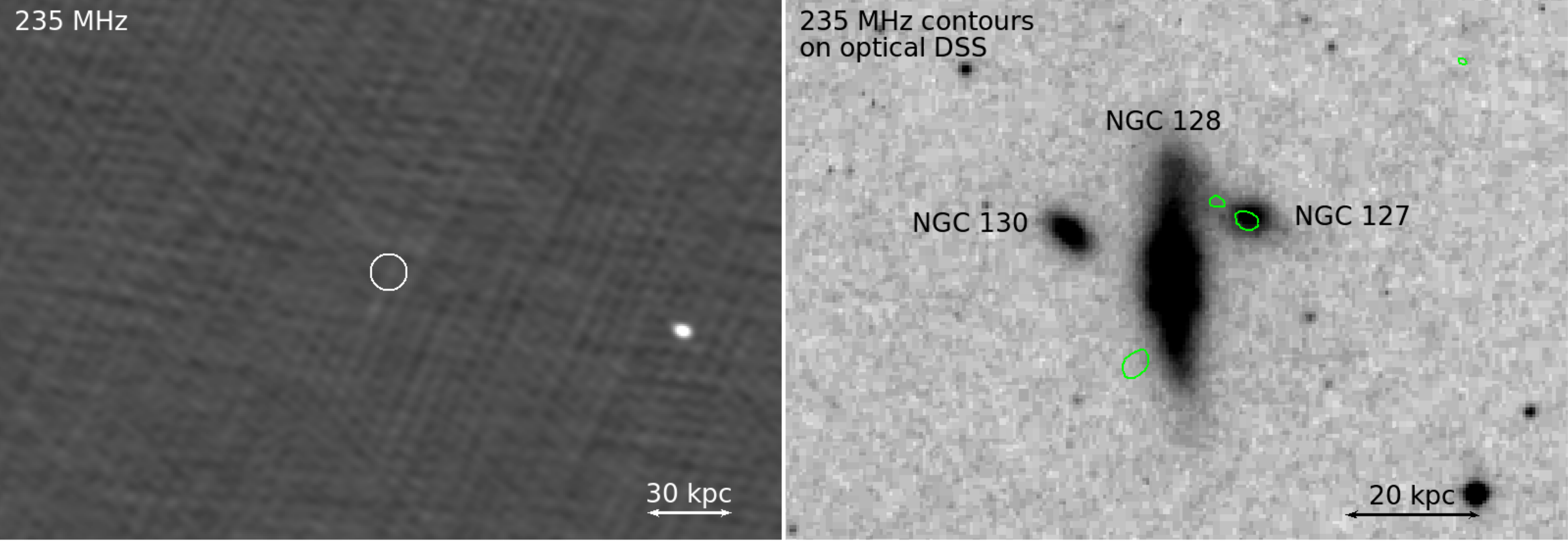}
\includegraphics[width=1.00\textwidth]{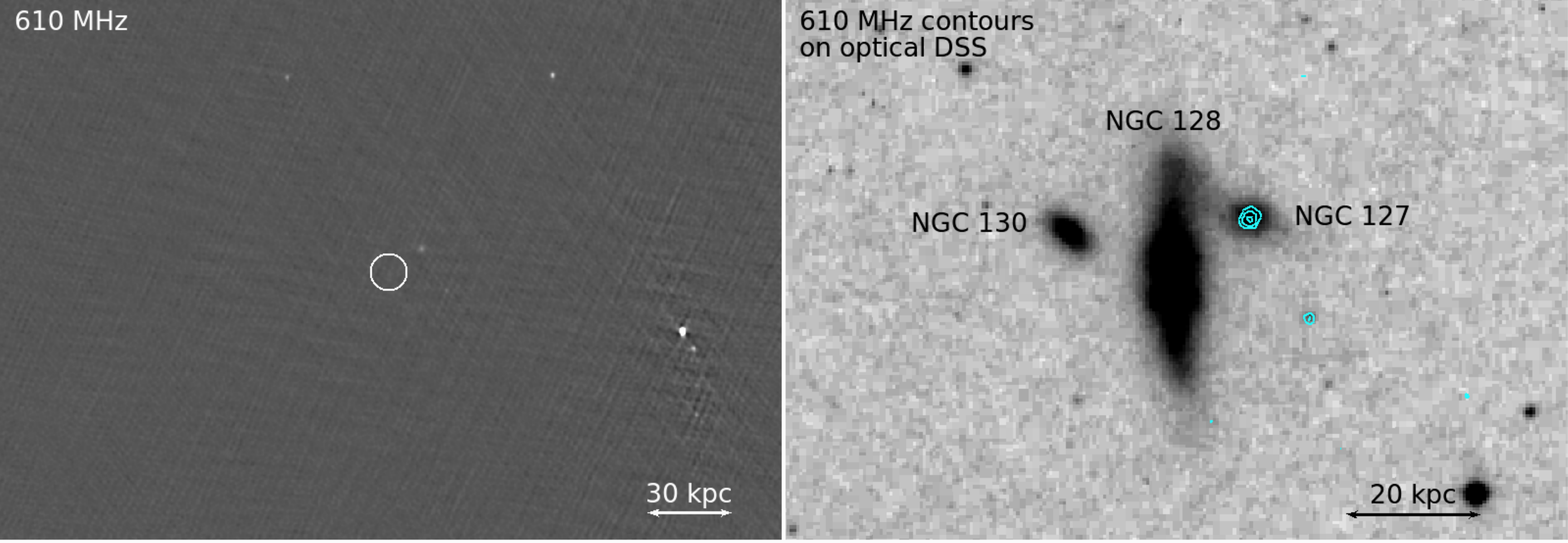}
\caption{LGG 6 / NGC 128. \textit{Top left:} GMRT 235~MHz image. \textit{Top right:} GMRT 235~MHz contours in green (1$\sigma$ = 1.90 mJy beam$^{-1}$), overlaid on the \textit{Digitized Sky Survey (DSS)} optical image. \textit{Bottom left:} GMRT 610~MHz image. \textit{Bottom right:} GMRT 610~MHz contours in cyan (1$\sigma$ = 160 $\mu$Jy beam$^{-1}$), overlaid on the \textit{Digitized Sky Survey (DSS)} optical image. In both panels the radio contours are spaced by a factor of two, starting from 3$\sigma$ level of significance. For this source the scale is 0.291 kpc arcsec$^{-1}$.}
\label{fig:128}
\vspace{-170mm}
\end{figure*}

A faint central point source was detected at both 1.4~GHz \citep{Brown11} and single dish 2.38~GHz \citep{Dressler78}, with flux densities of 1.5~mJy and 5$\pm$3~mJy respectively. However, we note that the Arecibo 2.38~GHz value is not trustworthy as its poor spatial resolution ($>$5~arcmin) means that it cannot separate NGC~128 from its neighbours. Despite the previous low power detection at 1.4~GHz, NGC~128 is undetected in our GMRT observations, but NGC~127 hosts a radio point source at both 610 and 235~MHz. Figure~\ref{fig:128} shows the radio contours that reveal the detection associated with NGC~127.

\subsection{LGG 12 / NGC 252} 
\label{AppA2}

\begin{figure*}
\vspace{3cm}
\centering
\includegraphics[width=1.00\textwidth]{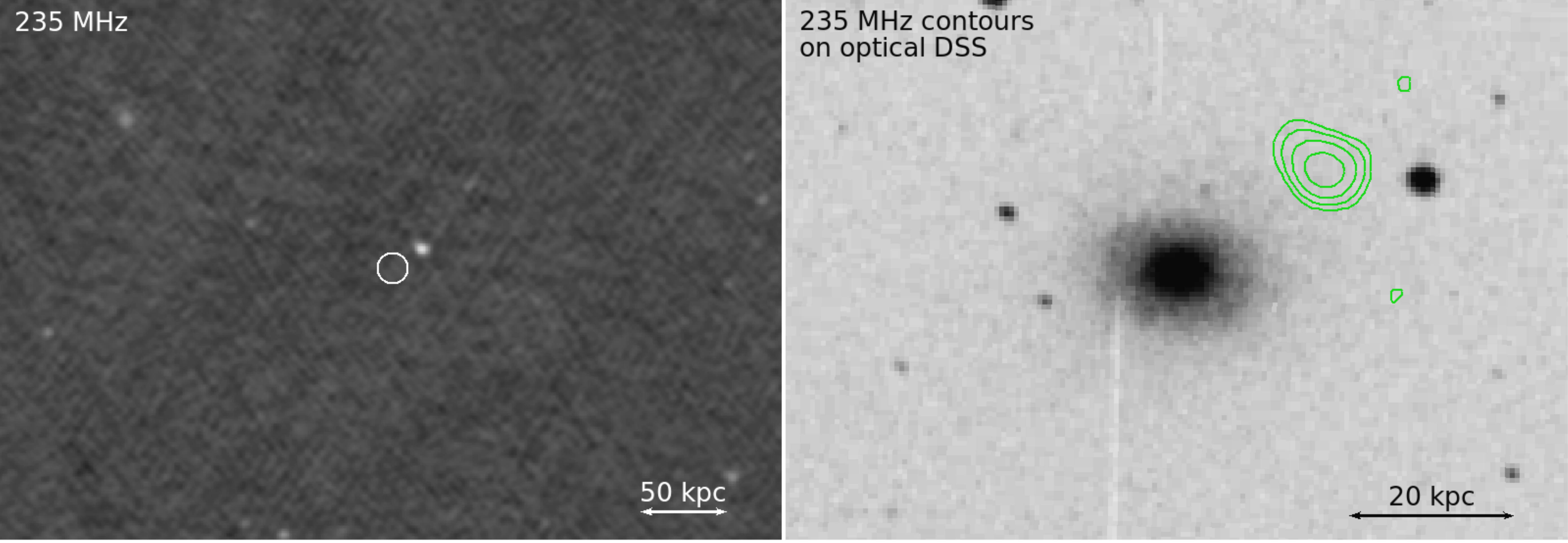}
\includegraphics[width=1.00\textwidth]{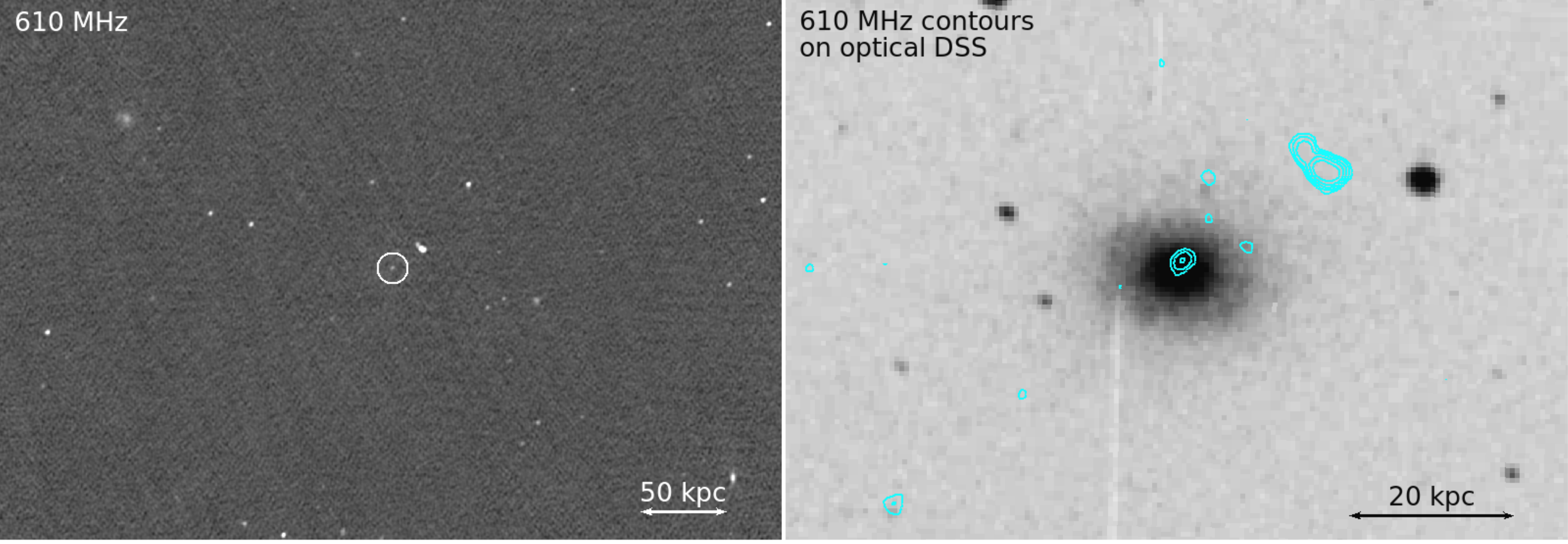}
\caption{LGG 12 / NGC 252. \textit{Top left:} GMRT 235~MHz image. \textit{Top right:} GMRT 235~MHz contours in green (1$\sigma$ = 0.60 mJy beam$^{-1}$), overlaid on the \textit{Digitized Sky Survey (DSS)} optical image. \textit{Bottom left:} GMRT 610~MHz image. \textit{Bottom right:} GMRT 610~MHz contours in cyan (1$\sigma$ = 60 $\mu$Jy beam$^{-1}$), overlaid on the \textit{Digitized Sky Survey (DSS)} optical image. In both panels the radio contours are spaced by a factor of two, starting from 3$\sigma$ level of significance. For this source the scale is 0.349 kpc arcsec$^{-1}$.}
\label{fig:252}
\end{figure*}

Previous 1.4~GHz \citep{Condon02} and single dish 2.38~GHz observations of NGC~252 found only a tentative detection with flux density of 2.5~mJy and 8$\pm$4 mJy \citep{Dressler78} respectively. We detect a weak radio point source associated with NGC~252 at 610~MHz (one of the weakest in the sample; see fig~\ref{fig:252}), and place an upper limit of $<$3.0~mJy at 5$\sigma$ level of significance for the 235~MHz flux (Table~\ref{Sourcetable}).

\subsection{LGG 23 / NGC 524} 
\label{AppA3}

\begin{figure*}
\vspace{3cm}
\centering
\includegraphics[width=1.00\textwidth]{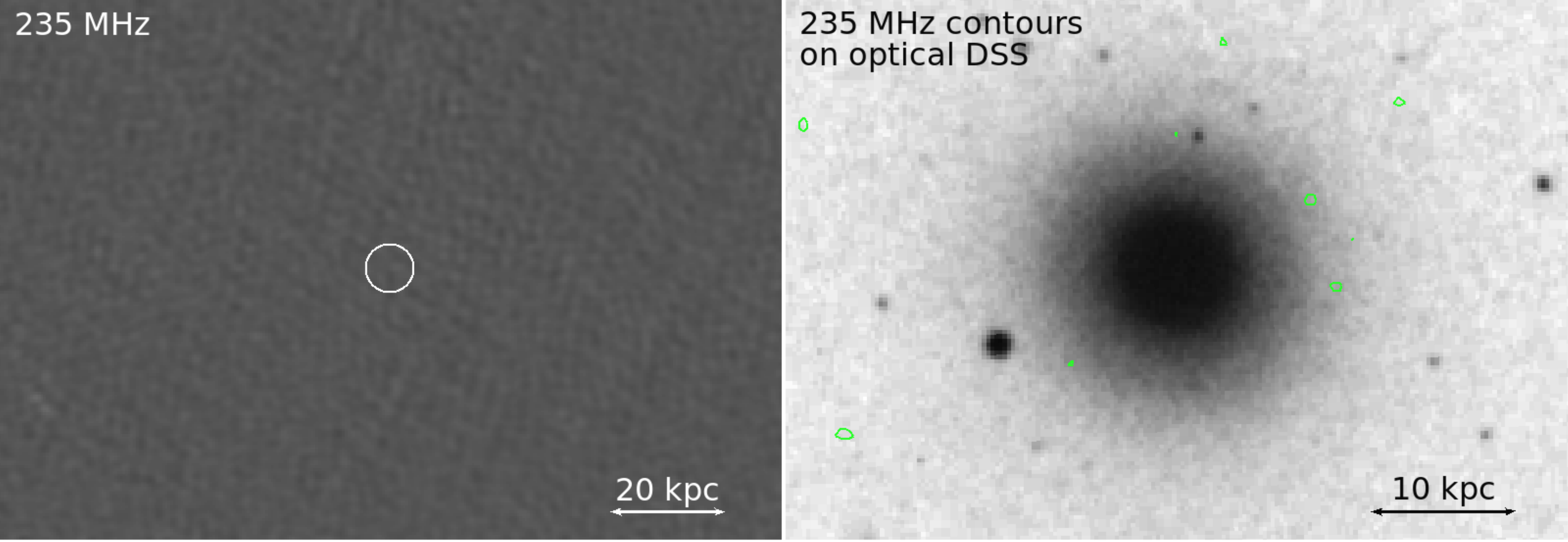}
\includegraphics[width=1.00\textwidth]{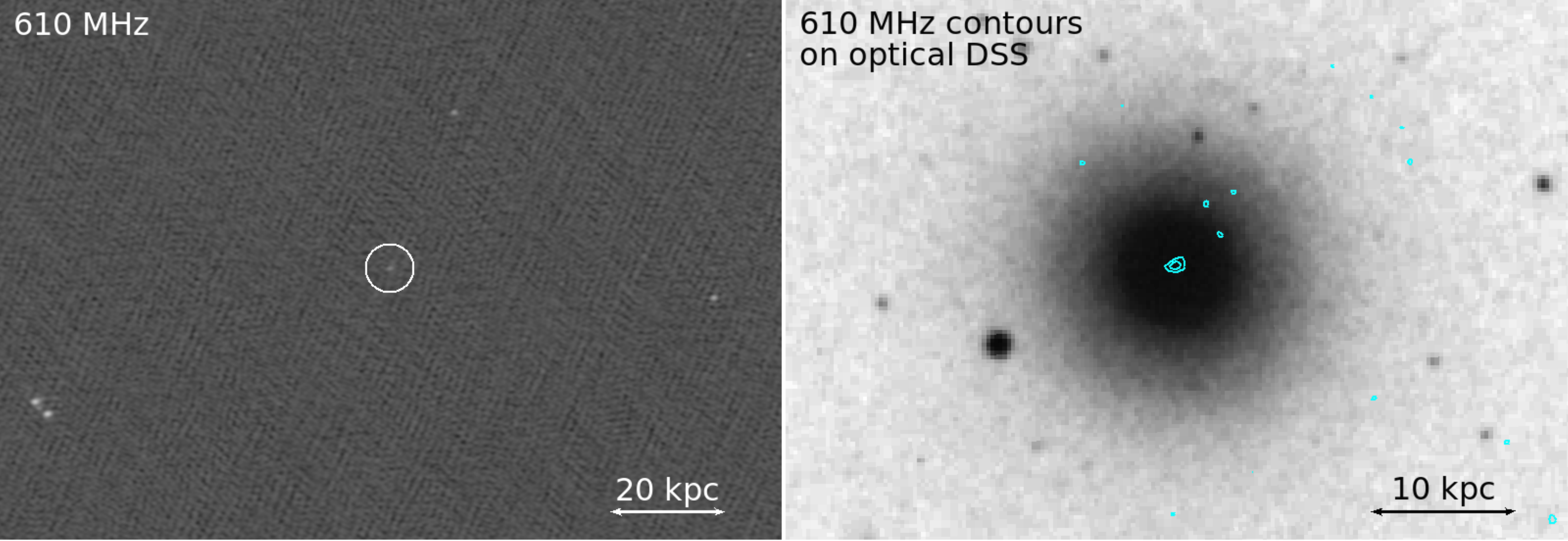}
\caption{LGG 14 / NGC 524. \textit{Top left:} GMRT 235~MHz image. \textit{Top right:} GMRT 235~MHz contours in green (1$\sigma$ = 2.10 mJy beam$^{-1}$), overlaid on the \textit{Digitized Sky Survey (DSS)} optical image. \textit{Bottom left:} GMRT 610~MHz image. \textit{Bottom right:} GMRT 610~MHz contours in cyan (1$\sigma$ = 180 $\mu$Jy beam$^{-1}$), overlaid on the \textit{Digitized Sky Survey (DSS)} optical image. In both panels the radio contours are spaced by a factor of two, starting from 3$\sigma$ level of significance. For this source the scale is 0.165 kpc arcsec$^{-1}$.}
\label{fig:524}
\end{figure*}

X-ray observations by the \textit{ROSAT} PSPC \citep{Mulchaey03} and more recently with \textit{Chandra}-ACIS \citep{Kharb12} show that NGC~524 group contains an extended X-ray hot gas component associated with NGC~524. Previous 1.4~GHz \citep[NVSS][]{Condon98} and single dish 2.38~GHz observations of NGC~524 detected a radio source of flux density 3.1~mJy and 10$\pm$5 mJy \citep{Dressler78} respectively. We detect a radio point source associated with NGC~524 only at 610~MHz (see fig~\ref{fig:524}), with an upper limit of $<$10.5~mJy at 5$\sigma$ significance for 235~MHz (Table~\ref{Sourcetable}).

\subsection{LGG 78 / NGC 1106} 
\label{AppA4}

\begin{figure*}
\vspace{3cm}
\centering
\includegraphics[width=1.00\textwidth]{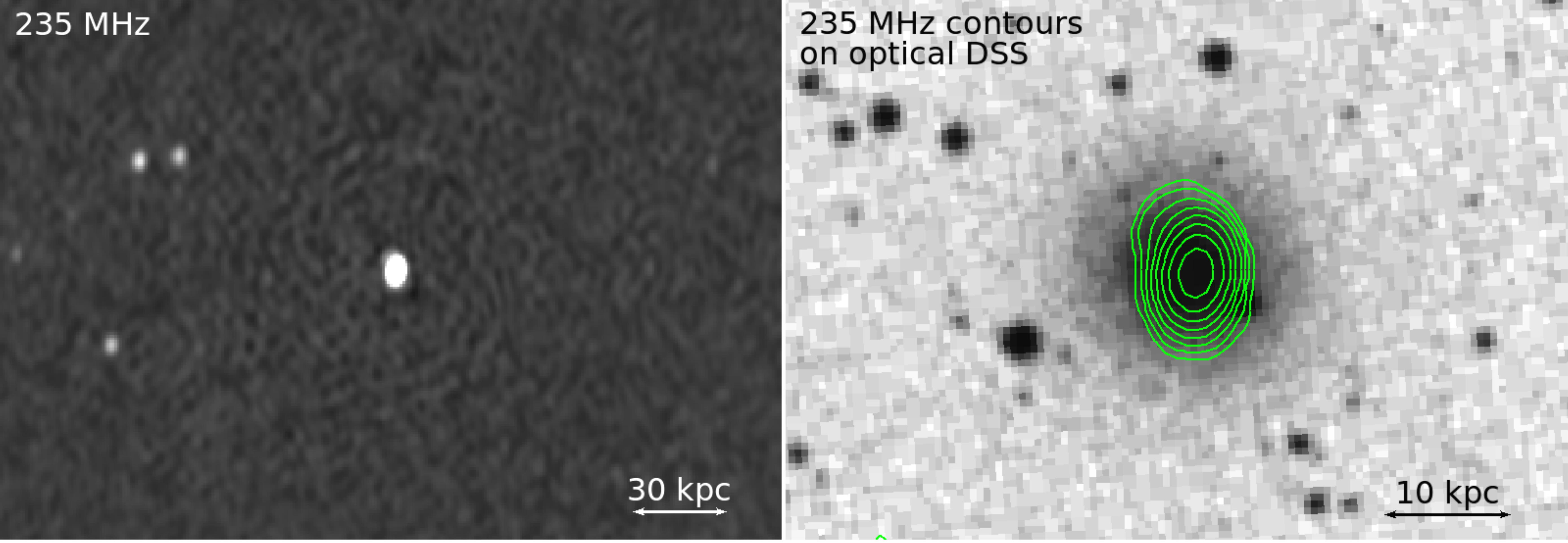}
\includegraphics[width=1.00\textwidth]{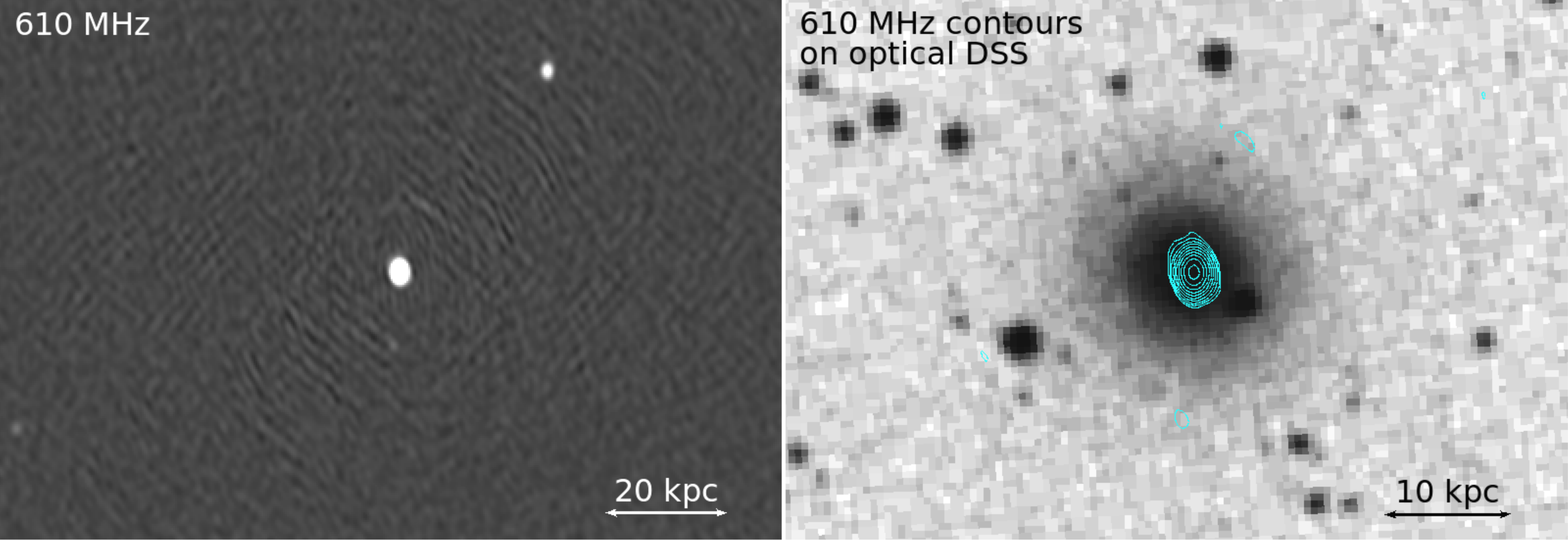}
\caption{LGG 78 / NGC 1106. \textit{Top left:} GMRT 235~MHz image. \textit{Top right:} GMRT 235~MHz contours in green (1$\sigma$ = 0.60 mJy beam$^{-1}$), overlaid on the \textit{Digitized Sky Survey (DSS)} optical image. \textit{Bottom left:} GMRT 610~MHz image. \textit{Bottom right:} GMRT 610~MHz contours in cyan (1$\sigma$ = 100 $\mu$Jy beam$^{-1}$), overlaid on the \textit{Digitized Sky Survey (DSS)} optical image. In both panels the radio contours are spaced by a factor of two, starting from 3$\sigma$ level of significance. For this source the scale is 0.310 kpc arcsec$^{-1}$.}
\label{fig:1106}
\end{figure*}

NGC~1106 is the BGE of the LGG~78 group and is identified as a Compton-thick dust obscured AGN \citet{Tanimoto16}. The NVSS \citep{Condon02} detected a strong point-like 1.4~GHz radio source with a flux density of 132~mJy. Our images at 610~MHz and 235~MHz (see figure~\ref{fig:1106}) show a compact radio point-like source at both GMRT frequencies that coincides with the central region of the optical galaxy. The spectral index of the associated AGN radio emission is $\alpha_{235}^{610}\sim0.7$.

\subsection{LGG 97 / NGC 1395} 
\label{AppA5}

LGG~97 is part of the Eridanus supergroup and its BGE, NGC~1395, is X-ray luminous \citep{OSullivan03}, with the extended diffuse X-ray emission being detected in both \textit{ROSAT} PSPC \citep[out to $\sim$135~kpc]{Omar05} and \textit{XMM-Newton} \citep[out to $\sim$66~kpc]{Escudero18} observations.

\begin{figure*}
\vspace{3cm}
\centering
\includegraphics[width=1.00\textwidth]{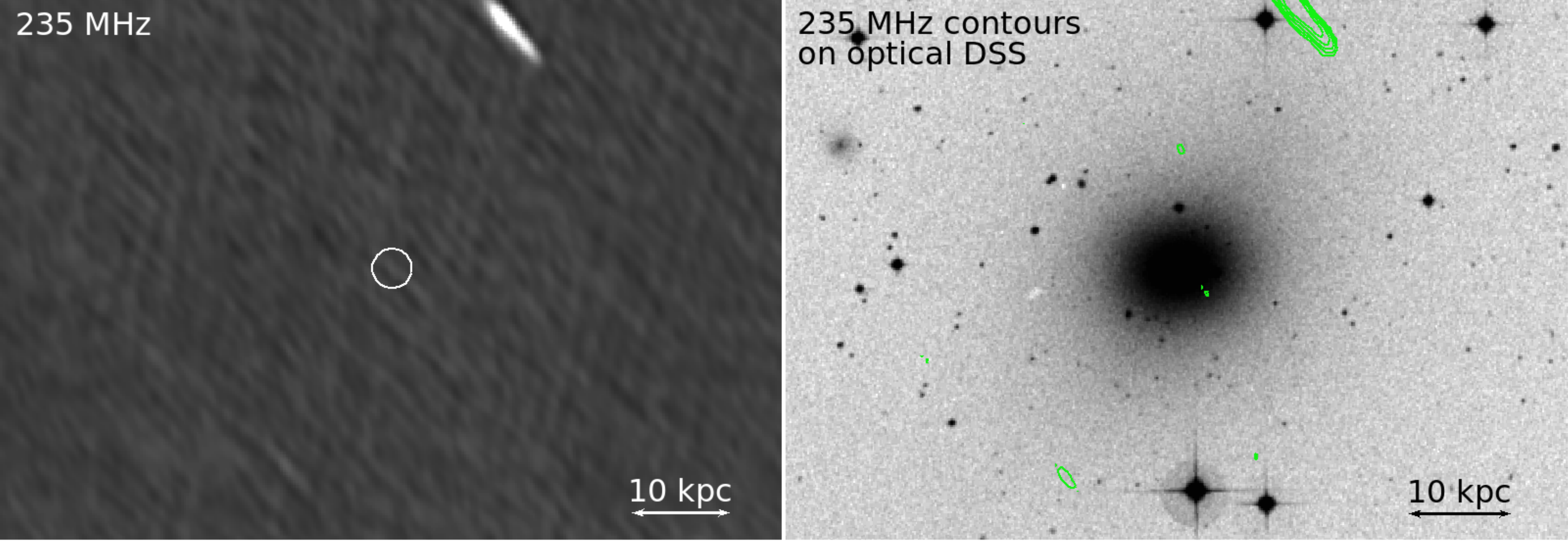}
\includegraphics[width=1.00\textwidth]{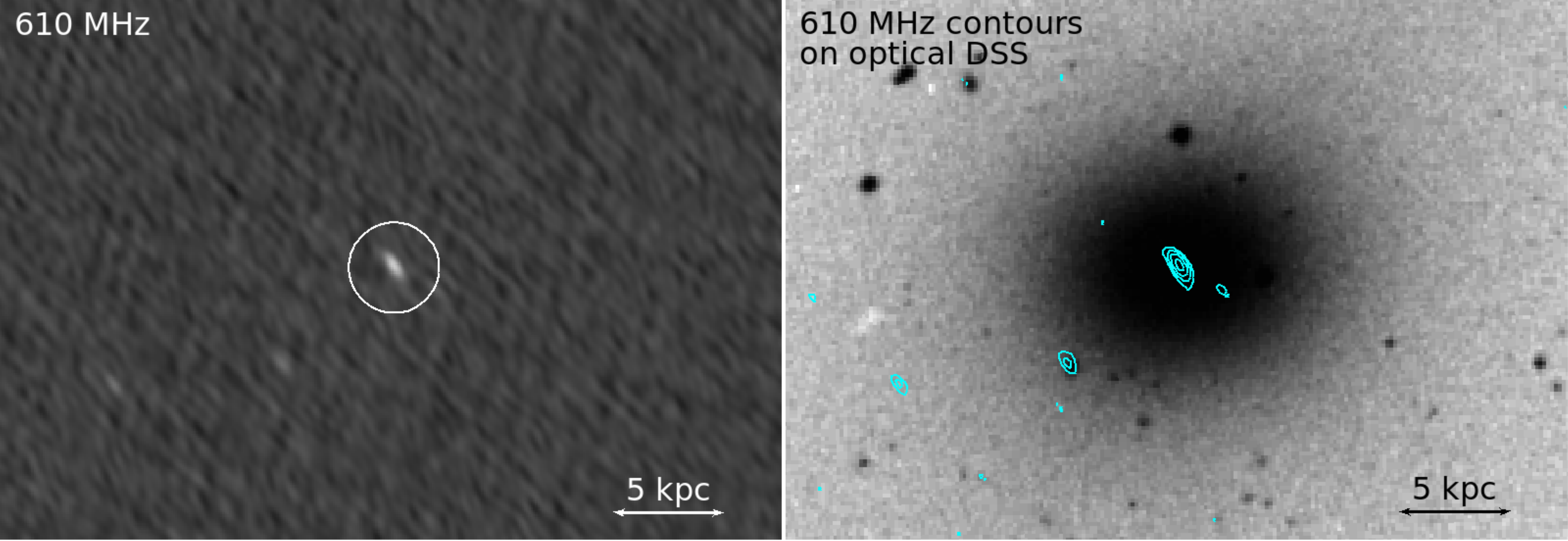}
\caption{LGG 97 / NGC 1395. \textit{Top left:} GMRT 235~MHz image. \textit{Top right:} GMRT 235~MHz contours in green (1$\sigma$ = 0.70 mJy beam$^{-1}$), overlaid on the \textit{Digitized Sky Survey (DSS)} optical image. \textit{Bottom left:} GMRT 610~MHz image. \textit{Bottom right:} GMRT 610~MHz contours in cyan (1$\sigma$ = 100 $\mu$Jy beam$^{-1}$), overlaid on the \textit{Digitized Sky Survey (DSS)} optical image. In both panels the radio contours are spaced by a factor of two, starting from 3$\sigma$ level of significance. For this source the scale is 0.102 kpc arcsec$^{-1}$.}
\label{fig:1395}
\end{figure*}

Previous 1.4~GHz observations detected a weak point radio source with a flux density of 1.1~mJy \citep{Brown11}. We detect a radio point-like source associated with NGC~1395 only at 610~MHz (see fig~\ref{fig:1395}), with an upper limit of $<$3.5~mJy (5$\sigma$ significance) at 235~MHz (Table~\ref{Sourcetable}).



\subsection{LGG 113 / NGC 1550} 
\label{AppA6}

NGC~1550 is the BGE of the X-ray bright LGG~113 group, is classified as an optically overluminous early-type galaxy \citep[OLEG;][]{Vikhlinin99} or potential fossil group, and is considered to be a product of multiple earlier mergers \citep{Kawaharada09}. Previous X-ray observations by \textit{ROSAT All-Sky Survey} (RASS; \citealt{Beuing99}) detected a very bright X-ray source (RX J0419+0225) associated with the system, with more recent X-ray observations by \textit{Chandra} \citep{Sun03} and \textit{XMM-Newton} \citep{Kawaharada09} showing that the hot IGM extends out to at least $\sim$200~kpc from the centre of the galaxy.

Previous 1.4~GHz observations from both NVSS \citep{Condon98} and \citet{Brown11} detected a relatively weak radio source with a flux density of 16.6~mJy and 17~mJy respectively. The source presents an asymmetric double peaked radio morphology with the first peak coincident with the optical centre of NGC~1550 and the second one being observed as a lobe-like extension to the west \citep{Dunn10}. 

\begin{figure*}
\vspace{3cm}
\centering
\includegraphics[width=1.00\textwidth]{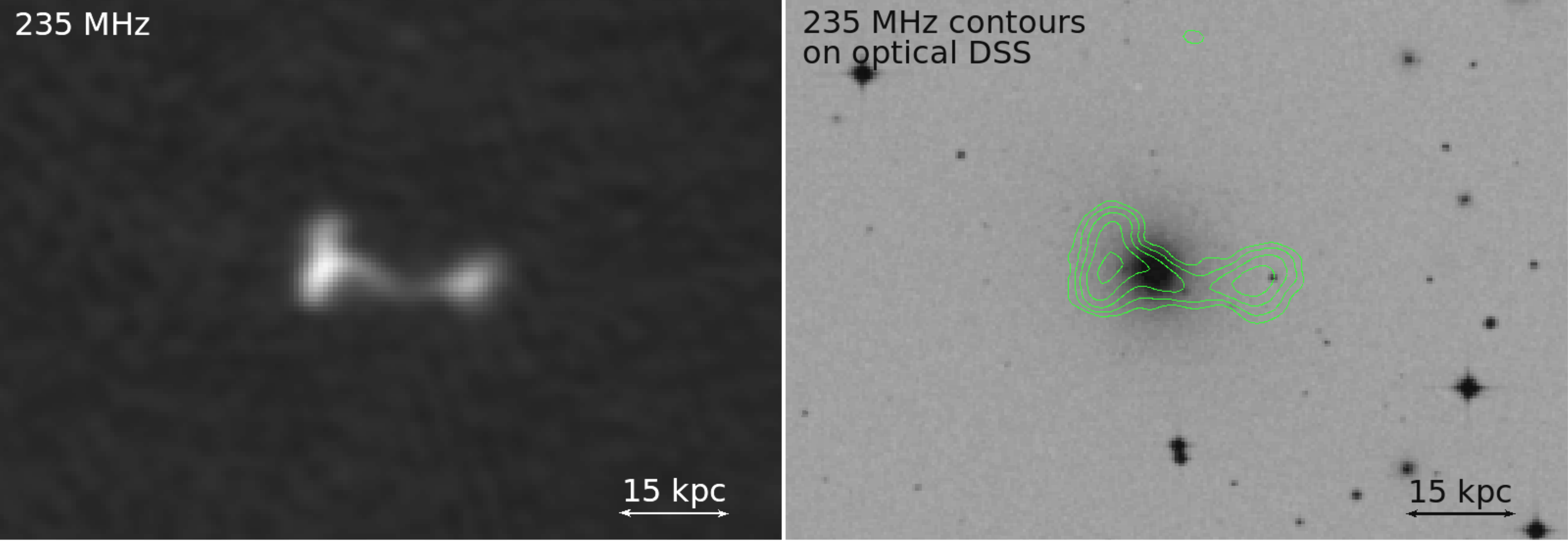}
\includegraphics[width=1.00\textwidth]{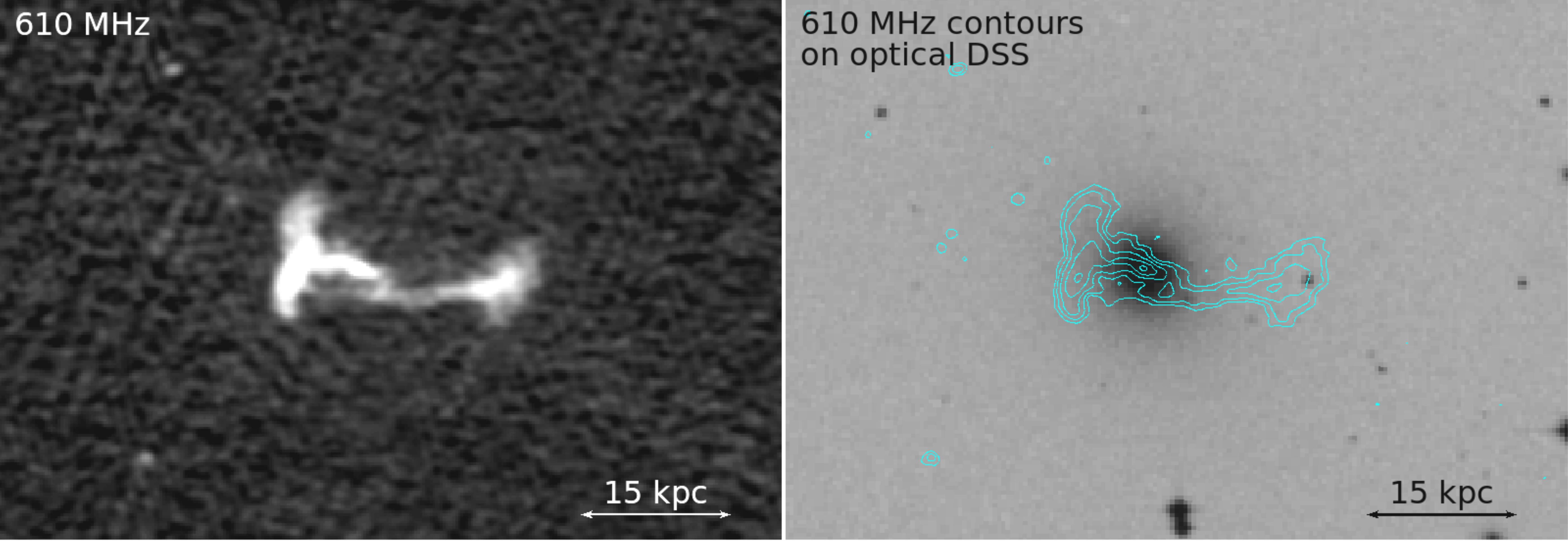}
\caption{LGG 113 / NGC 1550. \textit{Top left:} GMRT 235~MHz image. \textit{Top right:} GMRT 235~MHz contours in green (1$\sigma$ = 0.45 mJy beam$^{-1}$), overlaid on the \textit{Digitized Sky Survey (DSS)} optical image. \textit{Bottom left:} GMRT 610~MHz image. \textit{Bottom right:} GMRT 610~MHz contours in cyan (1$\sigma$ = 40 $\mu$Jy beam$^{-1}$), overlaid on the \textit{Digitized Sky Survey (DSS)} optical image. In both panels the radio contours are spaced by a factor of two, starting from 3$\sigma$ level of significance. For this source the scale is 0.257 kpc arcsec$^{-1}$.}
\label{fig:1550}
\end{figure*}

Our GMRT radio images of the source are shown in figure~\ref{fig:1550}. We detect NGC~1550 at both frequencies with much higher angular resolution and sensitivity than prior studies. We find an asymmetric jet-lobe structure along an east-west axis, with the centre of the eastern lobe close to the stellar body of the galaxy and the second lobe expanding $\sim$20~kpc to the west. At 610~MHz the western jet has a bend or kink to the south before continuing relatively straight to the lobe. This system will be examined in more detail in a future paper (Kolokythas et al., in prep.).

\subsection{LGG 126 / NGC 1779} 
\label{AppA7}

\begin{figure*}
\vspace{3cm}
\centering
\includegraphics[width=1.00\textwidth]{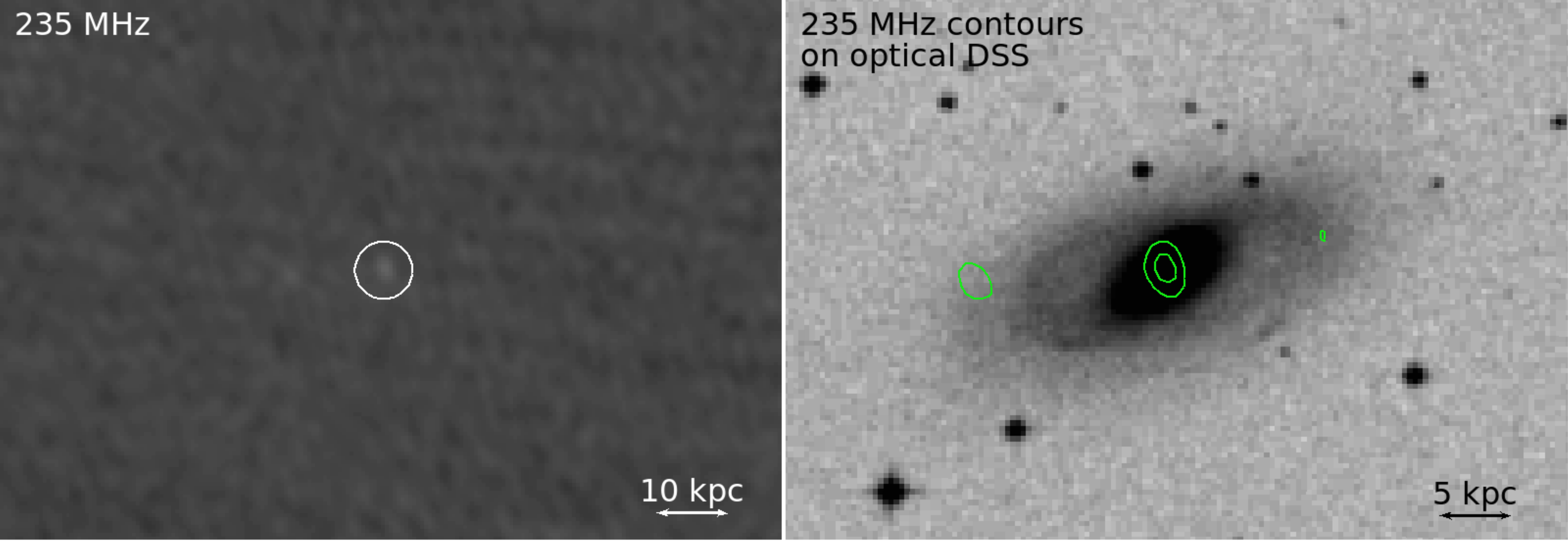}
\includegraphics[width=1.00\textwidth]{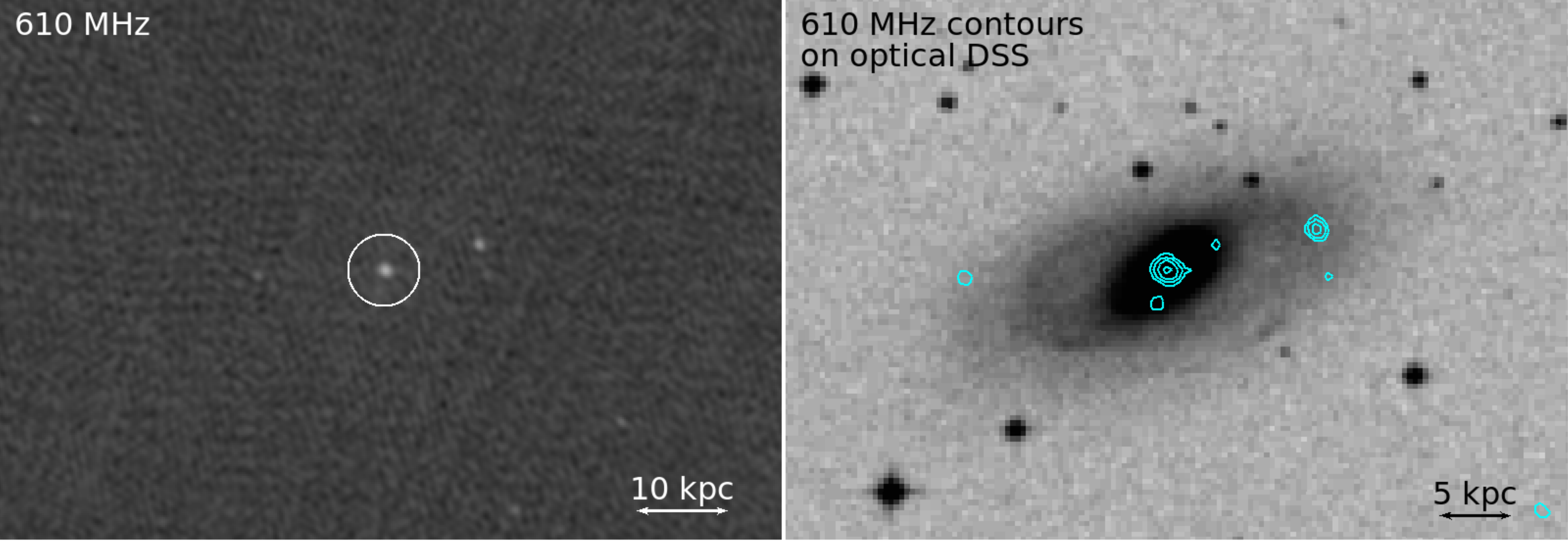}
\caption{LGG 126 / NGC 1779. \textit{Top left:} GMRT 235~MHz image. \textit{Top right:} GMRT 235~MHz contours in green (1$\sigma$ = 0.45 mJy beam$^{-1}$), overlaid on the \textit{Digitized Sky Survey (DSS)} optical image. \textit{Bottom left:} GMRT 610~MHz image. \textit{Bottom right:} GMRT 610~MHz contours in cyan (1$\sigma$ = 50 $\mu$Jy beam$^{-1}$), overlaid on the \textit{Digitized Sky Survey (DSS)} optical image. In both panels the radio contours are spaced by a factor of two, starting from 3$\sigma$ level of significance. For this source the scale is 0.218 kpc arcsec$^{-1}$.}
\label{fig:1779}
\end{figure*}

The NVSS shows a weak 1.4~GHz point source in NGC~1779 with a flux density of 5.4$\pm$0.6~mJy \citep{Condon98}. We also detect a weak radio point source associated with NGC~1779 at both GMRT frequencies (see fig~\ref{fig:1779}), with the spectral index in the 235$-$610~MHz range being $\sim$0.60 and in the 235$-$1400~MHz range appearing to be inverted ($\alpha_{235}^{1400}$$\sim$-0.09; see Table~\ref{Sourcetable}).

\subsection{LGG 138 / NGC 2292} 
\label{AppA8}

\begin{figure*}
\vspace{3cm}
\centering
\includegraphics[width=1.00\textwidth]{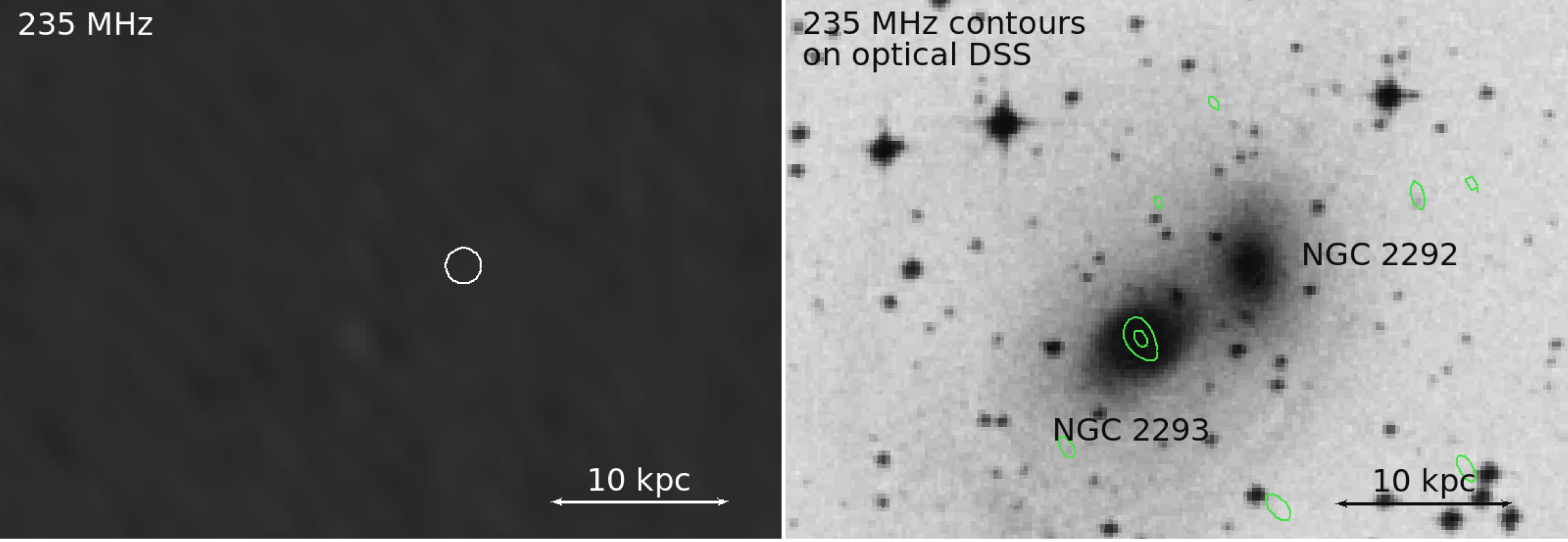}
\includegraphics[width=1.00\textwidth]{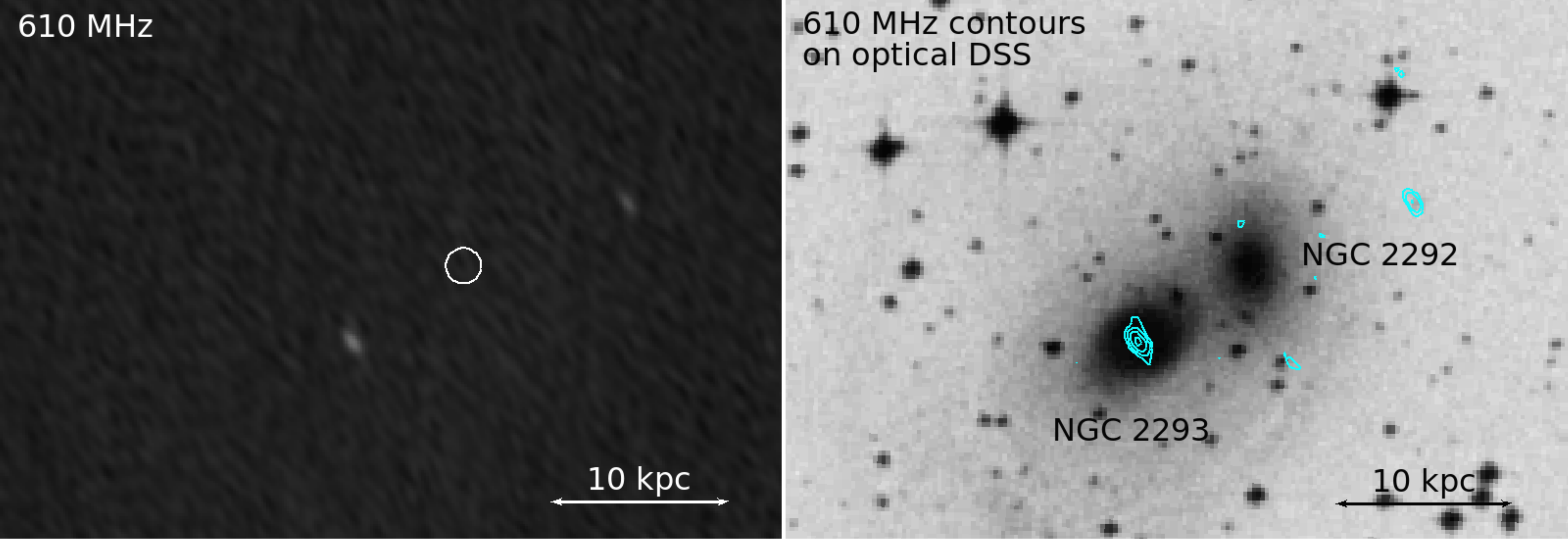}
\caption{LGG 138 / NGC 2292. \textit{Top left:} GMRT 235~MHz image. \textit{Top right:} GMRT 235~MHz contours in green (1$\sigma$ = 0.42 mJy beam$^{-1}$), overlaid on the \textit{Digitized Sky Survey (DSS)} optical image. \textit{Bottom left:} GMRT 610~MHz image. \textit{Bottom right:} GMRT 610~MHz contours in cyan (1$\sigma$ = 50 $\mu$Jy beam$^{-1}$), overlaid on the \textit{Digitized Sky Survey (DSS)} optical image. In both panels the radio contours are spaced by a factor of two, starting from 3$\sigma$ level of significance. For this source the scale is 0.145 kpc arcsec$^{-1}$.}
\label{fig:2292}
\end{figure*}

The S0 NGC~2292 is part of the interacting galaxy pair NGC~2292/2293. Earlier studies of this system \citep[e.g.,][]{Barnes99,Stickel05} have found an extended, dusty tidal H\textsc{i} ring surrounding the pair. The galaxy was not previously detected in radio continuum emission and our GMRT observations find no radio detection at either frequency (fig.~\ref{fig:2292}). The flux density upper limit set for this galaxy at 5$\sigma$ level of significance is 0.3~mJy at 610~MHz and 1.9~mJy at 235~MHz. However, we do detect a point source in the companion NGC~2293 at both 235~MHz ($\sim$2.3~mJy) and 610~MHz ($\sim$1.9~mJy).

\subsection{LGG 167 / NGC 2768} 
\label{AppA9}

\begin{figure*}
\vspace{3cm}
\centering
\includegraphics[width=1.00\textwidth]{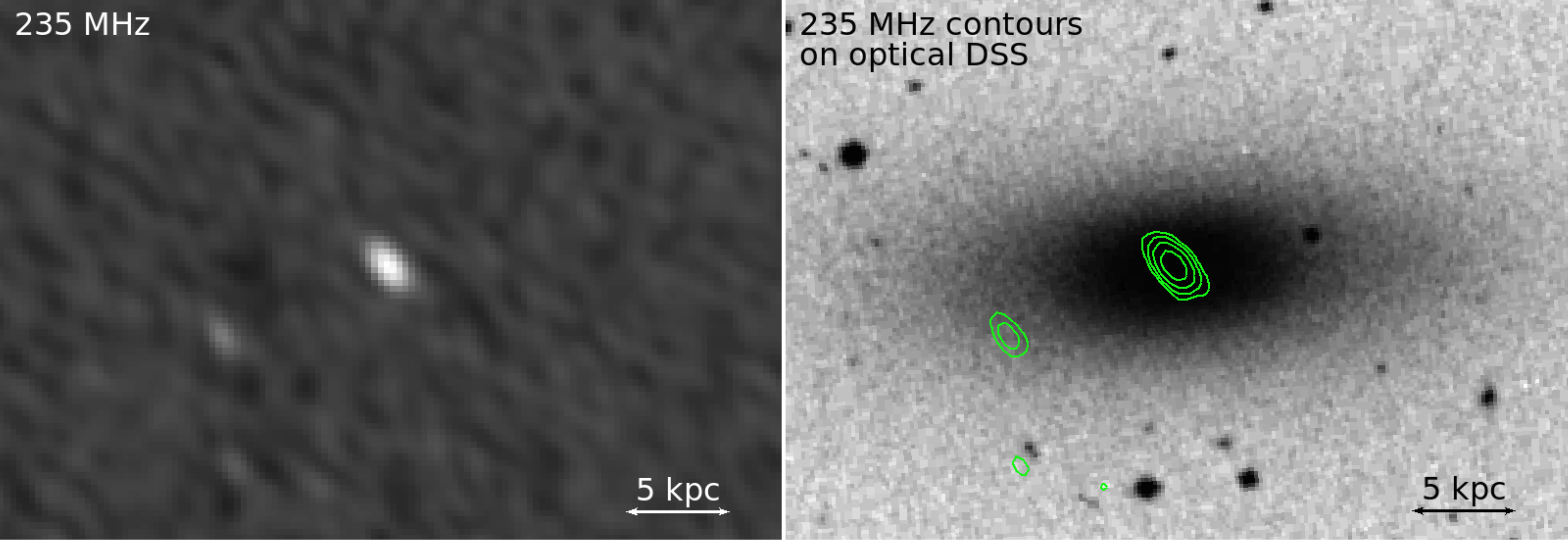}
\includegraphics[width=1.00\textwidth]{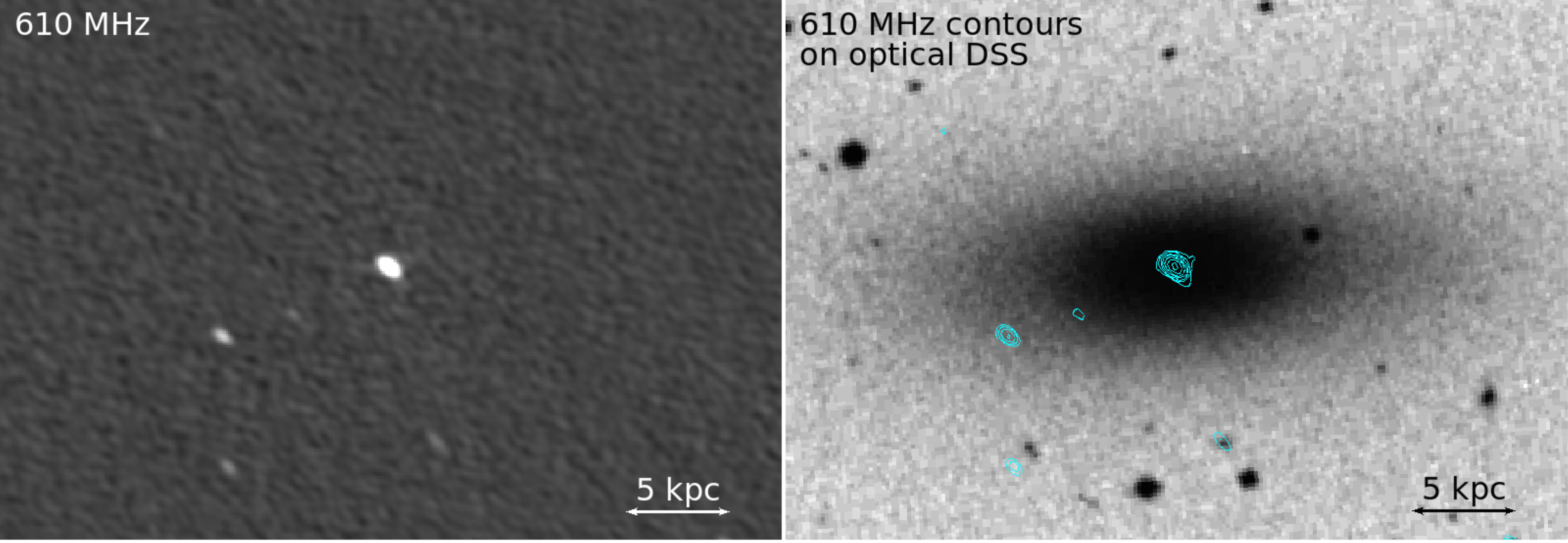}
\caption{LGG 167 / NGC 2768. \textit{Top left:} GMRT 235~MHz image. \textit{Top right:} GMRT 235~MHz contours in green (1$\sigma$ = 0.30 mJy beam$^{-1}$), overlaid on the \textit{Digitized Sky Survey (DSS)} optical image. \textit{Bottom left:} GMRT 610~MHz image. \textit{Bottom right:} GMRT 610~MHz contours in cyan (1$\sigma$ = 30 $\mu$Jy beam$^{-1}$), overlaid on the \textit{Digitized Sky Survey (DSS)} optical image. In both panels the radio contours are spaced by a factor of two, starting from 3$\sigma$ level of significance. For this source the scale is 0.112 kpc arcsec$^{-1}$.}
\label{fig:2768}
\end{figure*}

NGC~2768 hosts a polar ionized O[II] gas disk, rotating perpendicular to the stellar component and thus likely the product of a past merger \citep{Fried94}. Previous 1.4~GHz observations from the NVSS \citep{Condon98} and \citet{Brown11} detected a radio source with a flux density of 15.9$\pm$1~mJy and 14$\pm$1~mJy respectively. We detect a compact radio point source at both GMRT frequencies (see fig~\ref{fig:2768}). The spectral index of the associated compact radio emission appears to be flat in the 235$-$610~MHz regime, $\alpha_{235}^{610}\sim0.14$, and inverted in the 235$-$1400~MHz one, with $\alpha_{235}^{1400}\sim-0.04$ (Table~\ref{Sourcetable}).

\subsection{LGG 177 / NGC 2911} 
\label{AppA10}

\begin{figure*}
\vspace{3cm}
\centering
\includegraphics[width=1.00\textwidth]{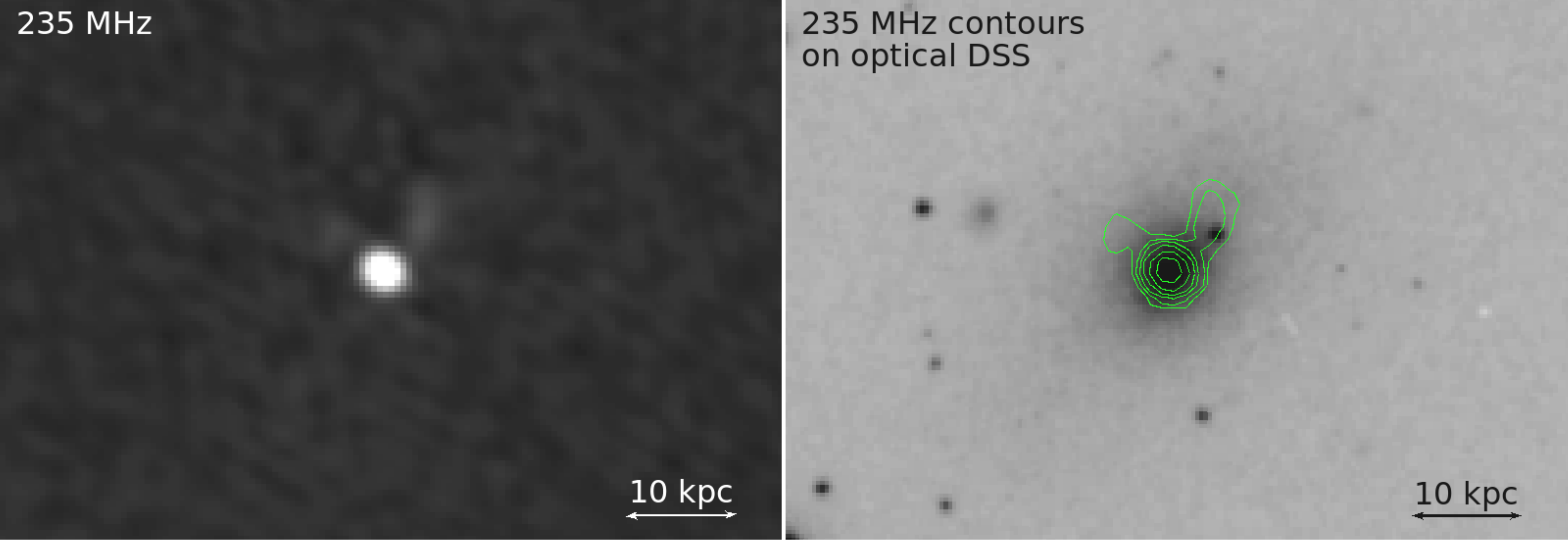}
\includegraphics[width=1.00\textwidth]{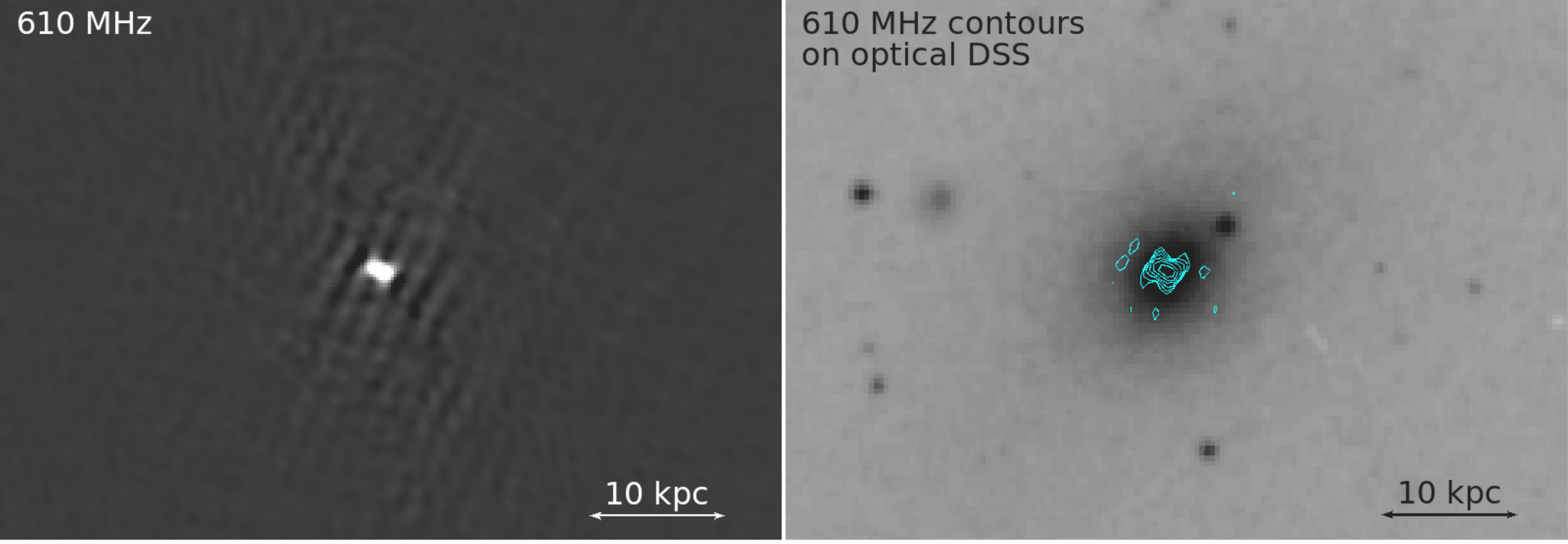}
\caption{LGG 177 / NGC 2911. \textit{Top left:} GMRT 235~MHz image. \textit{Top right:} GMRT 235~MHz contours in green (1$\sigma$ = 0.30 mJy beam$^{-1}$), overlaid on the \textit{Digitized Sky Survey (DSS)} optical image. \textit{Bottom left:} GMRT 610~MHz image. \textit{Bottom right:} GMRT 610~MHz contours in cyan (1$\sigma$ = 200 $\mu$Jy beam$^{-1}$), overlaid on the \textit{Digitized Sky Survey (DSS)} optical image. In both panels the radio contours are spaced by a factor of two, starting from 3$\sigma$ level of significance at 235~MHz and at 4$\sigma$ level of significance at 610~MHz due to the higher rms noise around the source at this frequency. For this source the scale is 0.218 kpc arcsec$^{-1}$.}
\label{fig:2911}
\end{figure*}

NGC~2911 is a low-ionization nuclear emission region (LINER) galaxy \citep{Prieto10}. VLA observations at 4.9 and 14.9~GHz with sub-arcsecond resolution  revealed a $\sim$500~pc jet with a steep spectral index \citep[$\alpha^{14.9}_{4.9}$$\sim$1.1][]{Mezcua14}. Lower resolution 1.4~GHz observations \citep{Brown11} reveal a radio source with a flux density of 56$\pm$2~mJy. We detect a radio point-like source associated with NGC~2911 at both GMRT frequencies (see fig~\ref{fig:2911}) presenting hints of extension. An elongated asymmetric structure of $\sim$7 kpc is observed only at 235~MHz, north-west from the centre whereas at 610~MHz the source appears to be slightly elongated but unresolved. We note here that the repetitive patterns observed around the source at 610~MHz, are artifacts owing to residual calibration errors. The spectral index of the associated compact radio emission appears to be flat in both 235$-$610~MHz ($\alpha_{235}^{610}\sim0.18$), and 235$-$1400~MHz regimes ($\alpha_{235}^{1400}\sim0.12$; Table~\ref{Sourcetable}).

\subsection{LGG 205 / NGC 3325} 
\label{AppA11}

\begin{figure*}
\vspace{3cm}
\centering
\includegraphics[width=1.00\textwidth]{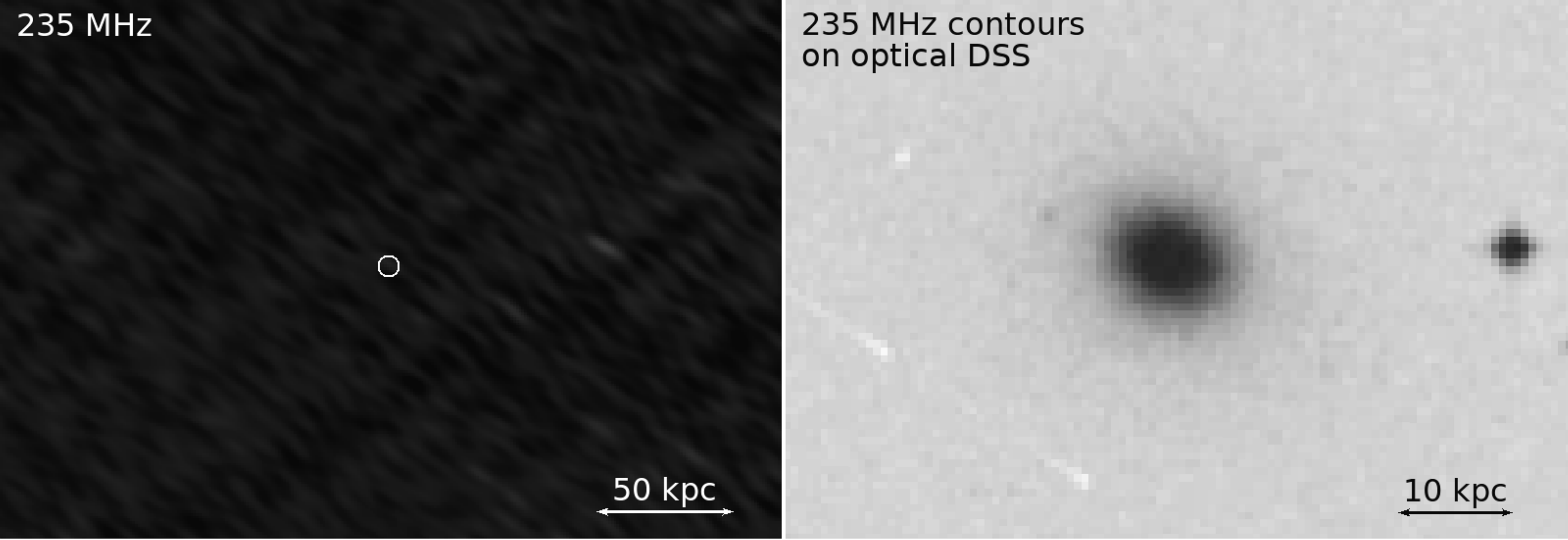}
\includegraphics[width=1.00\textwidth]{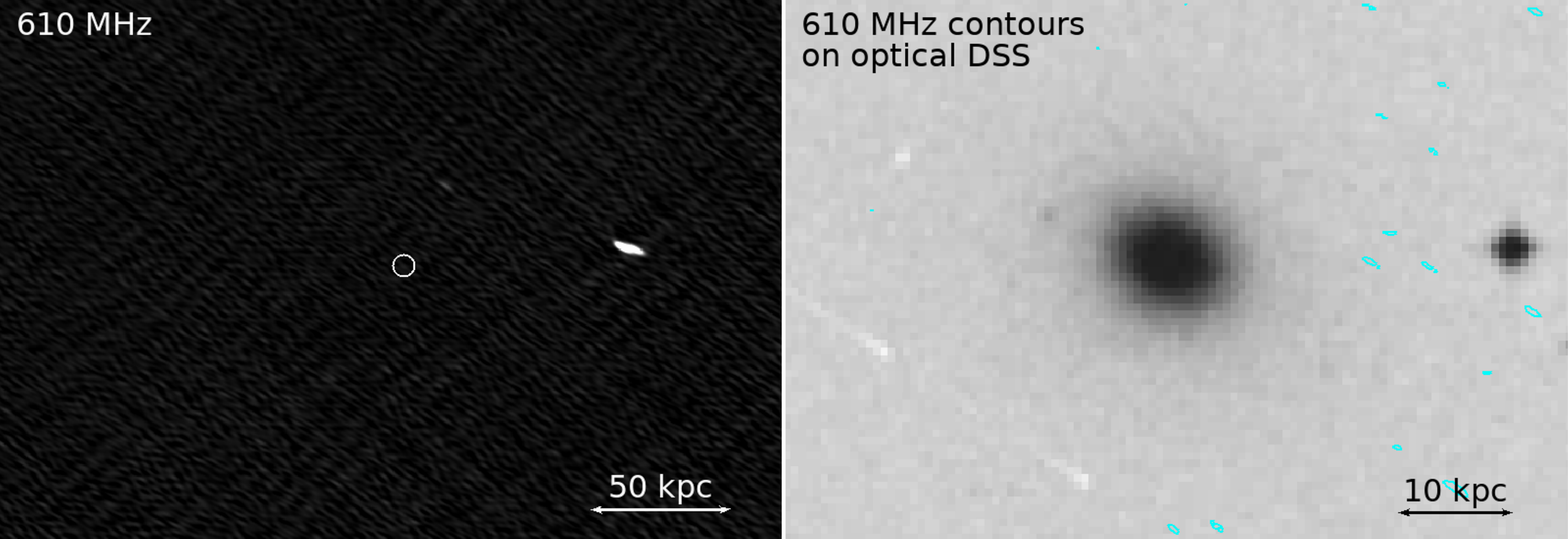}
\caption{LGG 205 / NGC 3325. \textit{Top left:} GMRT 235~MHz image. \textit{Top right:} GMRT 235~MHz contours in green (1$\sigma$ = 38 mJy beam$^{-1}$), overlaid on the \textit{Digitized Sky Survey (DSS)} optical image. \textit{Bottom left:} GMRT 610~MHz image. \textit{Bottom right:} GMRT 610~MHz contours in cyan (1$\sigma$ = 310 $\mu$Jy beam$^{-1}$), overlaid on the \textit{Digitized Sky Survey (DSS)} optical image. In both panels the radio contours are spaced by a factor of two, starting from 3$\sigma$ level of significance. The higher rms noise for this field at both frequencies is attributed to the short duration of the archival observations at hand. For this source the scale is 0.388 kpc arcsec$^{-1}$.}
\label{fig:3325}
\end{figure*}

Earlier studies of NGC~3325 found no detection in the radio band, with only a tentative single dish 2.38~GHz detection of 1$\pm$3~mJy \citep{Dressler78} being reported. The galaxy is undetected in our GMRT observations at both frequencies (fig.~\ref{fig:3325}). The flux density upper limit set for this galaxy at 5$\sigma$ significance is $<$1.55~mJy at 610~MHz and $<$191~mJy at 235~MHz. The extremely high upper limit set at 235~MHz is a result of the short observing time ($\sim$60 mins) of the available archival data.

\subsection{LGG 232 / NGC 3613} 
\label{AppA12}

\begin{figure*}
\vspace{3cm}
\centering
\includegraphics[width=1.00\textwidth]{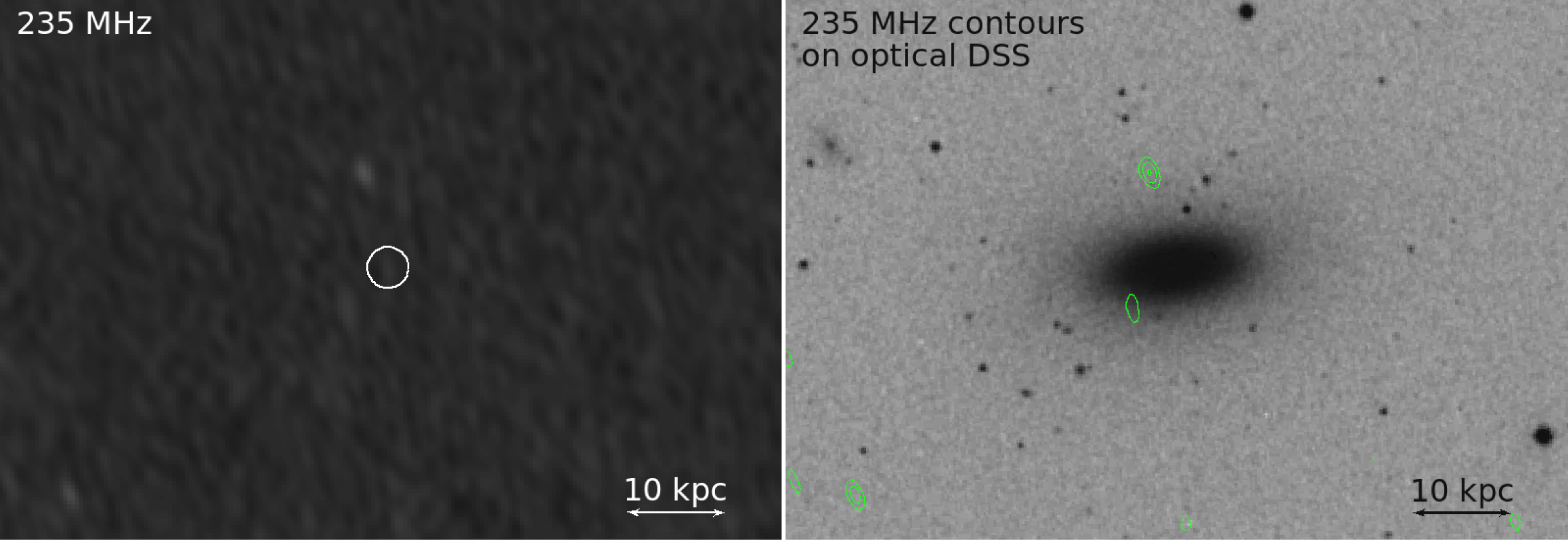}
\includegraphics[width=1.00\textwidth]{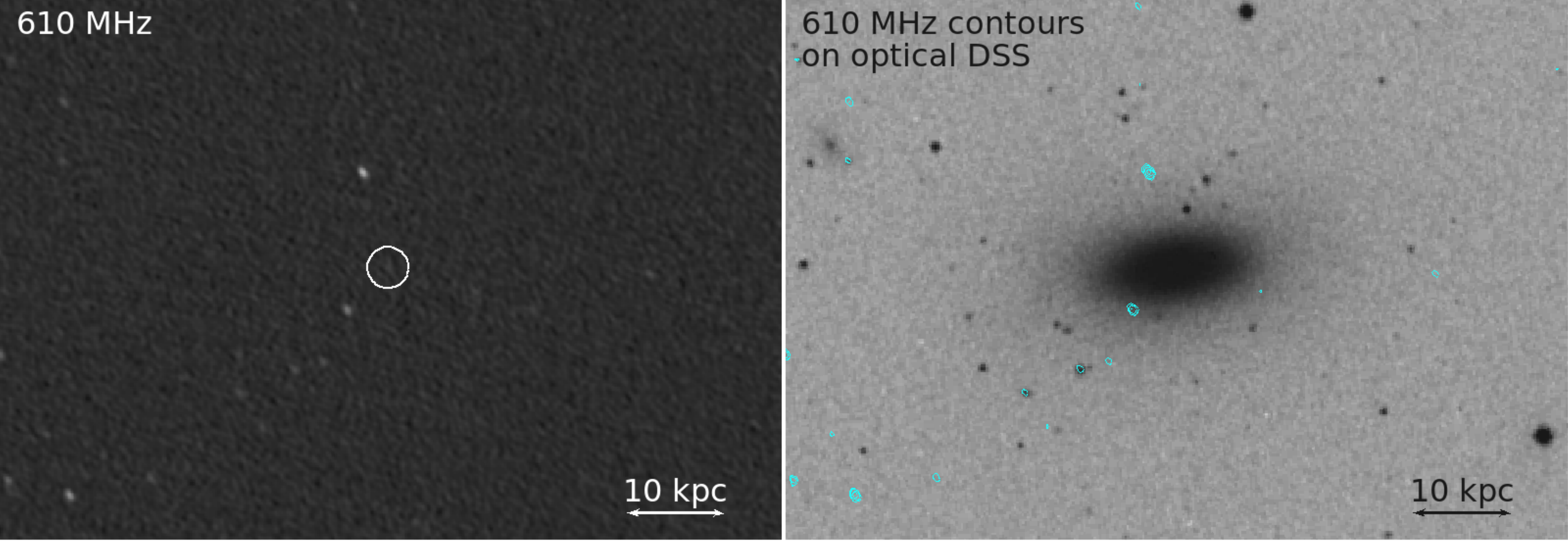}
\caption{LGG 232 / NGC 3613. \textit{Top left:} GMRT 235~MHz image. \textit{Top right:} GMRT 235~MHz contours in green (1$\sigma$ = 0.26 mJy beam$^{-1}$), overlaid on the \textit{Digitized Sky Survey (DSS)} optical image. \textit{Bottom left:} GMRT 610~MHz image. \textit{Bottom right:} GMRT 610~MHz contours in cyan (1$\sigma$ = 30 $\mu$Jy beam$^{-1}$), overlaid on the \textit{Digitized Sky Survey (DSS)} optical image. In both panels the radio contours are spaced by a factor of two, starting from 3$\sigma$ level of significance. For this source the scale is 0.156 kpc arcsec$^{-1}$.}
\label{fig:3613}
\end{figure*}

A very faint central radio source was detected at 1.4~GHz in NGC~3613 using the VLA \citep{Brown11}, with a flux density of 0.3$\pm$0.3~mJy, the faintest in our sample. Our GMRT observations find no radio detection at either frequency (fig.~\ref{fig:3613}). The flux density upper limit set for this galaxy at 5$\sigma$ significance is $<$0.2~mJy at 610~MHz and $<$1.1~mJy at 235~MHz.

\subsection{LGG 236 / NGC 3665} 
\label{AppA13}

NGC~3665 is the BGE of the X-ray luminous LGG~236 loose group of galaxies, which exhibits an unusually low velocity dispersion \citep[$\sim$217 km~s$^{-1}$][]{Heldson05}. The galaxy contains a significant cold gas reservoir \citep{Young11,Alatalo13} but has only low-level star formation \citep{Xiao18}. NGC~3665 hostss a relatively weak double-jet radio source \citep{Parma86,Nyland16} with its jet axis nearly perpendicular to the major axis of the central molecular gas disk \citep{Onishi17}.  

\begin{figure*}
\vspace{3cm}
\centering
\includegraphics[width=1.00\textwidth]{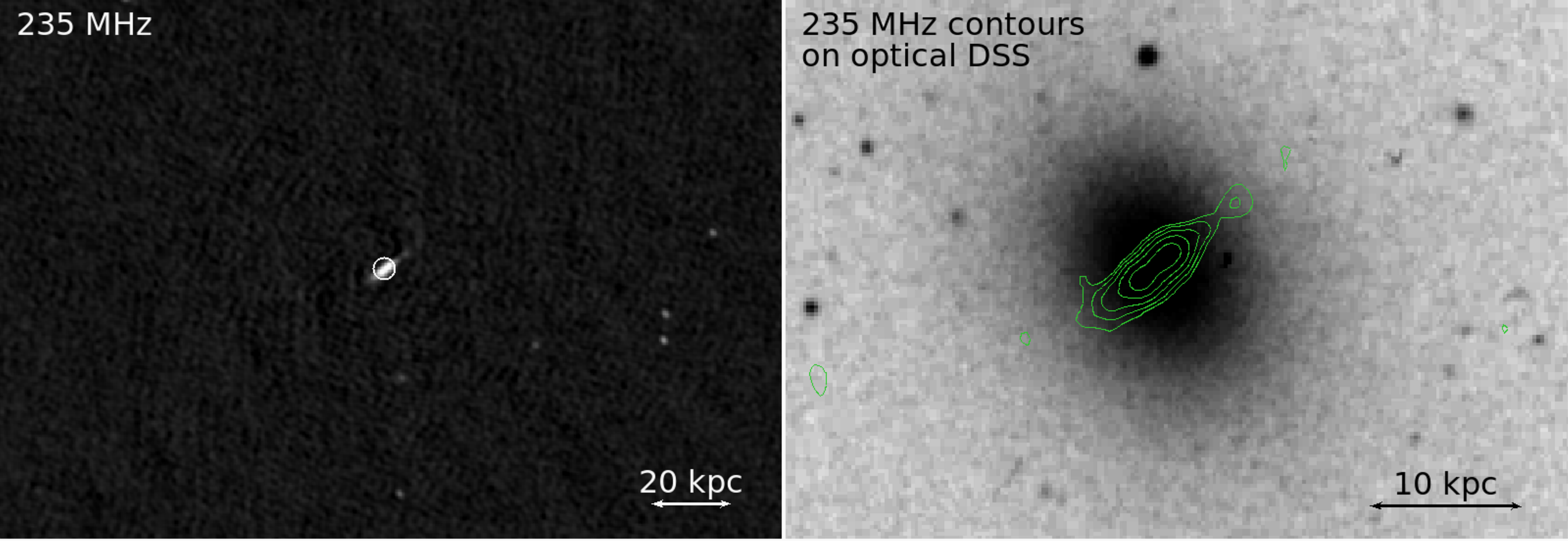}
\includegraphics[width=1.00\textwidth]{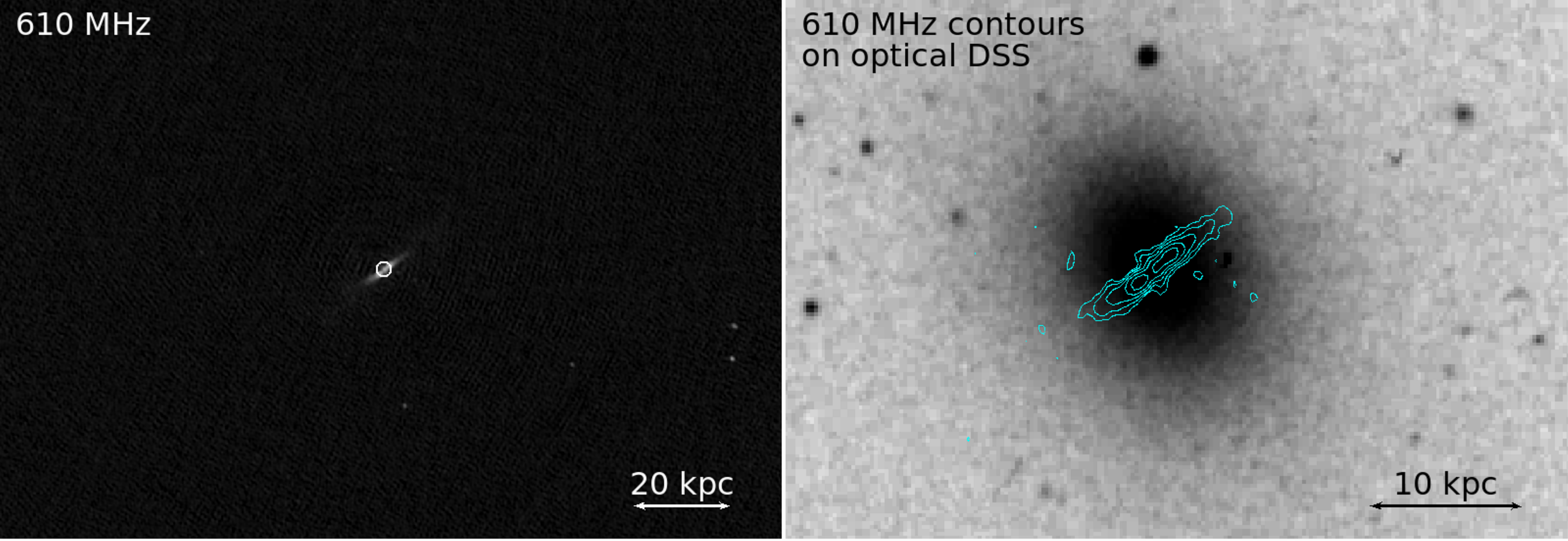}
\caption{LGG 236 / NGC 3665. \textit{Top left:} GMRT 235~MHz image. \textit{Top right:} GMRT 235~MHz contours in green (1$\sigma$ = 0.78 mJy beam$^{-1}$), overlaid on the \textit{Digitized Sky Survey (DSS)} optical image. \textit{Bottom left:} GMRT 610~MHz image. \textit{Bottom right:} GMRT 610~MHz contours in cyan (1$\sigma$ = 170 $\mu$Jy beam$^{-1}$), overlaid on the \textit{Digitized Sky Survey (DSS)} optical image. In both panels the radio contours are spaced by a factor of two, starting from 3$\sigma$ level of significance. For this source the scale is 0.156 kpc arcsec$^{-1}$.}
\label{fig:3665}
\end{figure*}

Previous 1.4~GHz observations from NVSS \citep{Condon98} detected a radio jet source with a flux density of 113.2$\pm$3.8~mJy. Our GMRT observations confirm the symmetric small-scale jet radio morphology at both frequencies, with the largest linear size of the source being $\sim$16~kpc at 235~MHz (see fig.~\ref{fig:3665}). At 235~MHz the source appears to be wider and more extended, whereas at 610~MHz it presents a symmetric thin-jet structure.

\subsection{LGG 255 / NGC 3923} 
\label{AppA14}

\begin{figure*}
\vspace{3cm}
\centering
\includegraphics[width=1.00\textwidth]{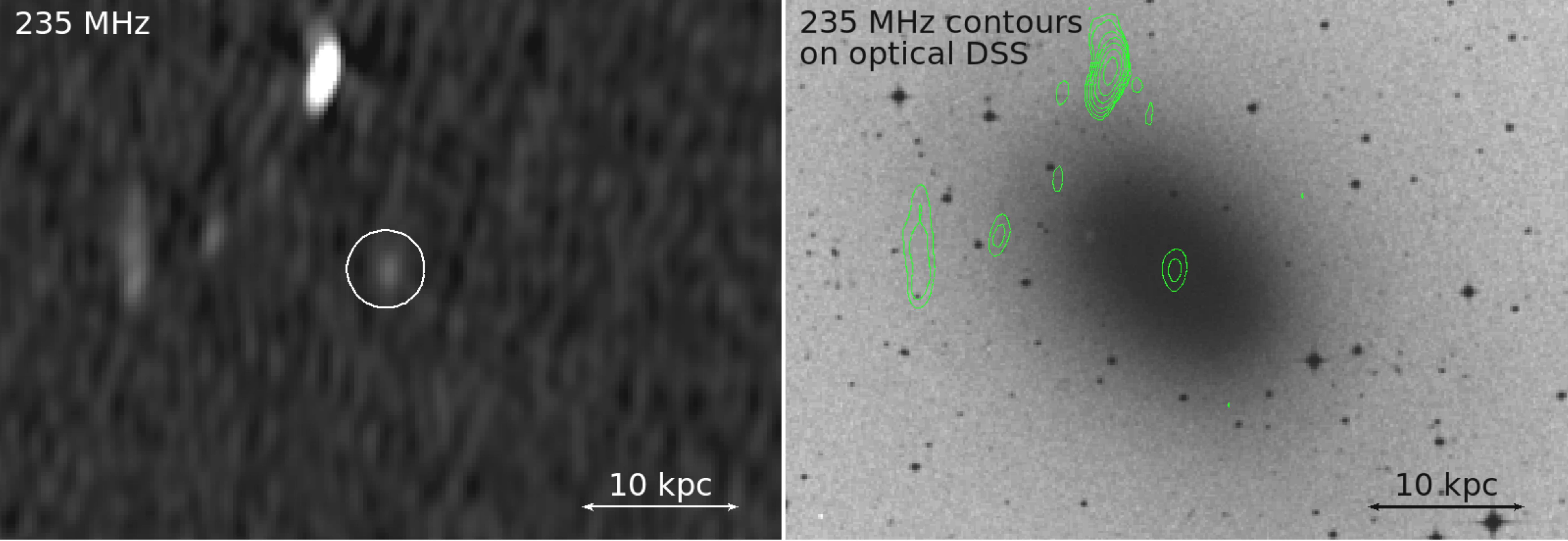}
\includegraphics[width=1.00\textwidth]{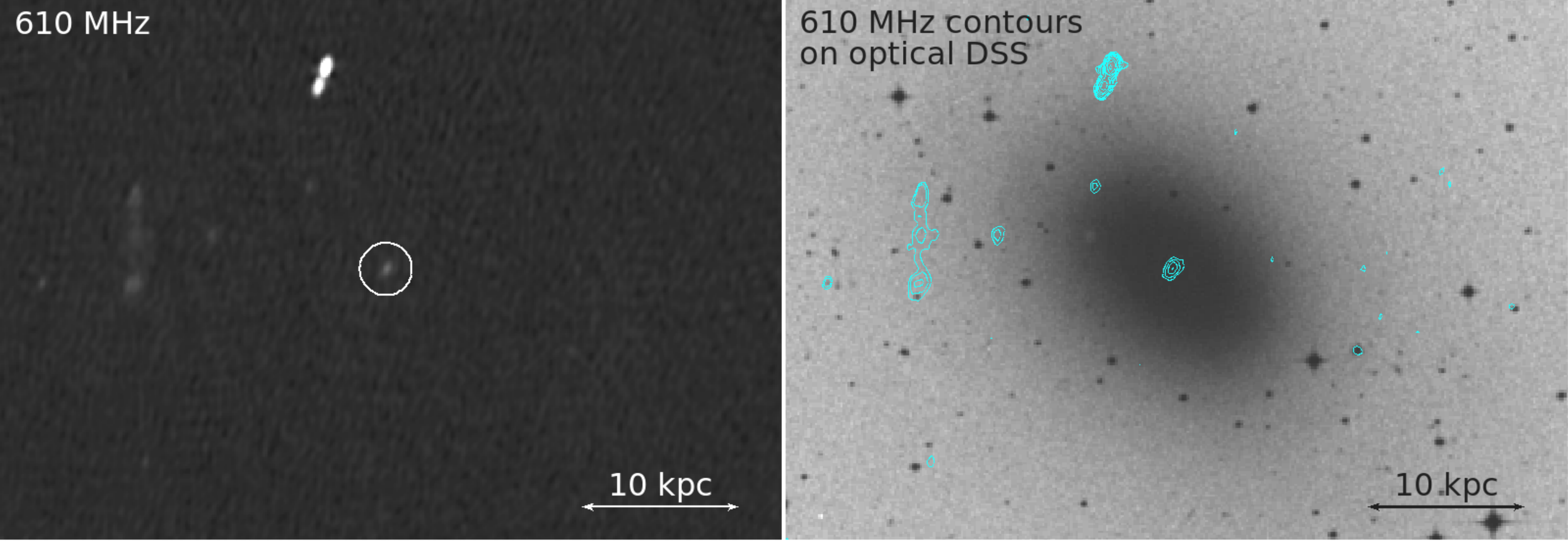}
\caption{LGG 255 / NGC 3923. \textit{Top left:} GMRT 235~MHz image. \textit{Top right:} GMRT 235~MHz contours in green (1$\sigma$ = 0.45 mJy beam$^{-1}$), overlaid on the \textit{Digitized Sky Survey (DSS)} optical image. \textit{Bottom left:} GMRT 610~MHz image. \textit{Bottom right:} GMRT 610~MHz contours in cyan (1$\sigma$ = 50 $\mu$Jy beam$^{-1}$), overlaid on the \textit{Digitized Sky Survey (DSS)} optical image. In both panels the radio contours are spaced by a factor of two, starting from 3$\sigma$ level of significance. For this source the scale is 0.097 kpc arcsec$^{-1}$.}
\label{fig:3923}
\end{figure*}

NGC~3923 is a shell galaxy, with multiple shells of stars in its halo and is considered to be a product of a minor merger, \citep[e.g.,][]{Dupraz86,Hernquist88}. Previous 1.4~GHz VLA \citep{Brown11} and single dish 5~GHz observations from Parkes radio telescope found only a tentative detection with a flux density of 1$\pm$0.5~mJy and 1.5$\pm$1.5 mJy \citep{Disney77} respectively. We detect a weak radio point source at both GMRT frequencies (fig~\ref{fig:3923}), with a spectral index of $\alpha_{235}^{610}\sim0.91$.

\subsection{LGG 314 / NGC 4697} 
\label{AppA15}

\begin{figure*}
\vspace{3cm}
\centering
\includegraphics[width=1.00\textwidth]{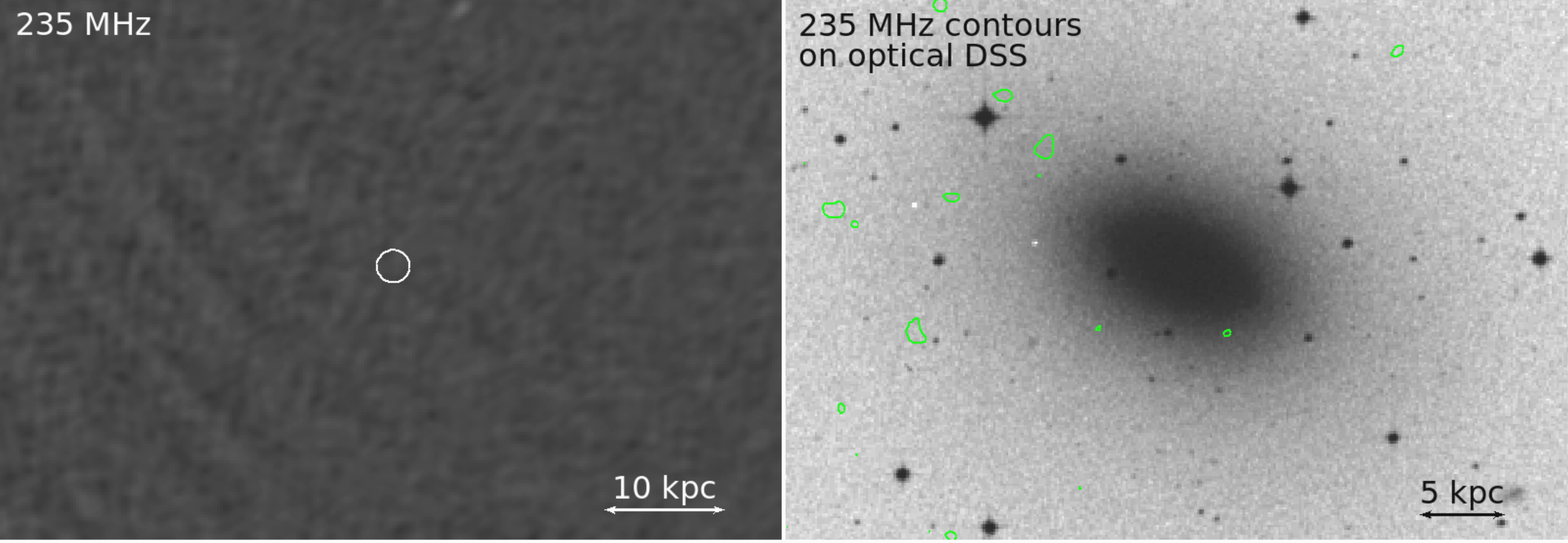}
\includegraphics[width=1.00\textwidth]{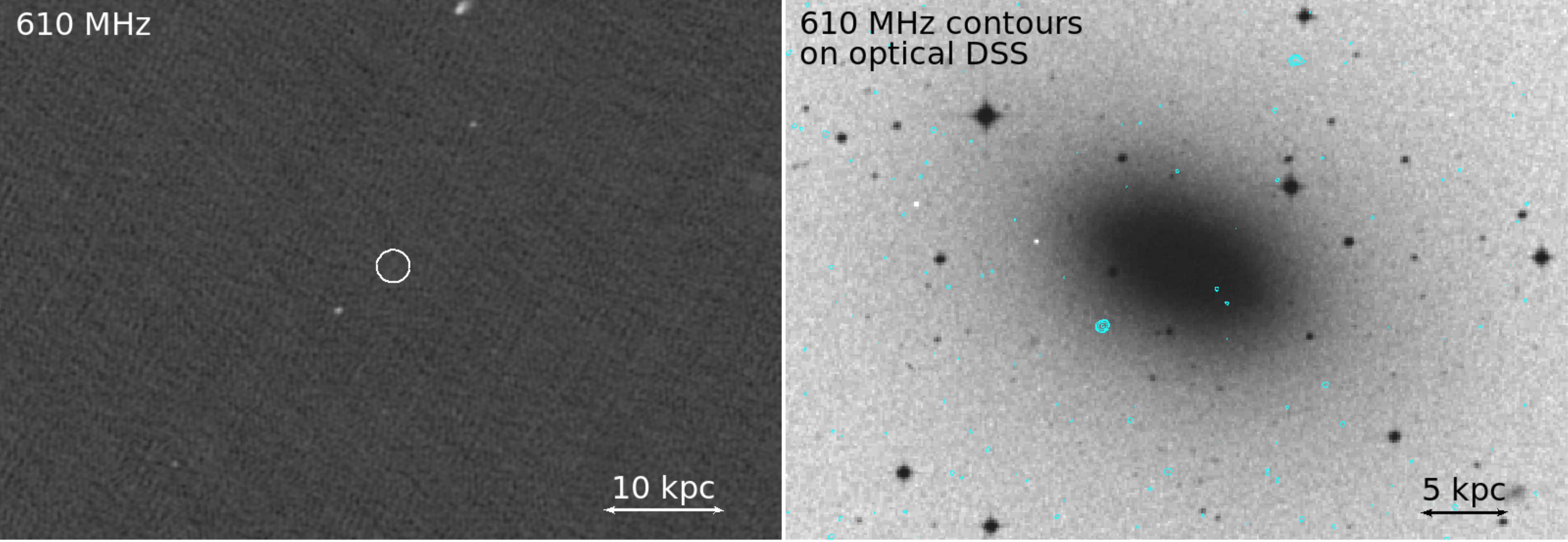}
\caption{LGG 314 / NGC 4697. \textit{Top left:} GMRT 235~MHz image. \textit{Top right:} GMRT 235~MHz contours in green (1$\sigma$ = 1.14 mJy beam$^{-1}$), overlaid on the \textit{Digitized Sky Survey (DSS)} optical image. \textit{Bottom left:} GMRT 610~MHz image. \textit{Bottom right:} GMRT 610~MHz contours in cyan (1$\sigma$ = 70 $\mu$Jy beam$^{-1}$), overlaid on the \textit{Digitized Sky Survey (DSS)} optical image. In both panels the radio contours are spaced by a factor of two, starting from 3$\sigma$ level of significance. For this source the scale is 0.087 kpc arcsec$^{-1}$.}
\label{fig:4697}
\end{figure*}

NGC~4697 is the BGE of the X-ray faint LGG~314 group of galaxies \citep[ROSAT;][]{Sansom00}. Previous 1.4~GHz VLA \citep{Brown11} observations detect only a very weak radio source with a flux density of 0.6$\pm$0.5~mJy. Our GMRT observations find no radio detection at either frequency (fig.~\ref{fig:4697}) for this galaxy, with 5$\sigma$ flux density upper limits of $<$0.35~mJy at 610~MHz and $<$5.7~mJy at 235~MHz.

\subsection{LGG 329 / NGC 4956} 
\label{AppA16}

\begin{figure*}
\vspace{3cm}
\centering
\includegraphics[width=1.00\textwidth]{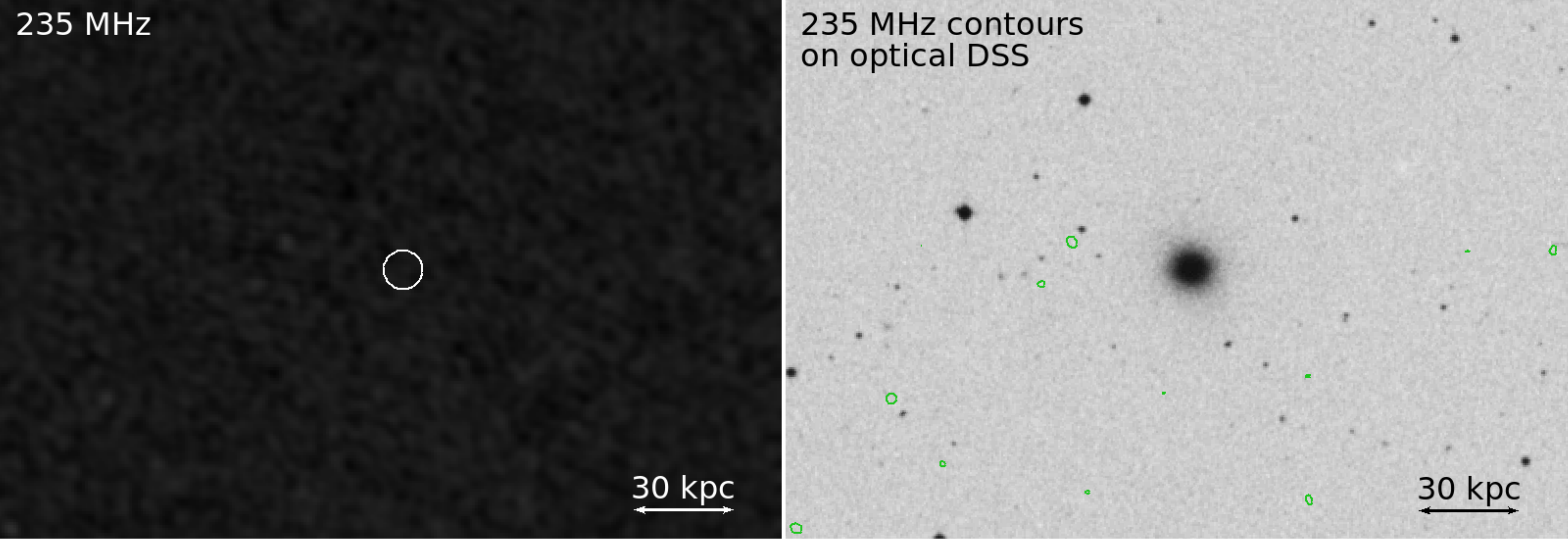}
\includegraphics[width=1.00\textwidth]{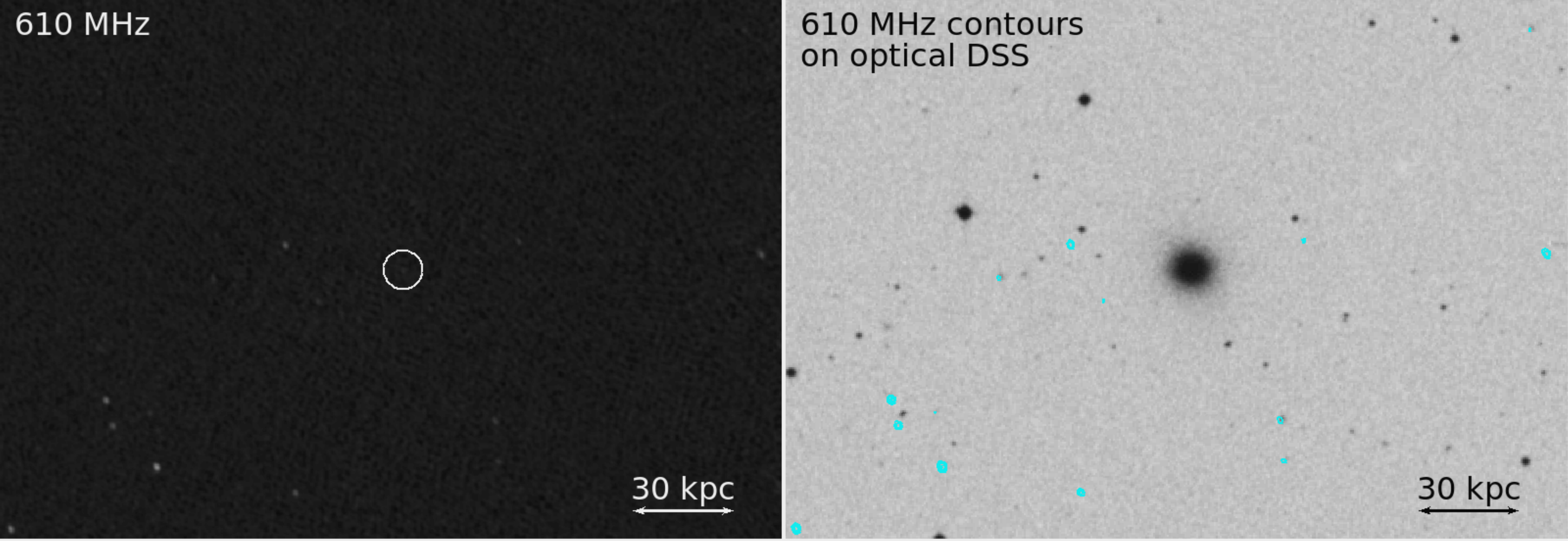}
\caption{LGG 329 / NGC 4956. \textit{Top left:} GMRT 235~MHz image. \textit{Top right:} GMRT 235~MHz contours in green (1$\sigma$ = 0.48 mJy beam$^{-1}$), overlaid on the \textit{Digitized Sky Survey (DSS)} optical image. \textit{Bottom left:} GMRT 610~MHz image. \textit{Bottom right:} GMRT 610~MHz contours in cyan (1$\sigma$ = 50 $\mu$Jy beam$^{-1}$), overlaid on the \textit{Digitized Sky Survey (DSS)} optical image. In both panels the radio contours are spaced by a factor of two, starting from 3$\sigma$ level of significance. For this source the scale is 0.344 kpc arcsec$^{-1}$.}
\label{fig:4956}
\end{figure*}

The S0 NGC~4956 was not previously detected in the radio band and our GMRT observations find no radio detection at either frequency (fig.~\ref{fig:4956}). The 5$\sigma$ flux density upper limits are $<$0.3~mJy at 610~MHz and $<$2.4~mJy at 235~MHz.

\subsection{LGG 341 / NGC 5061} 
\label{AppA17}

\begin{figure*}
\vspace{3cm}
\centering
\includegraphics[width=1.00\textwidth]{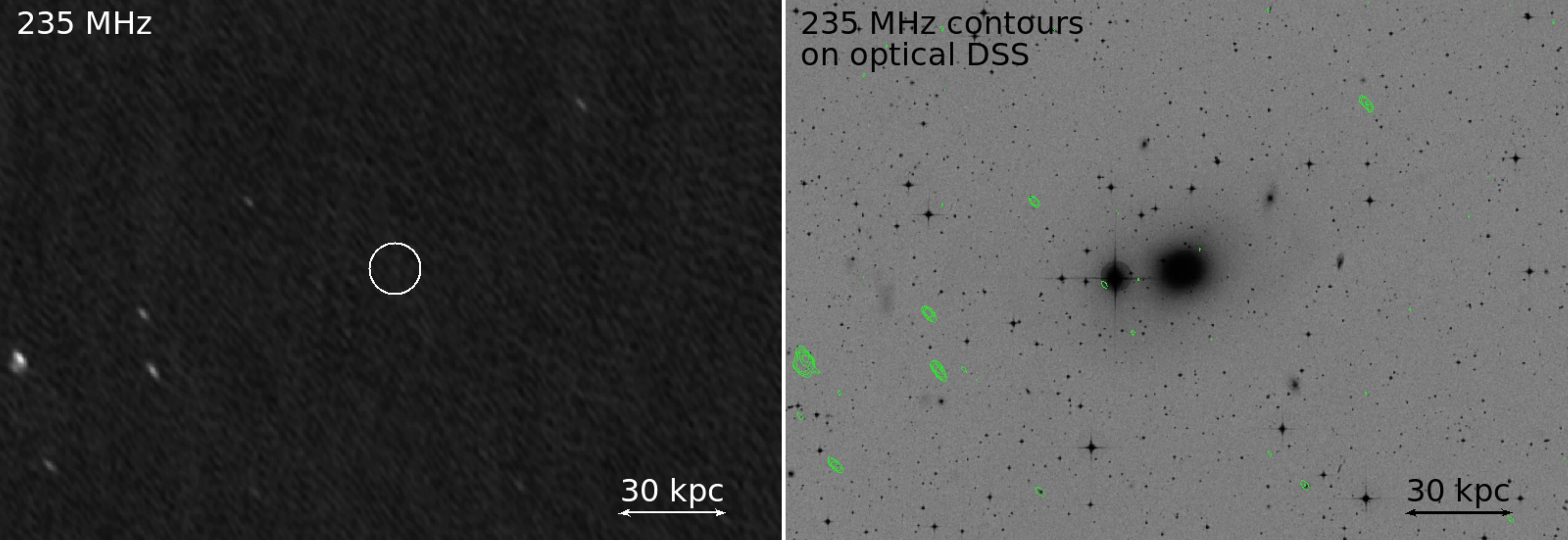}
\includegraphics[width=1.00\textwidth]{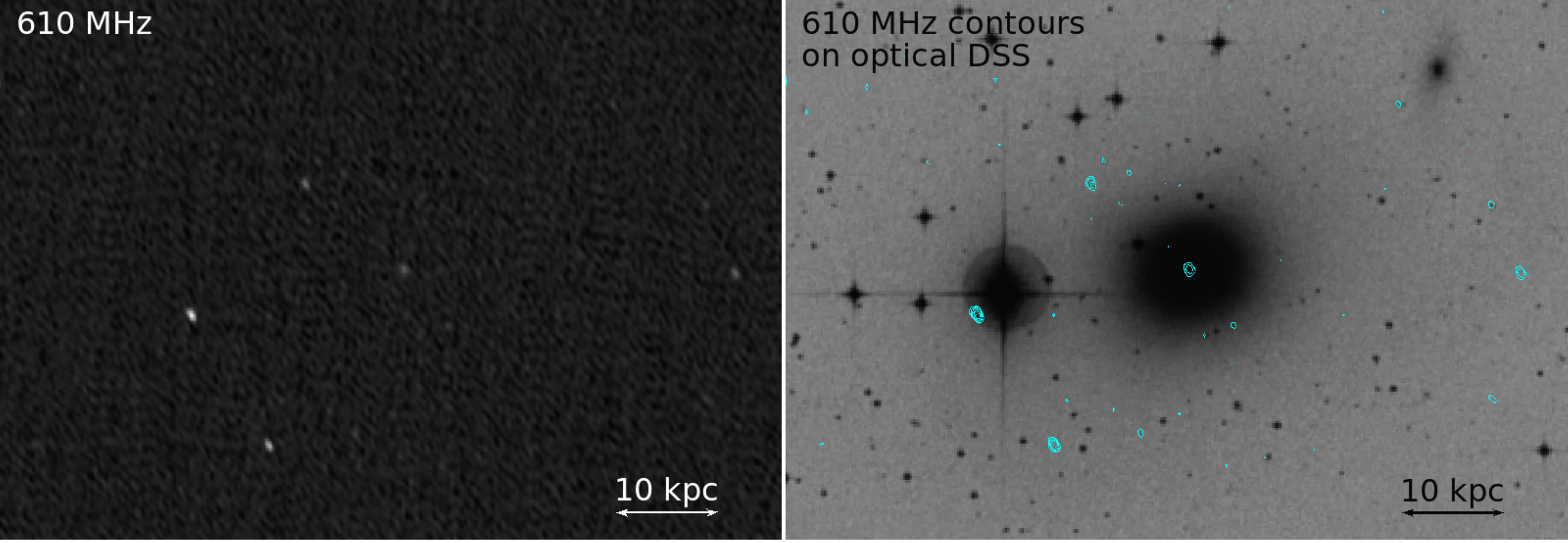}
\caption{LGG 341 / NGC 5061. \textit{Top left:} GMRT 235~MHz image. \textit{Top right:} GMRT 235~MHz contours in green (1$\sigma$ = 0.50 mJy beam$^{-1}$), overlaid on the \textit{Digitized Sky Survey (DSS)} optical image. \textit{Bottom left:} GMRT 610~MHz image. \textit{Bottom right:} GMRT 610~MHz contours in cyan (1$\sigma$ = 50 $\mu$Jy beam$^{-1}$), overlaid on the \textit{Digitized Sky Survey (DSS)} optical image. In both panels the radio contours are spaced by a factor of two, starting from 3$\sigma$ level of significance. For this source the scale is 0.136 kpc arcsec$^{-1}$.}
\label{fig:5061}
\end{figure*}

NGC~5061 is the dominant elliptical galaxy of the X-ray faint LGG~341 group of galaxies \citep{Beuing99}. It is thought to have formed through several dry minor and major merging events \citep{Dullo14}. The galaxy was not previously detected at 1.4~GHz, and only a single dish tentative detection from Parkes telescope at 5~GHz is reported, with a flux density of 1$\pm$2.8~mJy \citep{Disney77}. We detect a weak radio point source associated with NGC~5061 only at 610~MHz (fig~\ref{fig:5061}) with a 5$\sigma$ flux density upper limit of $<$2.5~mJy at 235~MHz.

\subsection{LGG 350 / NGC 5127} 
\label{AppA18}

NGC~5127 hosts the FR~I radio source B2~1321+31. Westerbork Synthesis Radio Telescope (WSRT) observations \citet{Klein95} revealed that the there is little variation in the 610$-$10700~MHz spectral along the jet major axis but the spectral index steepens from the jet centre to its edge. The mean spectral index of the jet is $\alpha_{610}^{10700}\sim0.56$ with the steepest indices found in the west lobe ($\alpha_{610}^{10700}\sim1.00$). 

\begin{figure*}
\vspace{3cm}
\centering
\includegraphics[width=1.00\textwidth]{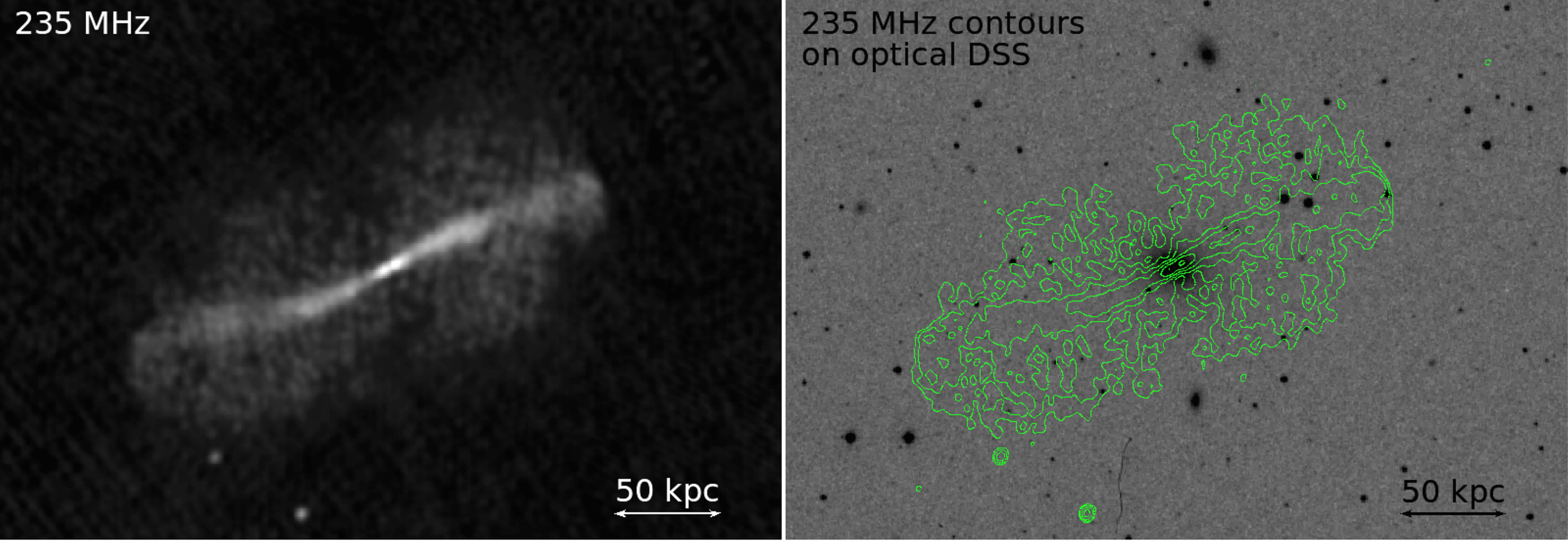}
\includegraphics[width=1.00\textwidth]{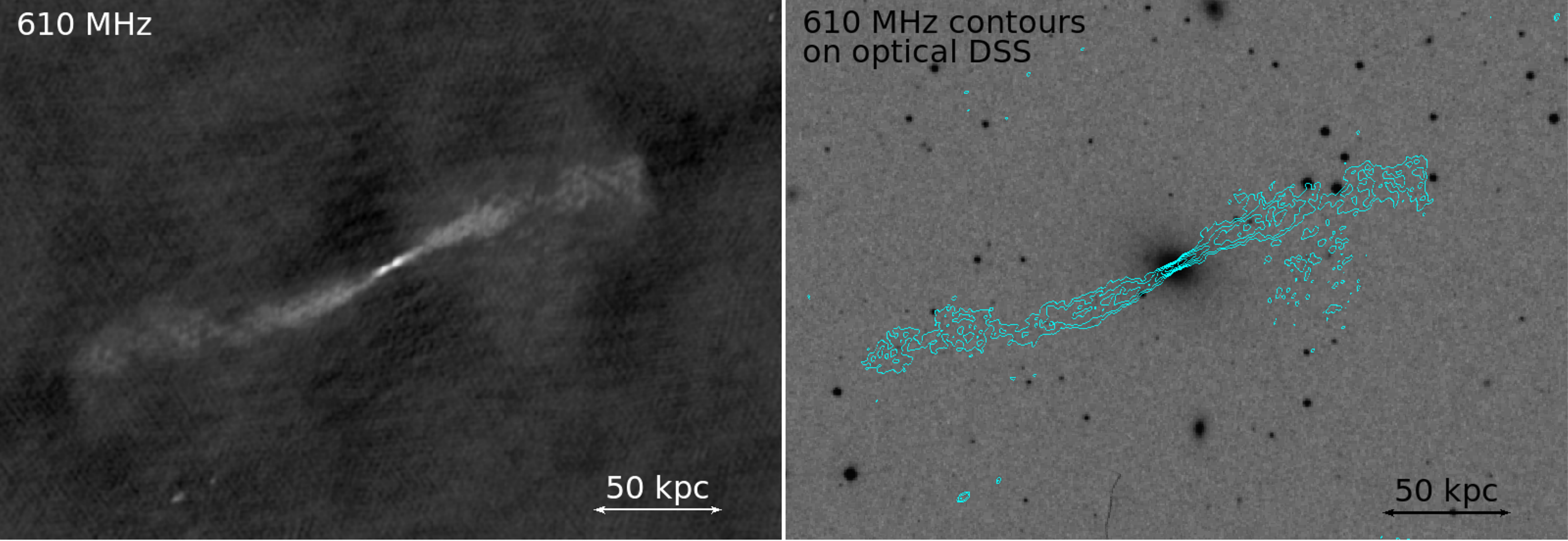}
\caption{LGG 350 / NGC 5127. \textit{Top left:} GMRT 235~MHz image. \textit{Top right:} GMRT 235~MHz contours in green (1$\sigma$ = 0.65 mJy beam$^{-1}$), overlaid on the \textit{Digitized Sky Survey (DSS)} optical image. \textit{Bottom left:} GMRT 610~MHz image. \textit{Bottom right:} GMRT 610~MHz contours in cyan (1$\sigma$ = 250 $\mu$Jy beam$^{-1}$), overlaid on the \textit{Digitized Sky Survey (DSS)} optical image. In both panels the radio contours are spaced by a factor of two, starting from 3$\sigma$ level of significance. For this source the scale is 0.349 kpc arcsec$^{-1}$.}
\label{fig:5127}
\end{figure*}

We detect the B2~1321+31 at both GMRT frequencies (fig~\ref{fig:5127}). It has a symmetric double jet radio morphology with the total size of the source being $\sim$244~kpc. At 235~MHz we see extensive lobes and diffuse radio emission that surrounds the main jet, whereas at 610~MHz the source appears to be thin, with only the linear jet emission being detected along with some clumpy emission in the west lobe. The eastern jet appears to bend to the north as it expands into the lobe.

\subsection{LGG 350 / NGC 5322} 
\label{AppA19}

NGC~5322 is a LINER \citep{Baldi09} with a central stellar mass deficit \citep{Carollo97}, a counter-rotating nuclear stellar and gaseous disk \citep[e.g.,][]{Bender88,Bender92, Dullo18}, and hosts a central radio jet source \citep{Feretti84,Hummel84}. In a recent study, \citet{Dullo18} suggested that the most likely formation scenario for NGC~5322 is through a sequence of several initial `dry' major mergers followed up by the subsequent accretion of a metal-rich gaseous satellite, with the latter event considered to have created the counter-rotating cold gas reservoir in the centre that fueled the AGN with gas and powered the radio source.

\begin{figure*}
\vspace{3cm}
\centering
\includegraphics[width=1.00\textwidth]{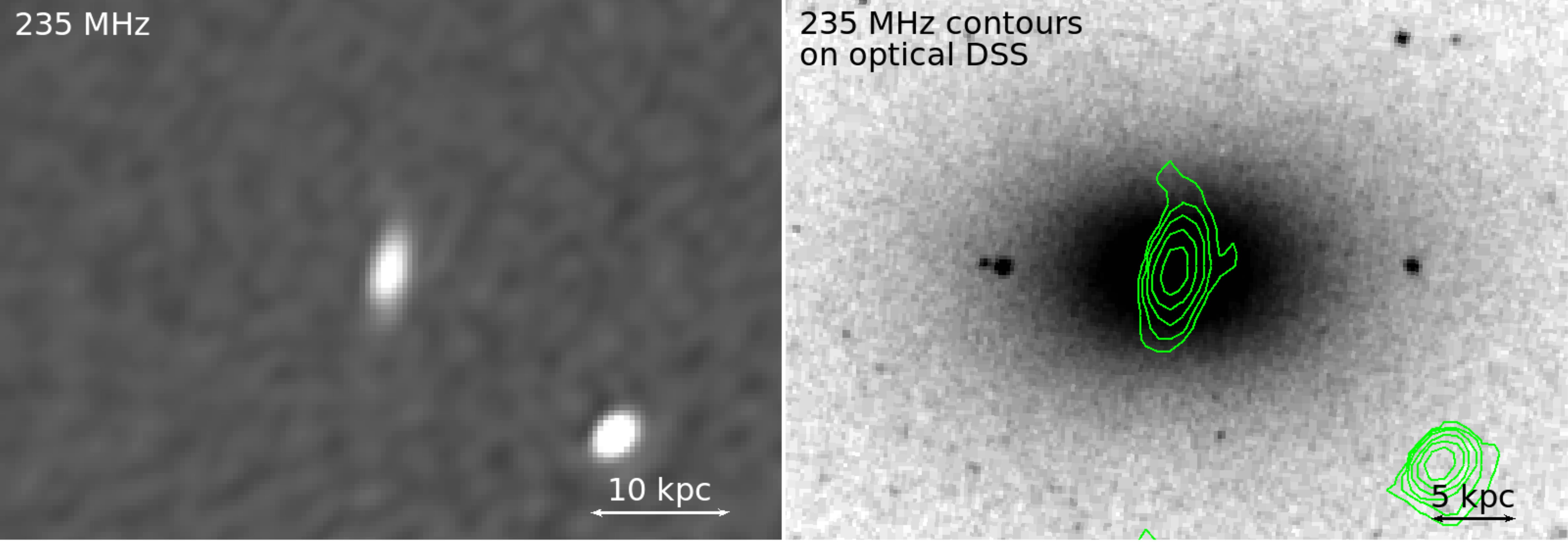}
\includegraphics[width=1.00\textwidth]{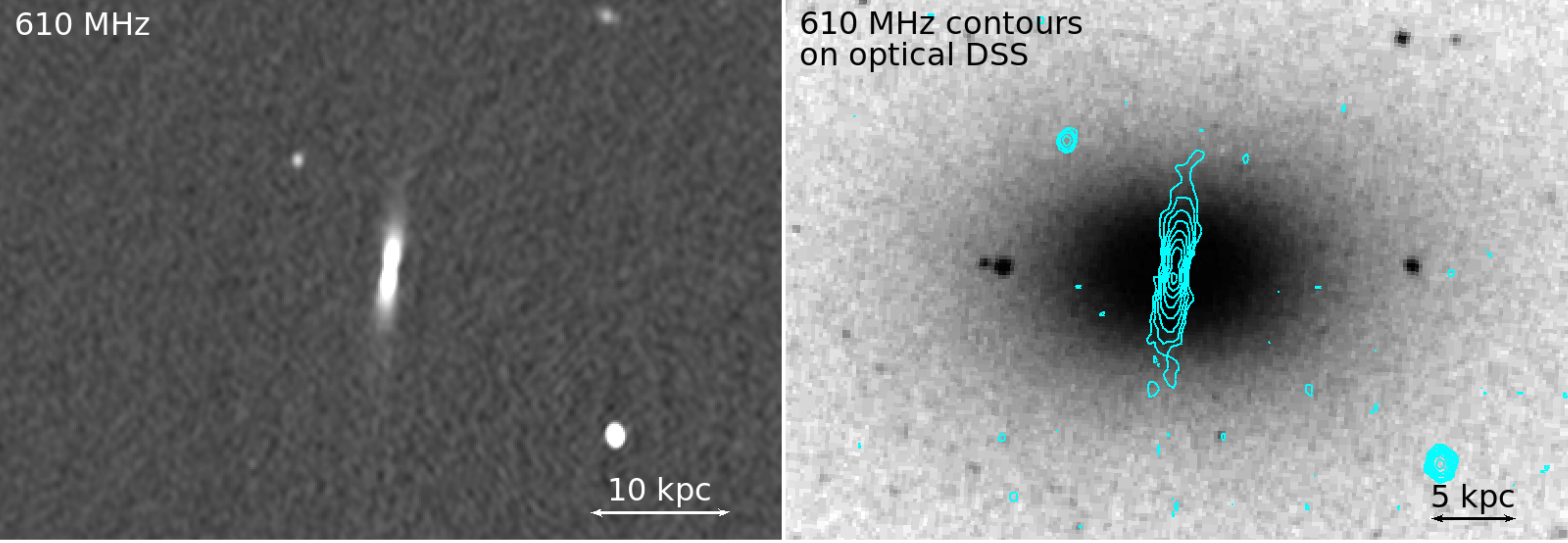}
\caption{LGG 360 / NGC 5322. \textit{Top left:} GMRT 235~MHz image. \textit{Top right:} GMRT 235~MHz contours in green (1$\sigma$ = 0.70 mJy beam$^{-1}$), overlaid on the \textit{Digitized Sky Survey (DSS)} optical image. \textit{Bottom left:} GMRT 610~MHz image. \textit{Bottom right:} GMRT 610~MHz contours in cyan (1$\sigma$ = 50 $\mu$Jy beam$^{-1}$), overlaid on the \textit{Digitized Sky Survey (DSS)} optical image. In both panels the radio contours are spaced by a factor of two, starting from 3$\sigma$ level of significance. For this source the scale is 0.141 kpc arcsec$^{-1}$.}
\label{fig:5322}
\end{figure*}

Previous 1.4~GHz VLA observations show an FR~I radio morphology  extending over $\sim$20'' \citep{Hummel84}. \citet{Dullo18} using sub-arcsecond resolution 1.5~GHz e-MERLIN observations report that the inner jet structure has the same position angle as the extended one observed by the VLA. In our GMRT observations the source has a symmetric small-scale jet morphology at both frequencies, with the largest linear size of the source being $\sim$15~kpc at 610~MHz (see fig.~\ref{fig:5322}). At 235~MHz the source is poorly unresolved and only slightly elongated, whereas at 610~MHz the thin-jet structure is much clearer, aligned roughly north-south. The small-scale jet presents a very flat spectral index of $\alpha_{235}^{610}\sim0.25$ indicating a young radio AGN.

\subsection{LGG 370 / NGC 5444} 
\label{AppA20}

\begin{figure*}
\vspace{3cm}
\centering
\includegraphics[width=1.00\textwidth]{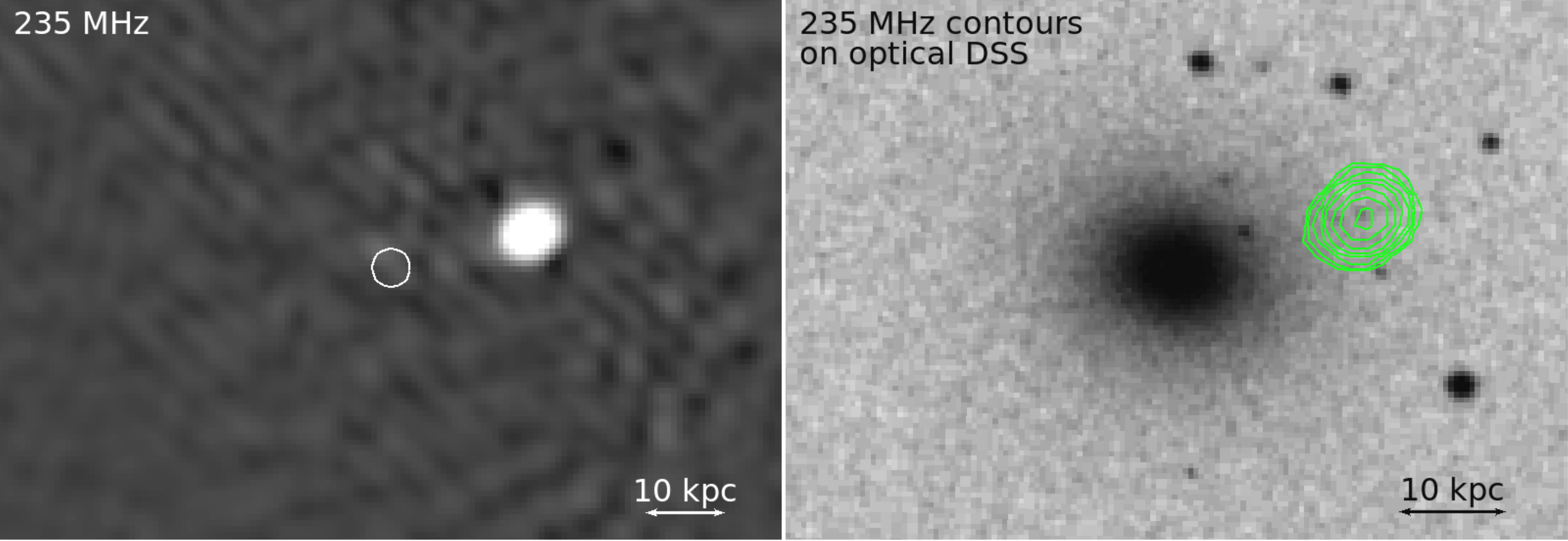}
\includegraphics[width=1.00\textwidth]{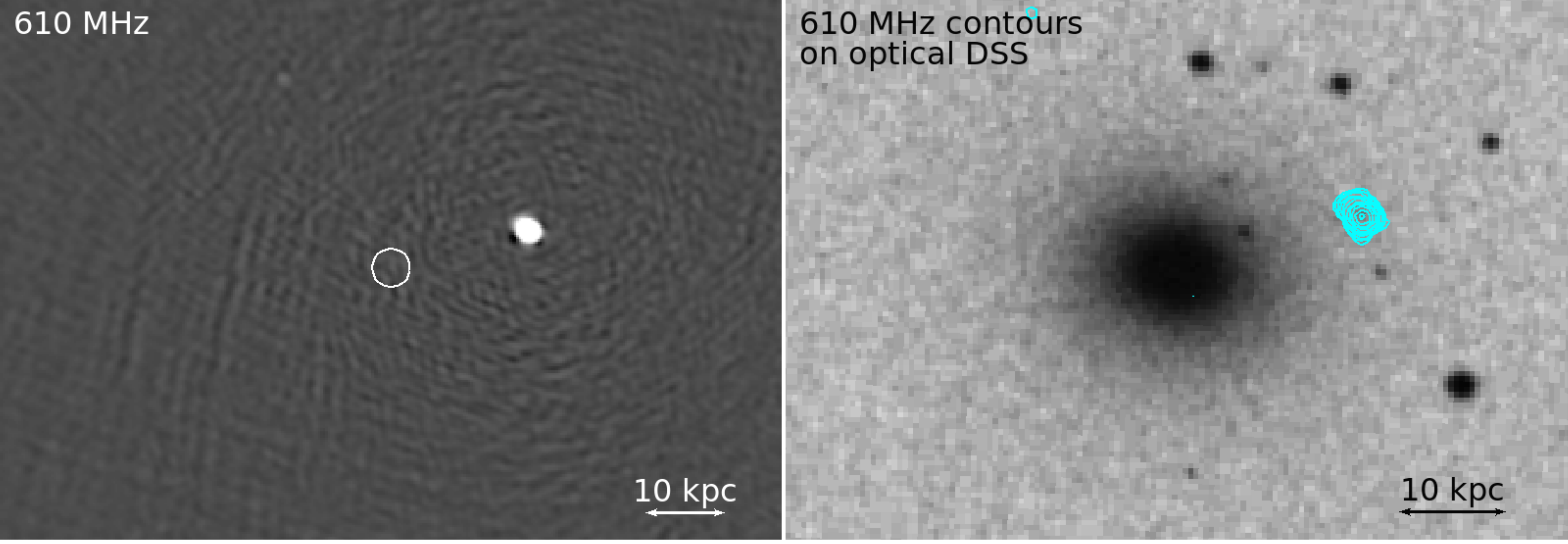}
\caption{LGG 370 / NGC 5444. \textit{Top left:} GMRT 235~MHz image. \textit{Top right:} GMRT 235~MHz contours in green (1$\sigma$ = 0.80 mJy beam$^{-1}$), overlaid on the \textit{Digitized Sky Survey (DSS)} optical image. \textit{Bottom left:} GMRT 610~MHz image. \textit{Bottom right:} GMRT 610~MHz contours in cyan (1$\sigma$ = 200 $\mu$Jy beam$^{-1}$), overlaid on the \textit{Digitized Sky Survey (DSS)} optical image. In both panels the radio contours are spaced by a factor of two, starting from 3$\sigma$ level of significance. For this source the scale is 0.291 kpc arcsec$^{-1}$.}
\label{fig:5444}
\end{figure*}

Previous 1.4~GHz VLA observations of NGC~5444 \citep{Brown11} report a detection with a flux density of 660$\pm$20~mJy. However, this flux actually arises from the nearby radio source 4C~+35.32, and not from NGC~5444 \citep[see][]{Brandie76}. Figure~\ref{fig:5444} shows that our GMRT observations find no radio detection at either frequency associated with NGC~5444. 4C~+35.32 is detected as a strong source with a projected separation of only $\sim$14~kpc ($\sim$1$'$) from NGC~5444. Owing to the presence of this source, the detection of the BGE here is dynamical range limited. Therefore, the upper limit for this galaxy is set from the maximum peak flux density at the position of NGC~5444, and is $<$1~mJy at 610~MHz and $<$12~mJy at 235~MHz.

\subsection{LGG 376 / NGC 5490} 
\label{AppA21}

\begin{figure*}
\vspace{3cm}
\centering
\includegraphics[width=1.00\textwidth]{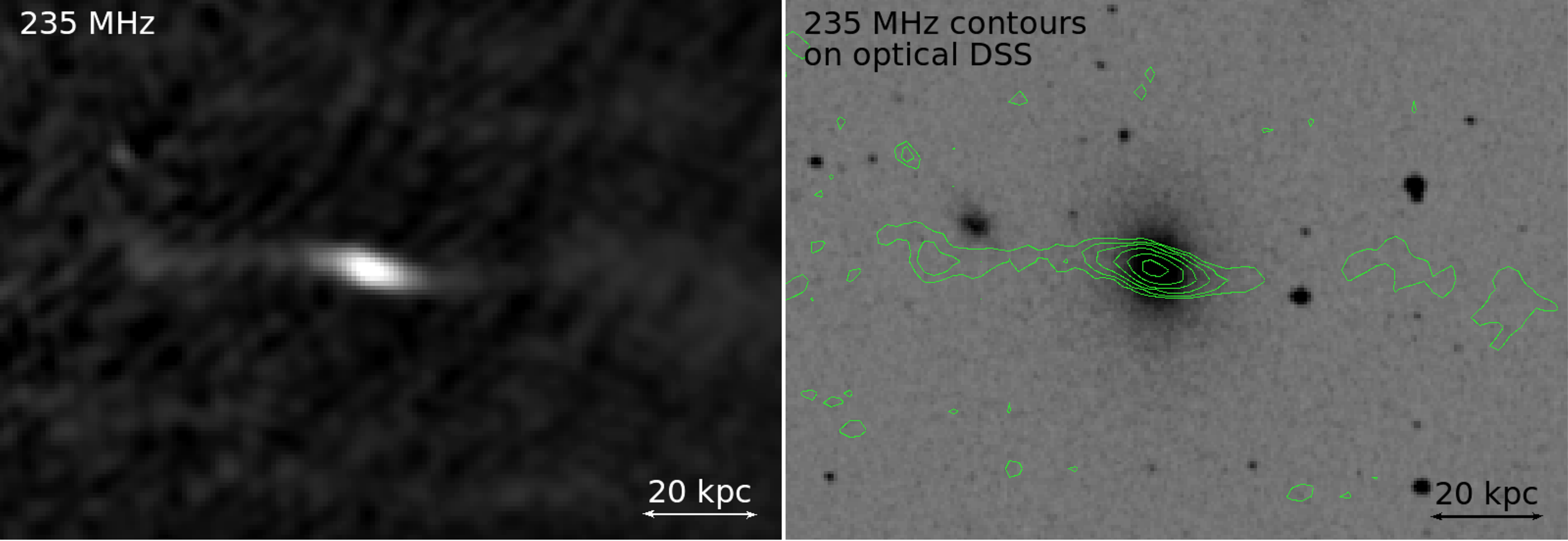}
\includegraphics[width=1.00\textwidth]{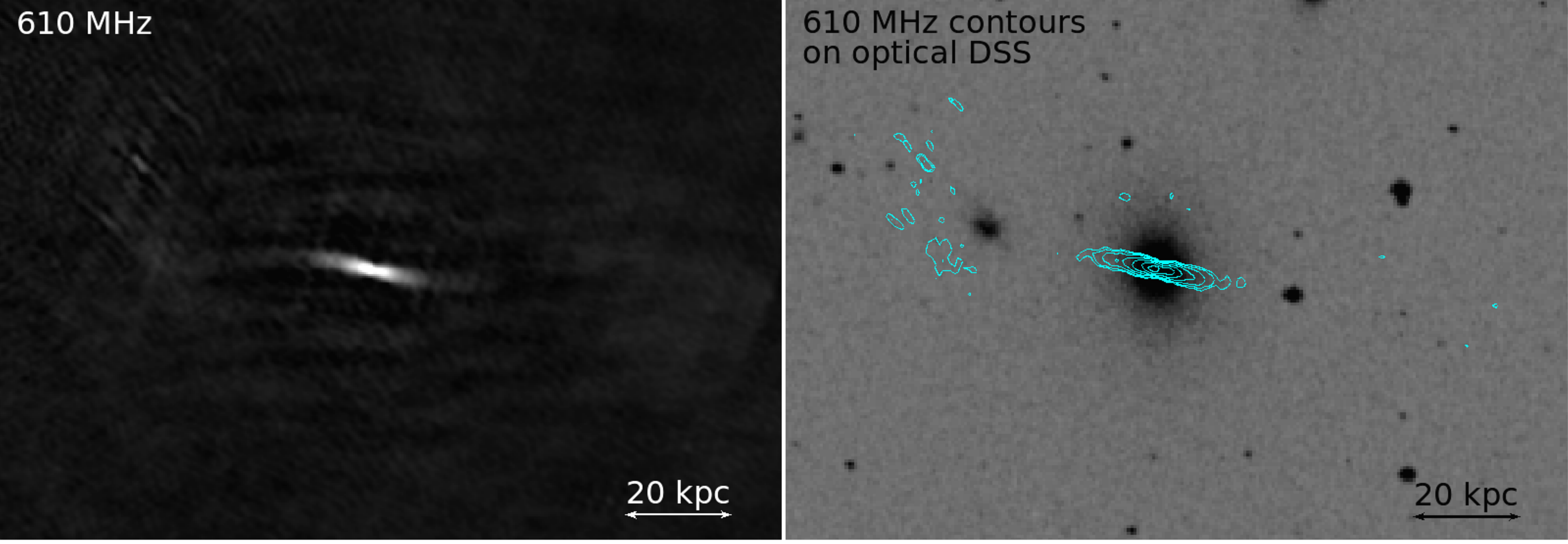}
\caption{LGG 376 / NGC 5490. \textit{Top left:} GMRT 235~MHz image. \textit{Top right:} GMRT 235~MHz contours in green (1$\sigma$ = 1.85 mJy beam$^{-1}$), overlaid on the \textit{Digitized Sky Survey (DSS)} optical image. \textit{Bottom left:} GMRT 610~MHz image. \textit{Bottom right:} GMRT 610~MHz contours in cyan (1$\sigma$ = 250 $\mu$Jy beam$^{-1}$), overlaid on the \textit{Digitized Sky Survey (DSS)} optical image. In both panels the radio contours are spaced by a factor of two, starting from 3$\sigma$ level of significance at 235~MHz and at 4$\sigma$ level of significance at 610~MHz due to the higher rms noise around the source at this frequency. For this source the scale is 0.344 kpc arcsec$^{-1}$.}
\label{fig:5490}
\end{figure*}

NGC~5490 hosts an FR~I radio source \citep{Condon88} in which high resolution (milliarcsecond) VLBA observations at 5~GHz detect a polarized core-jet structure towards the stronger eastern large-scale jet \citep{Xu00,Kharb12b}. VLA 1.4~GHz observations of the source \citep{Brown11} show a total flux density of 1300$\pm$100~mJy. Our GMRT observations detect the source at both frequencies but with different morphology (fig~\ref{fig:5490}). At 235~MHz we see an asymmetric radio core-jet morphology with only the eastern jet being visible as a continuous line of emission. The western jet is much shorter, but detached clumps of radio emission at larger radii presumably mark its path. At this frequency, the total size of the source is $\sim$120~kpc. At 610~MHz, we observe only the central $\sim$30~kpc jet which is linear and thin. The radio source presents an inverted spectral index of  $\alpha_{235}^{1400}\sim-0.07$ in the 235$-$1400~MHz range and a relatively flat spectral index of $\alpha_{235}^{610}\sim0.36$ in the 235$-$610~MHz one. The inverted spectral index can be attributed to the fact that we do not detect the same morphology at different frequencies.

\subsection{LGG 383 / NGC 5629} 
\label{AppA22}

\begin{figure*}
\vspace{3cm}
\centering
\includegraphics[width=1.00\textwidth]{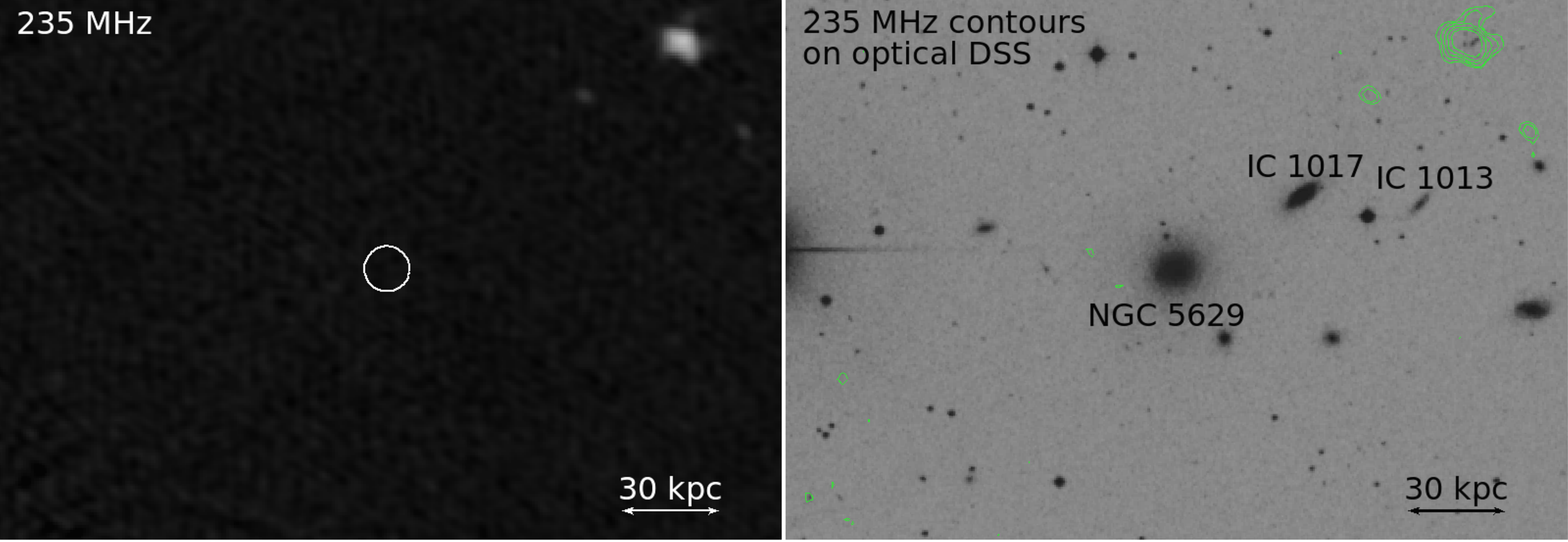}
\includegraphics[width=1.00\textwidth]{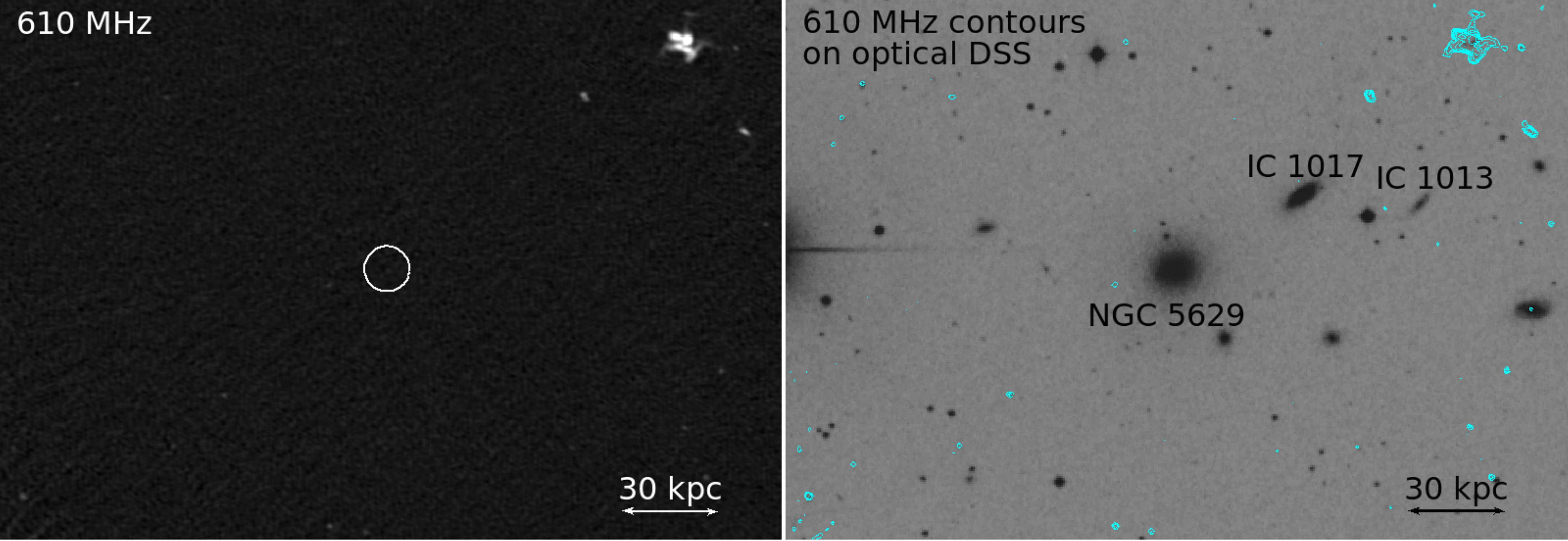}
\caption{LGG 383 / NGC 5629. \textit{Top left:} GMRT 235~MHz image. \textit{Top right:} GMRT 235~MHz contours in green (1$\sigma$ = 0.40 mJy beam$^{-1}$), overlaid on the \textit{Digitized Sky Survey (DSS)} optical image. \textit{Bottom left:} GMRT 610~MHz image. \textit{Bottom right:} GMRT 610~MHz contours in cyan (1$\sigma$ = 50 $\mu$Jy beam$^{-1}$), overlaid on the \textit{Digitized Sky Survey (DSS)} optical image. In both panels the radio contours are spaced by a factor of two, starting from 3$\sigma$ level of significance. For this source the scale is 0.325 kpc arcsec$^{-1}$.}
\label{fig:5629}
\end{figure*}

The S0 NGC~5629 was previously undetected in the radio band and our GMRT observations also find no radio detection at either frequency (fig.~\ref{fig:5629}). The 5$\sigma$ flux density upper limits for this galaxy are $<$0.3~mJy at 610~MHz and $<$2.0~mJy at 235~MHz.

\subsection{LGG 398 / NGC 5903} 
\label{AppA23}

\begin{figure*}
\vspace{3cm}
\centering
\includegraphics[width=1.00\textwidth]{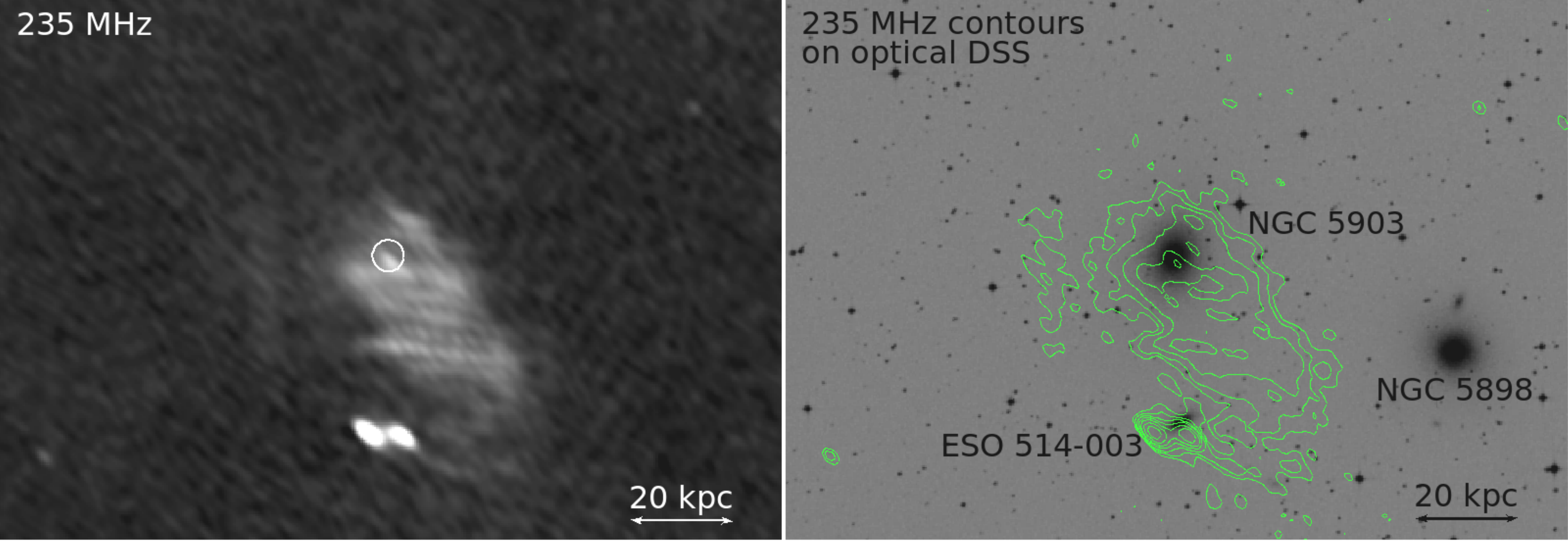}
\includegraphics[width=1.00\textwidth]{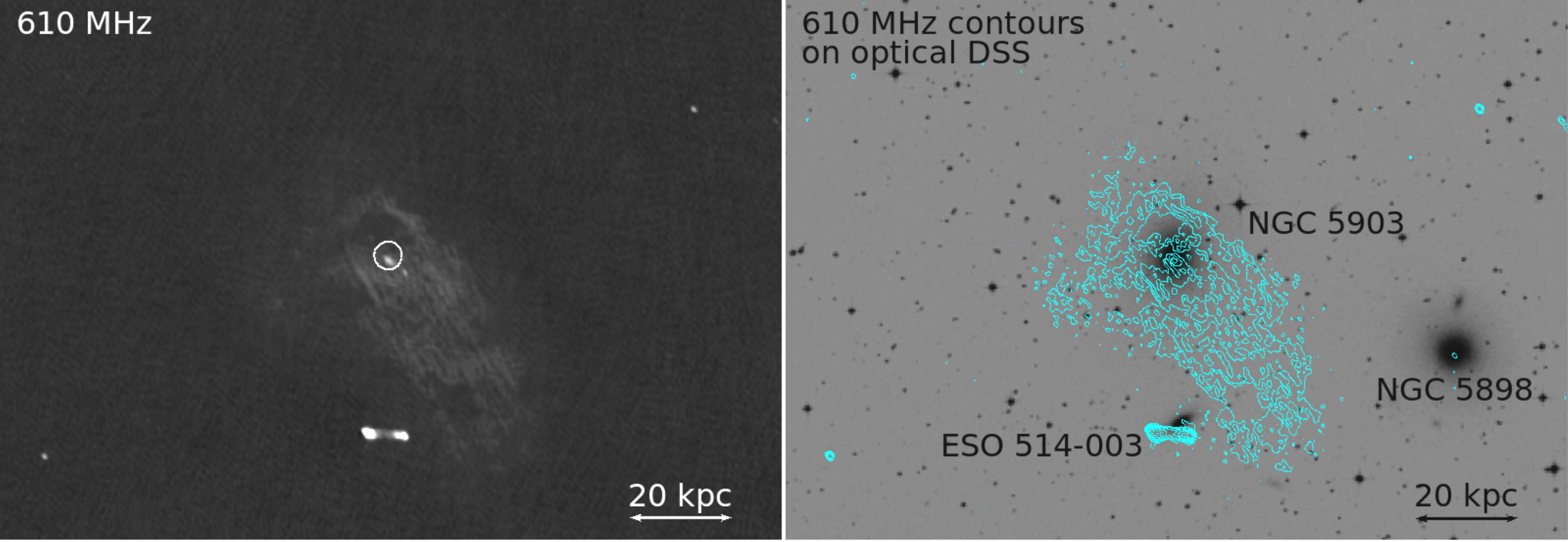}
\caption{LGG 398 / NGC 5903. \textit{Top left:} GMRT 235~MHz image. \textit{Top right:} GMRT 235~MHz contours in green (1$\sigma$ = 0.60 mJy beam$^{-1}$), overlaid on the \textit{Digitized Sky Survey (DSS)} optical image. \textit{Bottom left:} GMRT 610~MHz image. \textit{Bottom right:} GMRT 610~MHz contours in cyan (1$\sigma$ = 80 $\mu$Jy beam$^{-1}$), overlaid on the \textit{Digitized Sky Survey (DSS)} optical image. In both panels the radio contours are spaced by a factor of two, starting from 3$\sigma$ level of significance. For this source the scale is 0.175 kpc arcsec$^{-1}$.}
\label{fig:5903}
\end{figure*}

The radio continuum, H\textsc{i} and X-ray properties of the NGC~5903 group are described in detail in \citet{OSullivan5903}.
The group has a $\sim$0.65~keV IGM extending out to at least 145~kpc \citep{OSullivan5903}, and hosts an extended ($\sim$40 by 75~kpc) steep-spectrum diffuse radio source associated with NGC~5903 \citep{GopalKrishna78,Gopal12}. The spectral index of the radio emission is $\alpha_{150}^{612}\sim1.03\pm0.08$, suggesting a radiative age greater than 360~Myr. The most likely origin scenario for the radio structure is a combination of an interaction-triggered AGN outburst and high-velocity collisions between galaxies. Our images at 610~MHz and 235~MHz (see figure~\ref{fig:5903}) show the diffuse radio structure, with its brightest part coinciding with the central region of the galaxy.

\subsection{LGG 457 / NGC 7252} 
\label{AppA24}

NGC~7252 (also known as the `Atoms-for-Peace' galaxy) is the BGE in the X-ray faint LGG~457 group of galaxies \citep{Nolan04} and is identified as a Compton-thick dust obscured AGN \citep{Tanimoto16}. It is considered to be the remnant of an advanced merger between two gas-rich spiral galaxies and has two clearly visible and H\textsc{i}-rich tidal tails \citep[e.g.,][]{Schweizer82,Dupraz90,Wang92,Hibbard94}. The merger is estimated to have started $\sim$600$-$700 Myr ago, triggering a starburst \citep{HibbardMihos95,ChienBarnes10}, and NGC~7252 is expected to evolve into an elliptical galaxy after exhausting its cold gas reservoir \citep{Toomre77,Schweizer82}. 

\begin{figure*}
\vspace{3cm}
\centering
\includegraphics[width=1.00\textwidth]{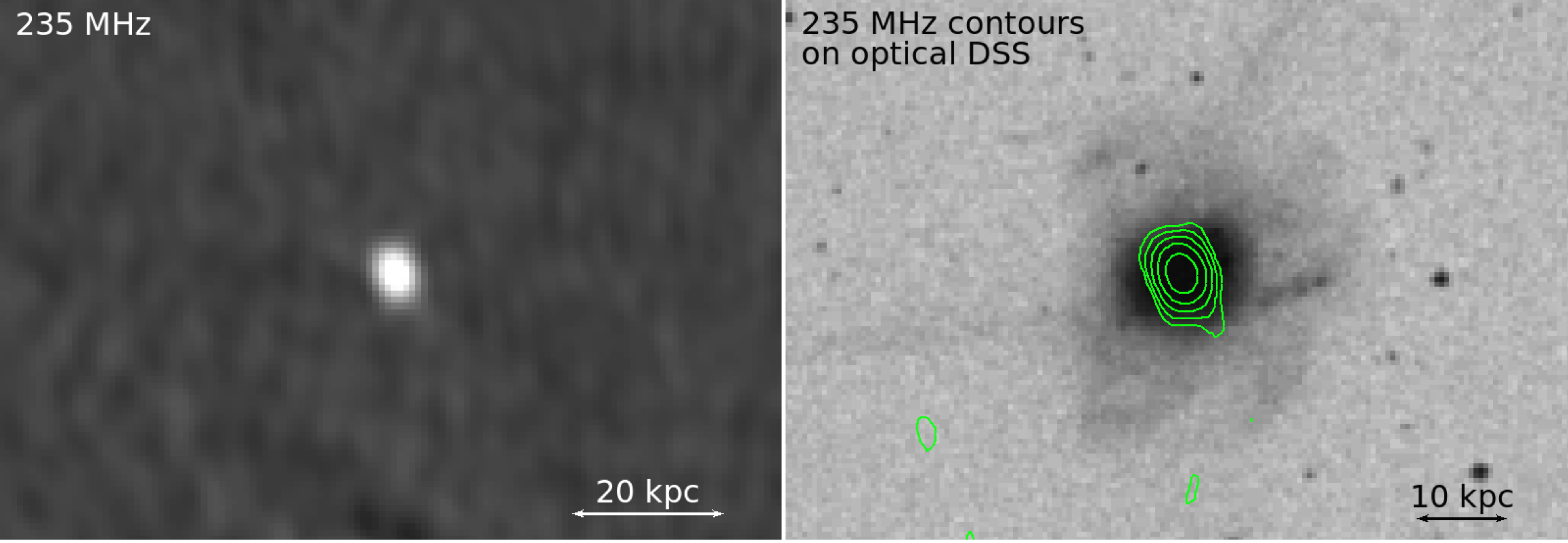}
\includegraphics[width=1.00\textwidth]{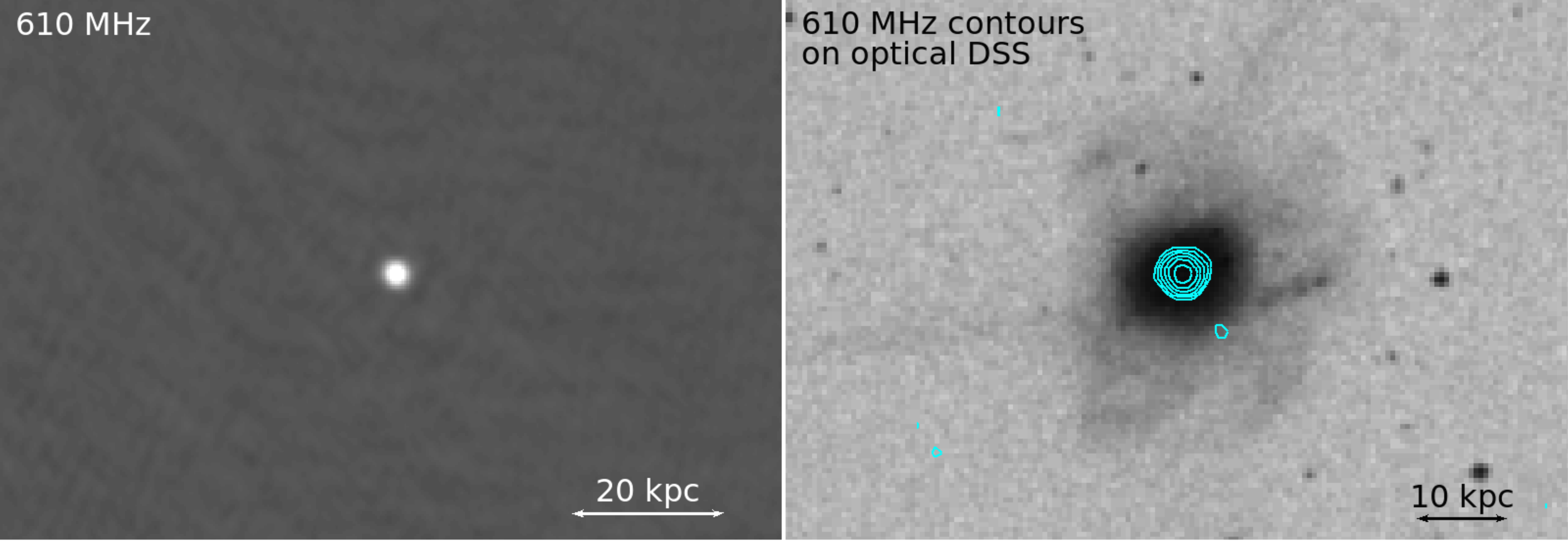}
\caption{LGG 457 / NGC 7252. \textit{Top left:} GMRT 235~MHz image. \textit{Top right:} GMRT 235~MHz contours in green (1$\sigma$ = 0.60 mJy beam$^{-1}$), overlaid on the \textit{Digitized Sky Survey (DSS)} optical image. \textit{Bottom left:} GMRT 610~MHz image. \textit{Bottom right:} GMRT 610~MHz contours in cyan (1$\sigma$ = 100 $\mu$Jy beam$^{-1}$), overlaid on the \textit{Digitized Sky Survey (DSS)} optical image. In both panels the radio contours are spaced by a factor of two, starting from 3$\sigma$ level of significance. For this source the scale is 0.320 kpc arcsec$^{-1}$.}
\label{fig:7252}
\end{figure*}

NVSS detected a point-like radio source with a flux density of 25.3$\pm$1.2~mJy \citep{Condon98}. Our images at 610~MHz and 235~MHz (see figure~\ref{fig:7252}) show a compact radio point-like source at both GMRT frequencies that coincides with the optical nucleus.

\subsection{LGG 463 / NGC 7377} 
\label{AppA25}

\begin{figure*}
\vspace{3cm}
\centering
\includegraphics[width=1.00\textwidth]{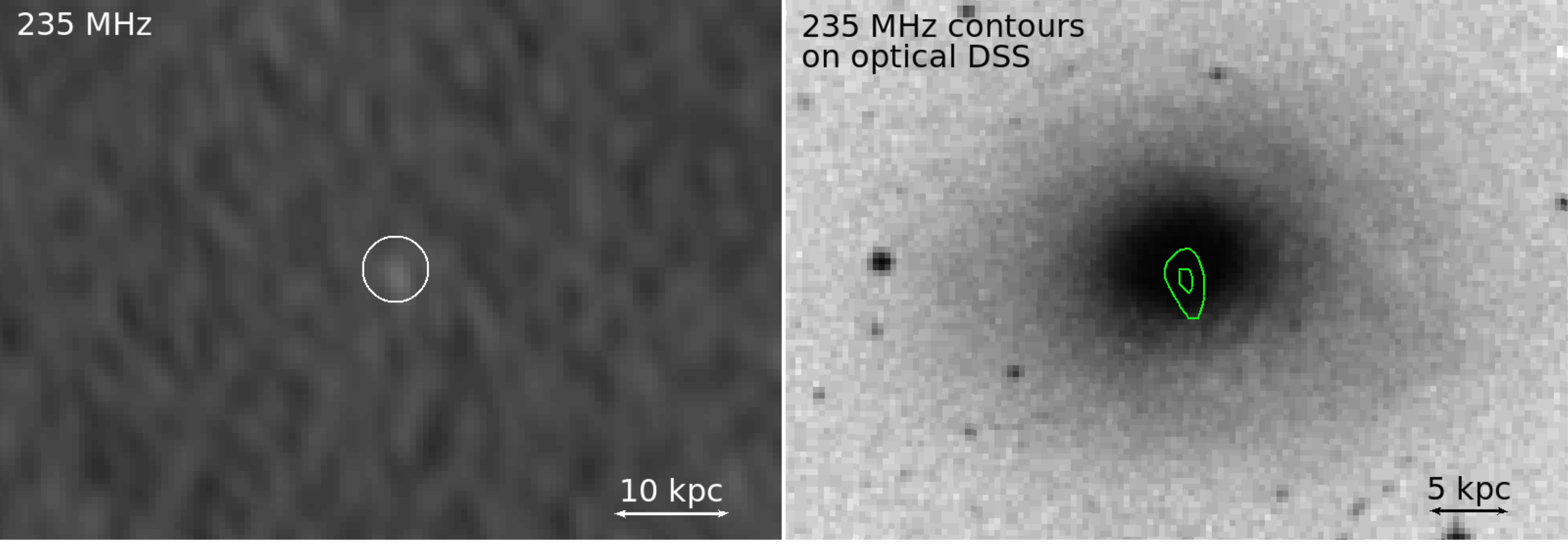}
\includegraphics[width=1.00\textwidth]{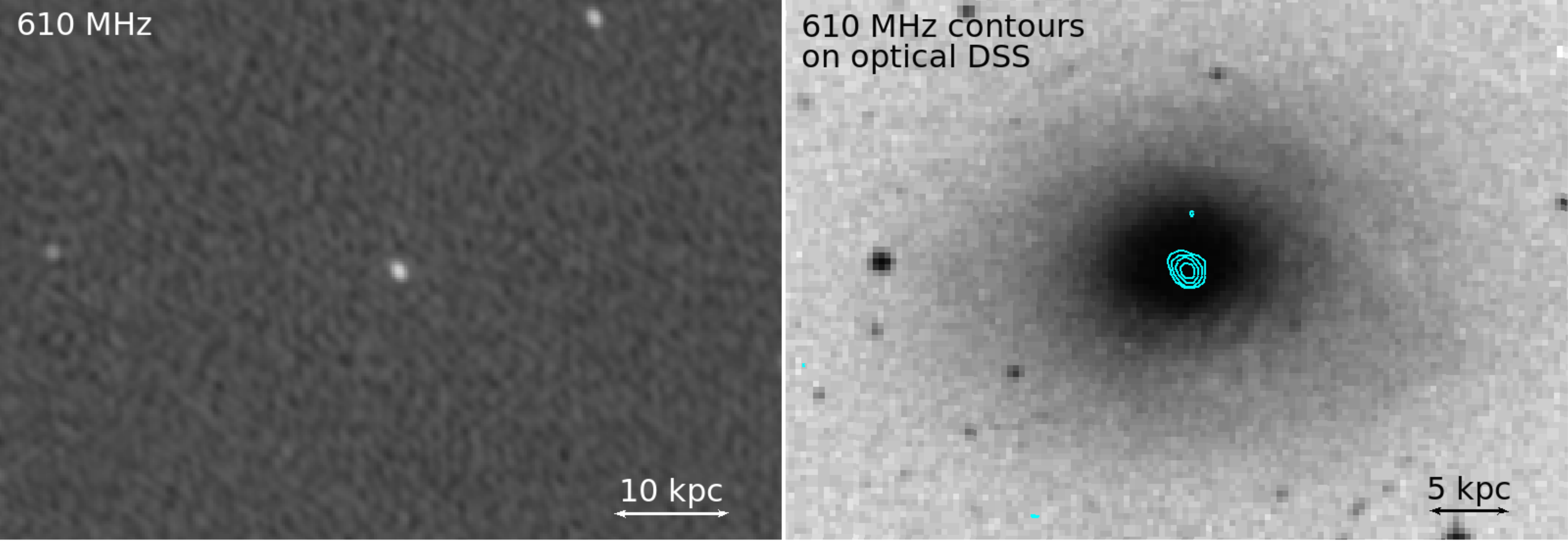}
\caption{LGG 463 / NGC 7377. \textit{Top left:} GMRT 235~MHz image. \textit{Top right:} GMRT 235~MHz contours in green (1$\sigma$ = 0.40 mJy beam$^{-1}$), overlaid on the \textit{Digitized Sky Survey (DSS)} optical image. \textit{Bottom left:} GMRT 610~MHz image. \textit{Bottom right:} GMRT 610~MHz contours in cyan (1$\sigma$ = 40 $\mu$Jy beam$^{-1}$), overlaid on the \textit{Digitized Sky Survey (DSS)} optical image. In both panels the radio contours are spaced by a factor of two, starting from 3$\sigma$ level of significance. For this source the scale is 0.223 kpc arcsec$^{-1}$.}
\label{fig:7377}
\end{figure*}

The NVSS shows NGC~7377 to host a weak radio point source with a flux density of 3.4$\pm$0.5~mJy \citep{Condon98}. In our GMRT observations we also detect a weak radio point source at both frequencies (see fig~\ref{fig:7377}). The spectral index in the 235$-$610~MHz range is $\sim$0.27 and in the 235$-$1400~MHz range appears to be inverted ($\alpha_{235}^{1400}$$\sim$-0.13; see Table~\ref{Sourcetable}).

\vspace{50mm}
\onecolumn
\section{GMRT RADIO CONTINUUM IMAGES}
\label{AppB}
In this appendix we provide GMRT radio and Digitized Sky Survey optical images for 25/27 of our BGEs analysed in this study. In all figures, the radio images are shown in the left-hand panels, and the optical images on the right. 235~MHz contours (green) are overlaid on the upper-right panels, 610~MHz (cyan) on the bottom-right. The white circle, when shown, indicates the optical position of the BGE.

\onecolumn
\section{BGE calculated properties for CLoGS sample}
\label{AppC}

{
\begin{table} 
\vspace{-1mm}
\caption{Central velocity dispersions and estimated BH masses for CLoGS BGEs. The columns list the BGE name, the central velocity dispersion $\sigma_o$, the references from which $\sigma_o$ was drawn, and the BH mass derived using the central velocity dispersion $\sigma_o$ $-$ BH mass relation from \citet{McConnellMa13}. \label{MBHsigma}}
\begin{center}
\begin{tabular}{lccc}
\hline 
 BGE & $\sigma_o$ (km s$^{-1}$)  & References (for $\sigma_o$)  &  M$_{BH\sigma}$  (10$^{9}$ M$_\odot$)\\ 
\hline
\multicolumn{4}{l}{\textsc{High--Richness Sub-sample}}\\
\hline
 NGC 193   & 197.6$\pm$4.8  & HyperLeda  & 0.23$^{+0.03}_{-0.03}$ \\
 NGC 410   & 311.0$\pm$6.0  &\citet{Loubser18}& 2.44$^{+0.84}_{-0.63}$ \\
 NGC 584   & 220.0$\pm$6.0  &\citet{Loubser18}& 0.40$^{+0.08}_{-0.06}$ \\
 NGC 677   & 228.4$\pm$20.2 & HyperLeda  & 0.49$^{+0.10}_{-0.08}$ \\
 NGC 777   & 321.0$\pm$9.0  &\citet{Loubser18}& 2.87$^{+1.04}_{-0.76}$ \\
 NGC 940   & 180.5$\pm$8.7  & HyperLeda  & 0.14$^{+0.02}_{-0.01}$ \\
 NGC 924   & 201.0$\pm$8.0  &\citet{Loubser18}& 0.25$^{+0.04}_{-0.03}$ \\
 NGC 978   & 231.8$\pm$5.3  & HyperLeda  & 0.53$^{+0.11}_{-0.09}$ \\
 NGC 1060  & 326.0$\pm$6.0  &\citet{Loubser18}& 3.11$^{+1.15}_{-0.84}$ \\
 NGC 1167  & 202.9$\pm$9.9  & HyperLeda  & 0.27$^{+0.04}_{-0.04}$ \\
 NGC 1453  & 312.0$\pm$6.0  &\citet{Loubser18}& 2.48$^{+0.86}_{-0.64}$ \\
 NGC 1587  & 239.0$\pm$6.0  &\citet{Loubser18}& 0.62$^{+0.14}_{-0.11}$ \\
 NGC 2563  & 287.0$\pm$6.0  &\citet{Loubser18}& 1.61$^{+0.49}_{-0.38}$ \\
 NGC 3078  & 243.3$\pm$6.4  & HyperLeda  & 0.68$^{+0.16}_{-0.13}$ \\
 NGC 4008  & 215.8$\pm$14.3 & HyperLeda  & 0.36$^{+0.07}_{-0.06}$ \\
 NGC 4169  & 183.5$\pm$4.7  & HyperLeda  & 0.16$^{+0.02}_{-0.02}$ \\
 NGC 4261  & 314.4$\pm$4.0  &\citet{Loubser18}& 2.56$^{+0.90}_{-0.67}$ \\
 ESO 507-25& 243.6$\pm$6.8  & HyperLeda  & 0.68$^{+0.16}_{-0.13}$ \\
 NGC 5044  & 224.9$\pm$9.1  & HyperLeda  & 0.45$^{+0.09}_{-0.07}$ \\
 NGC 5084  & 199.8$\pm$5.8  & HyperLeda  & 0.24$^{+0.04}_{-0.03}$ \\
 NGC 5153  & 182.6$\pm$5.1  & HyperLeda  & 0.15$^{+0.02}_{-0.02}$ \\
 NGC 5353  & 300.0$\pm$4.0  &\citet{Loubser18}& 2.02$^{+0.66}_{-0.50}$ \\
 NGC 5846  & 195.0$\pm$4.0  &\citet{Loubser18}& 0.22$^{+0.03}_{-0.03}$ \\
 NGC 5982  & 260.0$\pm$7.0  &\citet{Loubser18}& 0.96$^{+0.25}_{-0.20}$ \\
 NGC 6658  & 210.0$\pm$9.0  &\citet{Loubser18}& 0.32$^{+0.05}_{-0.05}$ \\
 NGC 7619  & 331.0$\pm$5.0  &\citet{Loubser18}& 3.37$^{+1.27}_{-0.92}$ \\
\hline
 \multicolumn{4}{l}{\textsc{Low--Richness Sub-sample}}\\
\hline
 NGC 128  & 216.4$\pm$14.0  & HyperLeda  & 0.37$^{+0.07}_{-0.06}$ \\
 NGC 252  & 223.6$\pm$12.4  & HyperLeda  & 0.44$^{+0.09}_{-0.07}$ \\
 NGC 315  & 339.0$\pm$8.0  &\citet{Loubser18}& 3.82$^{+1.48}_{-1.07}$ \\
 NGC 524  & 245.0$\pm$5.0  &\citet{Loubser18}& 0.71$^{+0.17}_{-0.13}$ \\
 NGC 1106 & 143.7$\pm$7.9  & HyperLeda  & 0.040$^{+0.001}_{-0.001}$ \\
 NGC 1395 & 240.1$\pm$4.2  & HyperLeda  & 0.63$^{+0.14}_{-0.12}$ \\
 NGC 1407 & 265.6$\pm$5.1  & HyperLeda  & 1.07$^{+0.29}_{-0.23}$ \\
 NGC 1550 & 300.4$\pm$5.3  & HyperLeda  & 2.04$^{+0.67}_{-0.50}$ \\
 NGC 1779 & 164.0$\pm$10.0  &\citet{Loubser18}& 0.09$^{+0.01}_{-0.01}$ \\
 NGC 2292 & 139.8$\pm$21.0  & HyperLeda  & 0.0400$^{+0.0004}_{-0.0004}$ \\
 NGC 2768 & 172.0$\pm$3.0  &\citet{Loubser18}& 0.11$^{+0.01}_{-0.01}$ \\
 NGC 2911 & 239.1$\pm$7.5  & HyperLeda  & 0.62$^{+0.14}_{-0.11}$ \\
 NGC 3325 & 168.9$\pm$12.1 & HyperLeda  & 0.10$^{+0.01}_{-0.01}$ \\
 NGC 3613 & 214.0$\pm$4.0  &\citet{Loubser18}& 0.35$^{+0.06}_{-0.05}$ \\
 NGC 3665 & 224.0$\pm$5.0  &\citet{Loubser18}& 0.44$^{+0.09}_{-0.07}$ \\
 NGC 3923 & 245.6$\pm$4.9  & HyperLeda  & 0.71$^{+0.17}_{-0.14}$ \\
 NGC 4697 & 165.2$\pm$1.6  & HyperLeda  & 0.09$^{+0.01}_{-0.01}$ \\
 NGC 4956 & 144.7$\pm$24.7 & HyperLeda  & 0.050$^{+0.001}_{-0.001}$ \\
 NGC 5061 & 188.0$\pm$4.8  & HyperLeda  & 0.18$^{+0.02}_{-0.02}$ \\
 NGC 5127 & 194.0$\pm$5.0  &\citet{Loubser18}& 0.21$^{+0.03}_{-0.03}$ \\
 NGC 5322 & 230.0$\pm$3.8  & HyperLeda  & 0.51$^{+0.11}_{-0.09}$ \\
 NGC 5444 & 220.7$\pm$33.1  & HyperLeda  & 0.41$^{+0.08}_{-0.07}$ \\
 NGC 5490 & 353.0$\pm$6.0  &\citet{Loubser18}& 4.71$^{+1.93}_{-1.37}$ \\
 NGC 5629 & 257.0$\pm$8.0  &\citet{Loubser18}& 0.93$^{+0.23}_{-0.18}$ \\
 NGC 5903 & 203.6$\pm$2.8  & HyperLeda  & 0.27$^{+0.04}_{-0.04}$ \\
 NGC 7252 & 156.8$\pm$22.0  & HyperLeda  & 0.070$^{+0.001}_{-0.001}$ \\
 NGC 7377 & 174.5$\pm$12.1  & HyperLeda  & 0.12$^{+0.01}_{-0.01}$ \\

\hline
\end{tabular}
\end{center}
\end{table}}

\vspace{8mm}

\bsp	
\label{lastpage}
\end{document}